\documentclass[sigconf]{acmart}

\usepackage{tikz}
\usepackage{amsmath}

\usepackage{amssymb}
\usepackage{multirow}
\usepackage{graphicx}

\usepackage[linesnumbered,ruled,vlined]{algorithm2e}

\newcommand{\stitle}[1]{\vspace*{0.4em}\noindent{\bf #1.\/}}
\newcommand{\sstitle}[1]{\vspace*{0.4em}\noindent{\bf #1:\/}}
\newcommand{\trim}{\vspace{-2mm}}

\newcommand{\squishlist}{
 \begin{list}{$\bullet$}
 { \setlength{\itemsep}{0pt}
   \setlength{\parsep}{3pt}
   \setlength{\topsep}{3pt}
   \setlength{\partopsep}{0pt}
   \setlength{\leftmargin}{1.2em}
   \setlength{\labelwidth}{1em}
   \setlength{\labelsep}{0.6em}
 }
}
\newcommand{\squishend}{
 \end{list}
}

\newcommand{\rvsn}[1]{{\color{black} #1}}

\newcommand{\sysname}{\textsf{AlayaLaser}}

\newcommand*\circled[1]{\tikz[baseline=(char.base)]{\node[shape=circle,draw,inner sep=0pt, minimum size=8pt, line width=0.3pt] (char) {\fontsize{7}{0}\selectfont{#1}};}}

\begin{document}
\fancyhead{}

\title{AlayaLaser: Efficient Index Layout and Search Strategy for Large-scale High-dimensional Vector Similarity Search}

\def\UrlFont{\fontsize{8.9pt}{10pt}\sffamily\selectfont}

\author{Weijian Chen}
\affiliation{%
  \institution{SUSTech}
  \country{}
  \def\UrlFont{\fontsize{8.9pt}{10pt}\sffamily\selectfont}
}
\email{chenwj2024@mail.sustech.edu.cn}

\author{Haotian Liu}
\affiliation{%
  \institution{AlayaDB AI}
  \country{}
  \def\UrlFont{\fontsize{8.9pt}{10pt}\sffamily\selectfont}
}
\email{haotian.liu@alayadb.ai}

\author{Yangshen Deng}
\affiliation{%
  \institution{University of Edinburgh}
  \country{}
  \def\UrlFont{\fontsize{8.9pt}{10pt}\sffamily\selectfont}
}
\email{yangshen.deng@ed.ac.uk}

\author{Long Xiang}
\affiliation{%
  \institution{AlayaDB AI}
  \country{}
  \def\UrlFont{\fontsize{8.9pt}{10pt}\sffamily\selectfont}
}
\email{long.xiang@alayadb.ai}

\author{Liang Huang}
\affiliation{%
  \institution{SUSTech}
   \country{}
   \def\UrlFont{\fontsize{8.9pt}{10pt}\sffamily\selectfont}
}
\email{huangl2025@mail.sustech.edu.cn}


\author{Bo Tang}
\authornote{Dr. Bo Tang is the corresponding author.}
\affiliation{%
  \institution{SUSTech}
   \country{}
   \def\UrlFont{\fontsize{8.9pt}{10pt}\sffamily\selectfont}
}
\email{tangb3@sustech.edu.cn}

\def\UrlFont{\sffamily}

\begin{abstract}

\rvsn{On-disk graph-based approximate nearest neighbor search (ANNS) is essential for large-scale, high-dimensional vector retrieval, yet its performance is widely recognized to be limited by the prohibitive I/O costs. 
Interestingly, we observed that the performance of on-disk graph-based index systems is compute-bound, not I/O-bound, with the rising of the vector data dimensionality (e.g., hundreds or thousands).
This insight uncovers a significant optimization opportunity: existing on-disk graph-based index systems universally target I/O reduction and largely overlook computational overhead, which leaves a substantial performance improvement space.

In this work, we propose \sysname{}, an efficient on-disk graph-based index system for large-scale high-dimensional vector similarity search.
In particular, we first conduct performance analysis on existing on-disk graph-based index systems via the adapted roofline model,
then we devise a novel on-disk data layout in \sysname{} to effectively alleviate the compute-bound, which is revealed by the above roofline model analysis, by exploiting SIMD instructions on modern CPUs. 
We next design a suite of optimization techniques (e.g., degree-based node cache, cluster-based entry point selection, and early dispatch strategy) to further improve the performance of \sysname{}.
We last conduct extensive experimental studies on a wide range of large-scale high-dimensional vector datasets to verify the superiority of \sysname{}.
Specifically, \sysname{} not only surpasses existing on-disk graph-based index systems but also matches or even exceeds the performance of in-memory index systems.
}
\end{abstract}

\maketitle

\section{Introduction}\label{sec:intro}

Large-scale, high-dimensional vector similarity search is a core subroutine in many applications, including recommendation systems~\cite{Lv2019SDM, Huang2013Learning, Covington2016Deep, wu2024survey}, computer vision~\cite{gordo2016deep, o2015introduction, babenko2014neural}, and Retrieval-Augmented Generation (RAG)~\cite{lewis2020retrieval, yu2024rankrag, Fan2024Survey, shao-etal-2023-enhancing} in Large Language Models.
Due to the curse of dimensionality, exact search becomes prohibitively expensive in high-dimensional spaces.
Consequently, approximate nearest neighbor search (ANNS) has emerged as the standard solution~\cite{Li2020Approximate, Wang2021comprehensive, gao2023high, simhadri2022results}.
The performance of ANNS solutions is typically assessed along two complementary dimensions: (i) search efficiency, measured by latency and throughput (queries per second, QPS), and (ii) search accuracy, commonly quantified by recall. 
A wide range of techniques has been proposed to optimize this trade-off, including hashing-based, cluster-based, and graph-based indexing methods. 
Among these, in-memory graph-based index methods have emerged as the state-of-the-art approach, consistently achieving superior recall while maintaining low latency and high throughput~\cite{fu2017fast, gou2025symphonyqg, aumuller2020ann, aguerrebere2023similarity, malkov2018efficient, fu2016efanna, fu2021high, munoz2019hierarchical, azizi2023elpis}.
However, the rapid growth of vector datasets has exposed a critical bottleneck: the memory demands of in-memory graph-based index far exceed the capacity of a single machine.
For example, SymphonyQG~\cite{gou2025symphonyqg}, despite its high search performance, encounters a hard memory constraint: it is unable to build its graph index for a 10.1 million 764-dimensional vector dataset (28.96 GB) within a single machine with 128 GB memory, as we will verify in Section~\ref{sec:exp}.

\begin{figure}
    \small
    \centering
    \includegraphics[width=0.99\linewidth]{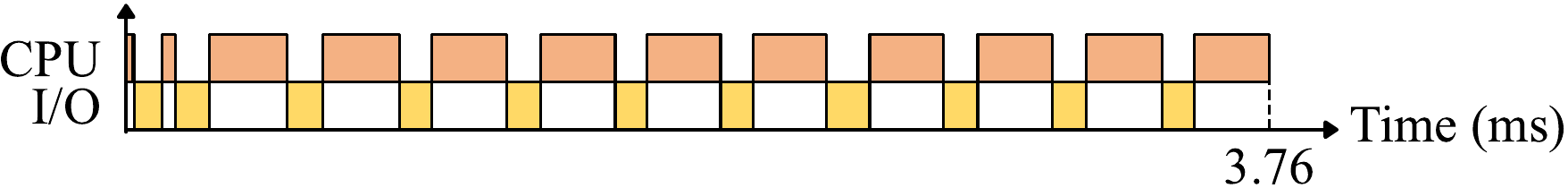}
    \\
    (a) DiskANN~\cite{jayaram2019diskann}
   \\
   \includegraphics[width=0.99\linewidth]{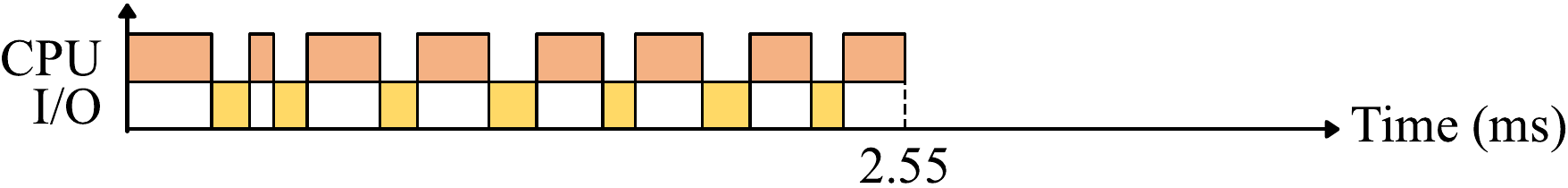}
   \\
   (b) Starling~\cite{wang2024starling}
    \\
    \includegraphics[width=0.99\linewidth]{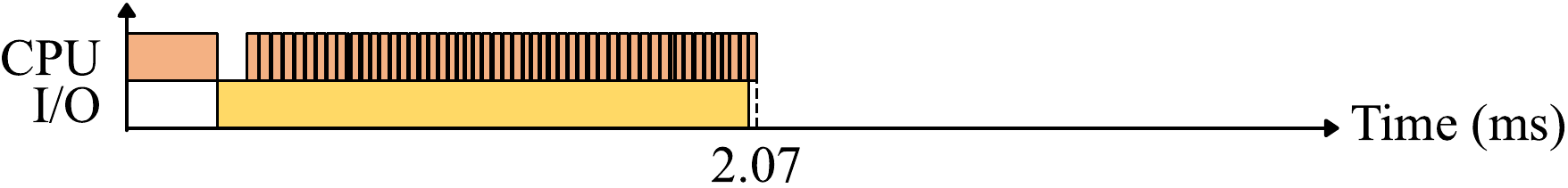}
    \\
    (c) PipeANN~\cite{guo2025achieving}
    \\
    \includegraphics[width=0.99\linewidth]{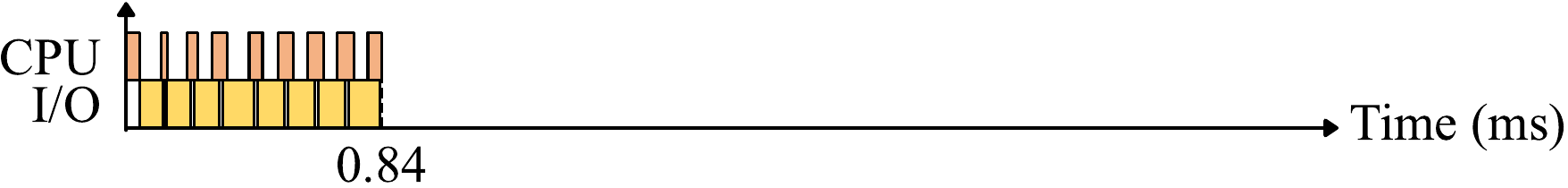}
    \\
    (d) \sysname{} (this work)
    \\
    \caption{Query latency profiling of on-disk graph-based index systems, Recall@10=0.90 on 960-dimensional GIST1M}
    \label{fig:cpu_io_behaviors}
    \trim 
\end{figure}

\rvsn{
Intuitively, on-disk graph-based index systems~\cite{jayaram2019diskann, wang2024starling, guo2025achieving} have been proposed to overcome the limitation of the prohibitive memory demands of in-memory graph-based indexes for efficient ANNS on large-scale high-dimensional vector dataset.
The general idea of existing on-disk graph-based index systems is that they store the underlying graph-based index structure to cost-effective Solid-State Drives (SSDs), which provide high I/O throughput and low random read latency~\cite{haas2020exploiting, Haas2023Storage}. 
In particular, DiskANN~\cite{jayaram2019diskann} is the first on-disk graph-based index system.
It introduces a new on-disk graph-based ANN index \textsf{Vamana}, which enables DiskANN to minimize the number of disk I/O reads. 
Moreover, \textsf{Vamana} can be combined with vector compression schemes (e.g., product quantization), where the compressed vectors are cached in memory for fast and approximate distance computation.
It is now well-established that I/O latency constitutes the fundamental performance bottleneck in DiskANN as graph traversal invokes expensive disk I/O accesses~\cite{wang2024starling, guo2025achieving}.
We profile the query latency of DiskANN on 960-dimensional GIST1M by setting Recall@10=0.90.
Figure~\ref{fig:cpu_io_behaviors}(a) shows the corresponding CPU computation costs and I/O costs from one of the profiled queries.
To reduce the I/O costs, Starling~\cite{wang2024starling} modifies the data layout of DiskANN via ``block-shuffling'' to improve on-disk data locality, and employs an in-memory navigation graph to shorten search paths.
Comparing the profiling result of DiskANN in Figure~\ref{fig:cpu_io_behaviors}(a), Starling successfully reduced the I/O costs to process the same query, see I/O cost in Figure~\ref{fig:cpu_io_behaviors}(b).
PipeANN~\cite{guo2025achieving}, in contrast, redesigns the search strategy in DiskANN by replacing the synchronous beam search with an asynchronous pipeline, which overlaps CPU computation cost and I/O cost, then hides I/O latency, see Figure~\ref{fig:cpu_io_behaviors}(c). 
}

\rvsn{Interestingly, as shown in Figures~\ref{fig:cpu_io_behaviors}(a), (b) and (c), we observed that the CPU costs contributed to a large proportion of the query latency in all existing on-disk graph-based index systems (i.e., DiskANN, Starling, and PipeANN).
We further systematically validate the above observation with an adapted roofline model. 
Contrary to current wisdom, i.e., the performance of on-disk graph-based index systems is limited by I/O costs, our analyzed results from the roofline model reveal the performance of these systems changes from \textsf{I/O-bound} to \textsf{compute-bound} with the rising of the vector dimensionality.
For example, the performance of DiskANN, Starling and PipeANN is actually I/O-bound on 96-dimensional DEEP10M, but it turned to compute-bound on 960-dimensional GIST1M, we will elaborate the details shortly in Section~\ref{sec:analysis}.}

\rvsn{
However, it is not trivial to overcome the compute-bound of existing on-disk graph-based index systems.
The technical challenges are from two-fold: (i) sequential computation pattern, and (ii) precision-storage dilemma.
In particular, the fast and approximate distance computation of DiskANN~\cite{jayaram2019diskann} relies on a sequential, step-by-step lookup table (LUT) accesses. Both Starling and PipeANN utilize the same mechanism of DiskANN.
However, this computation procedure cannot be effectively parallelized via SIMD instructions in modern CPUs.
Inspired by existing in-memory SIMD-optimized graph-based index systems (e.g., SymphonyQG~\cite{gou2025symphonyqg}), a straight-forward way to exploit SIMD instruction in modern CPUs is appending the quantization compressed vectors to the end of the data layout in DiskANN.
Unfortunately, it brings precision-storage dilemma, i.e., there is a tradeoff between the precision of compressed vector and the storage cost.
Specifically, low precision quantization compressed vectors consume less space but lose search accuracy, and high precision ones maintain good search accuracy but consume more space, see the detailed analysis in Section~\ref{sec:challenges}.

In this work, we propose \sysname{}, which consists of a novel index \textsf{\underline{La}}yout and an optimized \textsf{\underline{Se}}a\textsf{\underline{r}}ch strategy, for efficient ANNS.
In particular, a novel \textsf{SIMD-friendly layout} is devised in \sysname{} to address the above two technical challenges.
Each data vector will divide into two components: principal component and residual component via Principal Component Analysis (PCA). 
Only the quantization compressed vectors (we also refer it as quantized codewords) of principal component will be appended into the data layout, which addresses the precision-storage dilemma. 
Furthermore, the quantized codewords of all neighbors for each data vector are arranged in an interleaved fashion, facilitating efficient utilization of SIMD registers.
Interestingly, the SIMD-friendly layout in \sysname{} not only alleviates the compute-bound of on-disk graph-based index systems, but also unlocks a novel and unique opportunity for us to improve the search performance as the quantized codewords of data vectors do not need to store in memory cache any more and the available memory budget can be further utilized.
Hence, we propose in-degree based node cache scheme to fully utilize the memory budget in \sysname{}, which effectively reduces the number of I/O accesses and overlaps CPU computation costs and I/O costs, simultaneously.
As shown in Figure~\ref{fig:cpu_io_behaviors}(d), the CPU costs of \sysname{} are significantly reduced due to the effectiveness of the new SIMD-friendly layout.

We further accelerate the performance of \sysname{} by optimizing its search strategy.
First, we design cluster-based entry point selection method to shorten search paths.
Second, an adaptive beam expansion strategy is employed to avoid wasteful node exploration during graph traversal. 
Third, we identify the root cause of the long-tail I/O latency of modern SSDs, which is overlooked in existing work,
and devise an asynchronous early dispatch mechanism to handle slow I/O requests.
The efficiency of the optimized search strategy in \sysname{} is confirmed by the profiling result in Figure~\ref{fig:cpu_io_behaviors}(d).
As depicted in Figures~\ref{fig:exp-intro-latency}(a) and (b), 
the Recall@10-latency curve of our \sysname{} not only outperforms the representative on-disk index system DiskANN~\cite{jayaram2019diskann}, but also surpasses the widely-used in-memory index HNSWlib~\cite{malkov2018efficient} on both GIST1M and Cohere, respectively.

\begin{figure}
\small
\centering
\begin{tabular}{cc}
    \includegraphics[width=0.45\columnwidth]{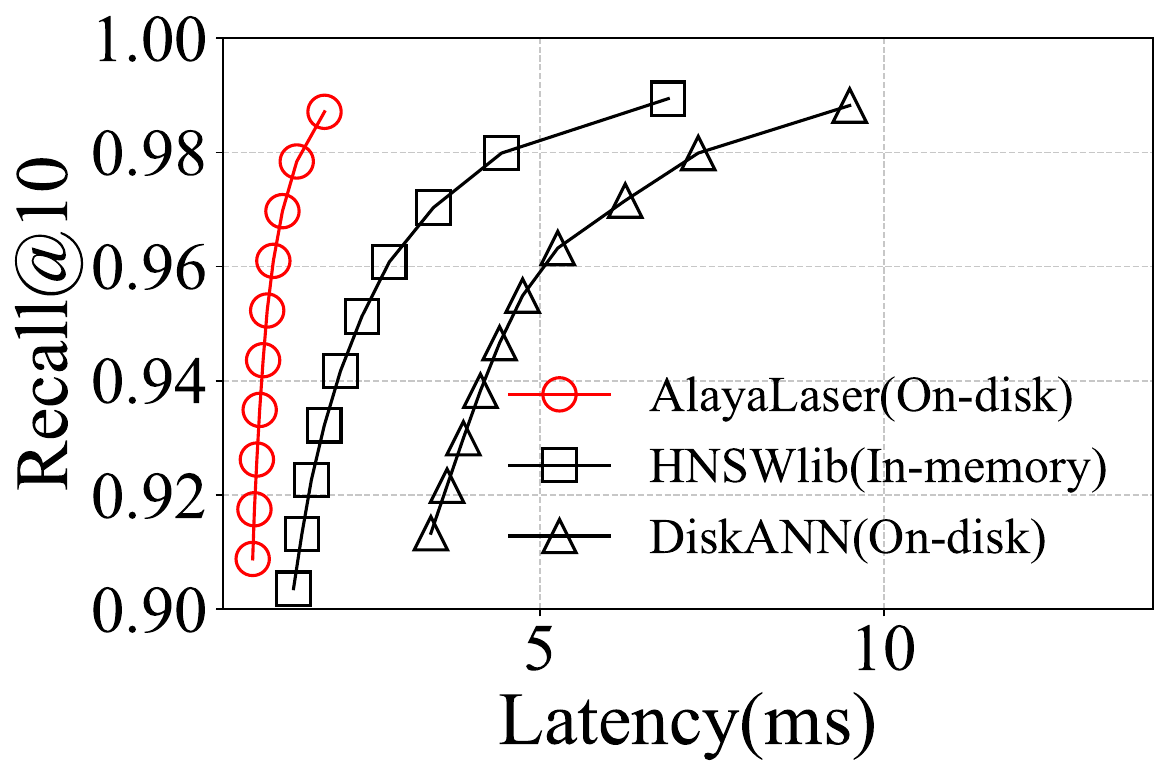} &
    \includegraphics[width=0.45\columnwidth]{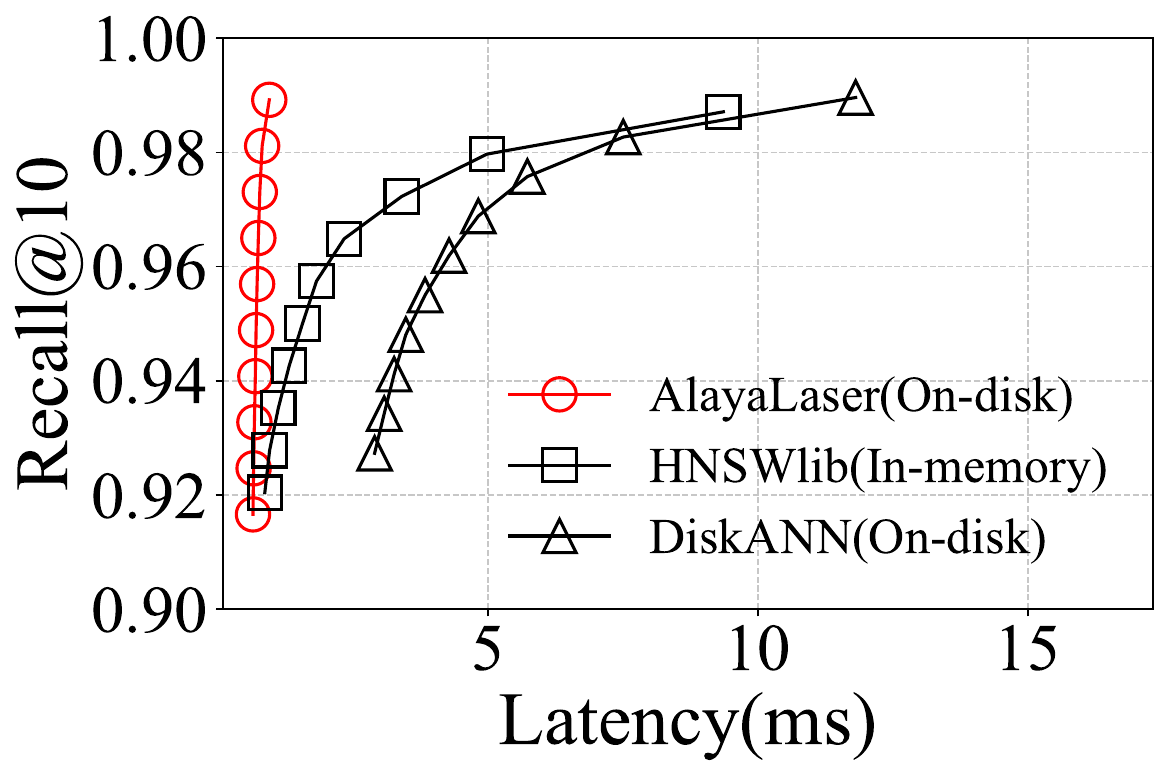}
    \\
    (a) GIST1M (960D) & (b) Cohere (768D)
\end{tabular}
\caption{Overall performance evaluation of \sysname{}} \label{fig:exp-intro-latency}
\trim 
\end{figure}

The major contributions of this work are summarized as follows:

\squishlist
    \item We reveal a fundamental paradigm shift in the performance characteristics of on-disk graph-based index systems. Contrary to the prevailing I/O-centric view, our adapted roofline model verifies that CPU computation, not disk I/O latency, has become the dominant bottleneck for large-scale, high-dimensional ANNS, a finding previously unrecognized in the literature.
    
    \item We address the emergent CPU bottleneck through a novel SIMD-friendly index layout in \sysname{}. It not only resolves the technical challenges in on-disk graph-based index systems, but also enables an efficient in-degree node cache scheme.
    
    \item We design a suite of optimization techniques (e.g., cluster-based entry point selection, adaptive beam expansion, and early dispatch mechanism) on top of SIMD-friendly layout in \sysname{} to further improve the search performance.
    
    \item We conduct comprehensive experiments to evaluate the superiority of \sysname{}, which is substantially outperforms state-of-the-art on-disk graph-based index systems. Counterintuitively, the performance of \sysname{} is comparable to or even slightly better than the leading in-memory ANNS systems (e.g., HNSWlib).
\squishend
}

The remainder of this paper is organized as follows. 
Section~\ref{sec:preliminary} introduces the preliminaries and analyzes the performance bottleneck via the adapted roofline model. 
Section~\ref{sec:layout} presents the novel SIMD-friendly index layout of \sysname{} and in-degree based node cache scheme. 
Section~\ref{sec:search} elaborates on the optimizations of the search strategy in \sysname{}. 
Section~\ref{sec:exp} conducts extensive experimental evaluations to verify the effectiveness of \sysname{}.
Sections~\ref{sec:related_work} summarize the related work,
and Section~\ref{sec:conclusion} makes a conclusion and highlights the promising future work directions.

\section{Preliminaries and Analysis}\label{sec:preliminary}
Let $\mathcal{D}$ be a dataset of $d$-dimensional vectors.
For a query vector $q$, vector similarity search identifies $k$ vectors in $\mathcal{D}$ with the smallest Euclidean distances to $q$.
In this work, we design an on-disk, graph-based index system that enables efficient ANNS on large-scale high-dimensional vector dataset $\mathcal{D}$ with a limited memory budget $B$.
Specifically, the size of vector dataset $\mathcal{D}$ vastly exceeds the available memory budget $B$.
The objective of this work is to maximize search performance (i.e., high throughput and low latency), while maintaining good result accuracy (i.e., high recall) for high-dimensional vectors (e.g., those with hundreds or thousands of dimensions).
In this section, we first present the general idea of existing on-disk graph-based index systems for high-dimensional vector similarity search in Section~\ref{sec:existingwork}, then identify the performance bottleneck of these systems on two vector datasets with different dimensionalities via adapted roofline model in Section~\ref{sec:analysis}.

\subsection{Existing On-disk Graph-based Indexes}\label{sec:existingwork}
DiskANN~\cite{jayaram2019diskann} is the first, also the \emph{de facto} standard, on-disk graph-based index system that enables efficient ANNS over large-scale high-dimensional vector datasets.
DiskANN consists of (i) the new graph-based ANNS index structure \textsf{Vamana}, and (ii) the off-the-shelf vector compression schemes (e.g., product quantization).

\rvsn{
\stitle{Data layout of DiskANN} 
DiskANN first constructs the new on-disk graph-based index Vamana, which initializes the index graph by randomly adding $R$ out-neighbors for each data vector, then prunes and reconnects the edges among these vectors to ensure the quality of the graph-based index. 
The constructed full graph-based index along with the full-precision raw vectors are stored on SSD disk, as shown in Figure~\ref{fig:diskann}.
DiskANN then uses product quantization (PQ) to compress raw vector and caches compressed vectors (a.k.a. quantized codewords) in memory.
Taking the vectors in Figure~\ref{fig:diskann} as an example, the 8-dimensional data vector space is divided into four 2-dimensional sub-vector spaces, we denote them as a, b, c, and d.
For each sub-vector space, PQ clusters the corresponding 2-dimensional data vectors into 4 clusters by running 4-Means algorithm.
Thus, each full precision data vector can be compressed into 8 bits quantized codewords via the PQ compression scheme.
For example, the compressed vector 2 in memory is an 8-bits string ``01 10 11 00'', where the first 2-dimensional of the data vector 2 is represented by ``01'' in the sub-vector space a.

\begin{figure}
    \small
    \centering
    \includegraphics[width=0.95\columnwidth]{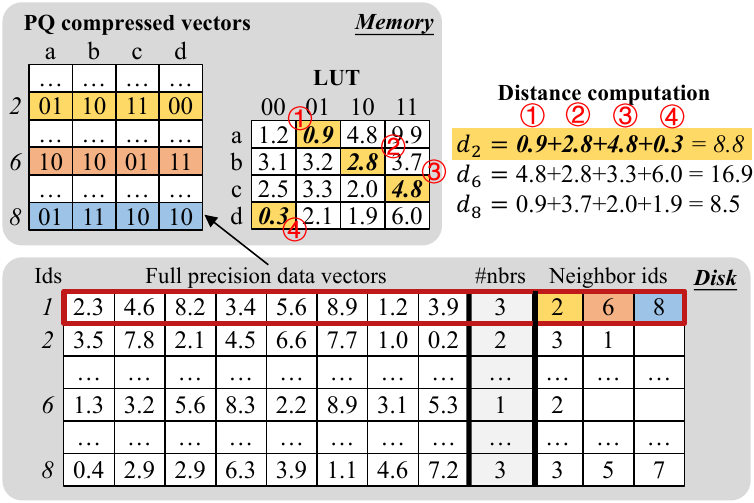}
    \trim \trim 
    \caption{Data layout and distance computation on DiskANN}
    \label{fig:diskann}
    \trim \trim 
\end{figure}
}

\stitle{Distance computation}
For a given query $q$, DiskANN generates a lookup table (LUT) in memory to store the distances of query sub-vector and the cluster centroids of every data sub-vector space, see the in-memory LUT in Figure~\ref{fig:diskann}.
The LUT can be used to obtain the approximate distance between query vector and candidate data vectors and guide the best vectors (and neighbors) to read from disk.
For example, the approximate distance between query $q$ and data vector $2$ can be computed by summing the values which are obtained by accessing \circled{1} to \circled{4} of LUT.
The approximate distance can be refined by computing the exact distance between the query vector and full precision data vectors, and the final ANNS results can be returned by re-ranking these candidates with exact distances.

\stitle{Beam search strategy}
To reduce the number of I/Os, DiskANN devises beam search strategy, which fetches the neighborhoods of a small number of the closest points (a.k.a., beam width $BW$) in one shot as reading a few random sectors from an SSD takes almost the same amount of time as reading one sector.

Recently, several work~\cite{wang2024starling,guo2025achieving} are built upon DiskANN to improve the search performance.
In particular, {Starling}~\cite{wang2024starling} introduced a block-shuffling technique to improve on-disk data locality.
Moreover, it enhances the search strategy of DiskANN by utilizing an in-memory navigation graph to shorten search paths.
The core idea of {PipeANN}~\cite{guo2025achieving} is replacing the synchronous beam search strategy by an asynchronous pipeline, which overlaps the computation cost and I/O cost, then hides the I/O latency.

\rvsn{
\subsection{Roofline Model Analysis}\label{sec:analysis}
The roofline model~\cite{10.1145/1498765.1498785} typically used to identify performance bottleneck of a system on a given hardware platform, e.g., whether it is \textsf{compute-bound} (limited by processor speed) or \textsf{memory-bound} (limited by memory bandwidth).
Since we focus on on-disk graph-based index systems for ANNS, we adapt the classic roofline model to analyze whether the system is \textsf{compute-bound} or \textsf{I/O-bound}.
In particular, in the adapted roofline model, the x-axis is the ratio of floating-point operations performed per byte read from the SSD (in FLOP/byte). 
The y-axis is the same as the classic roofline model, which shows the performance of the processor in GFLOP/s.
Our experimental studies are conducted on a machine with a 24-core/48-thread CPU (Intel Xeon Gold 5318Y @ 2.10GHz) and NVMe SSDs (SAMUSUNG PM9A3).
We evaluate the performance of three on-disk graph-based index systems for ANNS (i.e., DiskANN, Starling, and PipeANN) on two widely-used vector datasets: the 96-dimensional DEEP10M~\cite{7780595} and the 960-dimensional GIST1M~\cite{TexMexCorpus}.
Without loss of generality, all I/O reads in these systems are bypassing the effect of operating system caching.

\begin{figure}
   \small
   \centering
   \includegraphics[width=0.96\columnwidth]{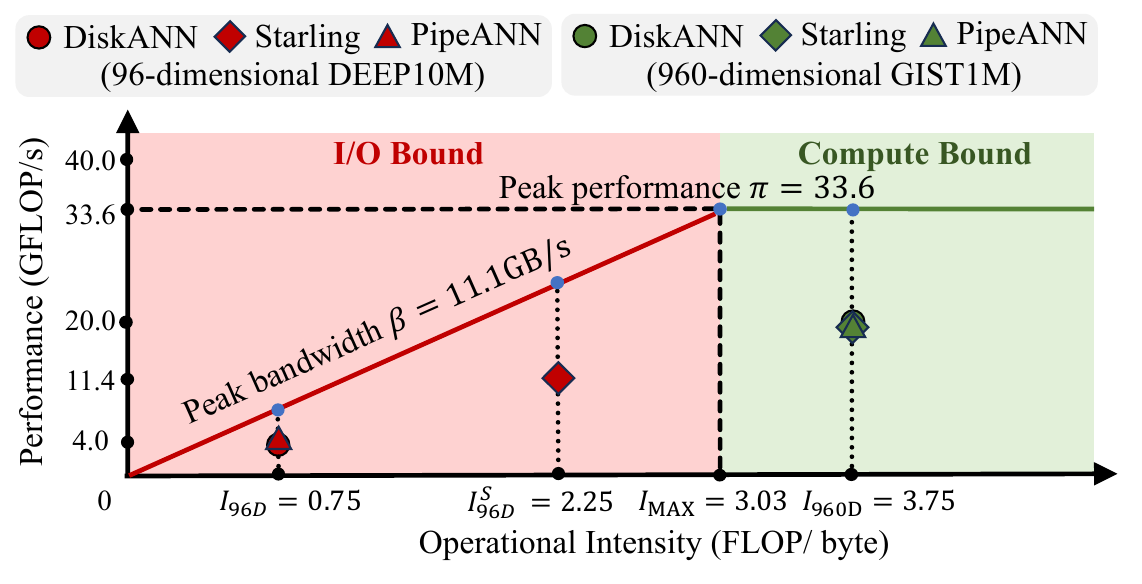}
   \trim 
   \caption{Roofline model analysis}    \label{fig:roofline_model}
   \trim 
\end{figure}

\stitle{Roofline model formulation}
In the roofline model, there are two parameters: (i) the peak performance $\pi$ and (ii) the peak bandwidth $\beta$ of the given NVMe SSD, and one variable: the operational intensity $\mathcal{I}$.
We first analyze the peak performance of these three evaluated on-disk graph-based index systems on the tested machine.
In particular, we derive it by investigating the operations of approximate distance computation in memory.
As shown in Figure~\ref{fig:diskann}, the approximate distance between the given query $q$ and each data vector $i$ is computed by three steps:
(i) reading PQ compressed bits in each sub-vector, (ii) looking up the corresponding distance from LUT, and (iii) accumulating these distance to the estimated approximate distance $d_i$.
Suppose both the compressed vectors and LUT are in L1 cache, the total latency is 3 clock cycles (i.e., 2 L1 loads and 1 FP add).
Thus, the peak throughput per thread is  2.1 GHz / 3 cycles = 0.7 GFLOP/s as the CPU frequency is 2.1 GHz.
Consequently, the peak performance for the 24-core/48-thread CPU is $\pi$ = 0.7 GFLOP/s × 48 =  33.6 GFLOP/s, as the compute roof shown in Figure~\ref{fig:roofline_model}.
We next derive the \textit{maximum operational intensity} $\mathcal{I}_{max}$ in the roofline model.
The maximum I/O bandwidth of the used NVMe SSDs is the peak bandwidth $\beta$ = 11.1 GB/s. 
The ridge point (where the diagonal and horizontal roof meet) defines the \textit{maximum operational intensity} $\mathcal{I}_{max}$ =  $\pi / \beta$ = 33.6 GFLOP/s / 11.1 GB/s $\approx$ 3.03 FLOP/byte, see $\mathcal{I}_{max}$ in the x-axis of Figure~\ref{fig:roofline_model}.
Thus, if the operational intensity of the evaluated system $\mathcal{I} < \mathcal{I}_{max}$, it is I/O-bound, see the left part in Figure~\ref{fig:roofline_model}.
Conversely, if $\mathcal{I} > \mathcal{I}_{max}$, it is compute-bound, as the right part shown in Figure~\ref{fig:roofline_model}.

\stitle{Operational intensity of on-disk graph-based indexes}
We then analyze the operational intensity of existing on-disk graph-based index systems (i.e., DiskANN, Starling, PipeANN) on different datasets with the given hardware configuration on the used machine.
In fact, the operational intensity $\mathcal{I}$ of an on-disk graph-based index system is determined by:
$\mathcal{I} = \frac{\text{\# of FLOPs per disk page}}{\text{disk page size}}.$

\sstitle{DEEP10M} For 96-dimensional DEEP10M, each data vector is divided into 48 sub-vectors and each has at most 64 neighbors as out-neighbor $R$ is 64.
Processing a 4KB disk page in DiskANN and PipeANN for a node involves 64 × 48 = 3,072 FLOPs.
Thus, the arithmetic intensity of DiskANN and PipeANN on 96-dimensional DEEP10M is $\mathcal{I}_{96D}$ = 3,072 FLOPs / 4096 bytes = 0.75 FLOP/byte, see $\mathcal{I}_{96D}$ at x-axis of Figure~\ref{fig:roofline_model}.
The block-shuffling technique in Starling~\cite{wang2024starling} enables it processes 3 nodes per disk page on DEEP10M.
Thus, the arithmetic intensity of Starling on DEEP10M is  $\mathcal{I}_{96D}^{S}$ = 3,072 x 3 FLOPs / 4096 bytes = 2.25 FLOP/byte, as shown in Figure~\ref{fig:roofline_model}.
All three on-disk graph-based index systems are I/O-bound at 96-dimensional DEEP10M as both $\mathcal{I}_{96D} < \mathcal{I}_{max}$ and $\mathcal{I}_{96D}^{S} < \mathcal{I}_{max}$.

\sstitle{GIST1M} For 960-dimensional GIST1M, each data vector is divided into 480 sub-vectors and each has at most 64 neighbors.
Processing an 8KB disk page for a node and its neighbors requires 64 × 480 = 30,720 FLOPs.
Thus, the arithmetic intensity of all three solutions (i.e., DiskANN, Starling, and PipeANN) on 960-dimensional GIST1M is $\mathcal{I}_{960D}$ = 30,720 FLOPs / 8192 bytes $\approx$ 3.75 FLOP/Byte.
The reason why the arithmetic intensity of Starling is the same as its of DiskANN and PipeANN on GIST1M as each 8KB disk page only can store 1 data node with its neighbors in all these three on-disk graph-based index systems.
All these solutions are compute-bound at 960-dimensional GIST1M as $\mathcal{I}_{960D} > \mathcal{I}_{max}$.

\stitle{Achieved performance evaluation}
We measure the achieved performance of these three on-disk graph-based index systems on two vector datasets with the used machine.
We use all 48 available threads to maximize the system utilization.
As depicted in Figure~\ref{fig:roofline_model}, the achieved performance of DiskANN, Starling and PipeANN on 96-dimensional DEEP10M are 3.96 GFLOP/s, 11.37 GFLOP/s and 5.09 GFLOP/s, respectively.
On DEEP10M, the performance of DiskANN and Starling is limited by the actual SSD bandwidth as it only achieved 6.5 GB/s, which is far below its peak bandwidth $\beta$ = 11.1 GB/s. 
The reason is that the beam search width $BW$ is configured as 8, which limits the depth of I/O queue in them~\cite{Haas2023Storage}. 
PipeANN achieves a higher performance than DiskANN by employing a deeper I/O queue (i.e., 32) in its pipeline, which utilizes the available SSD bandwidth better.
On 960-dimensional GIST1M, the achieved performance of DiskANN, Starling and PipeANN are 18.94 GFLOP/s, 18.85 GFLOP/s, and 18.28 GFLOP/s, respectively.
All of these solutions are significantly below the peak performance $\pi$ = 33.6 GFLOP/s.
The reason is that the peak performance is derived at the ideal situation, i.e., it only takes 3 cycles for each FLOP as it assumes all the data are in L1 cache.
However, the PQ compressed vector of each data vector in 960-dimensional GIST1M is impossible to entirely store on L1 cache, which means it takes more CPU cycles for each FLOP then the peak performance $\pi$ will be smaller than the maximum 33.6 GFLOP/s in real-world systems.

\begin{table}
\centering 
\small
\trim 
\caption{Summary of on-disk graph-based index systems}\label{tab:analysis_existing_studies}
\trim 
\begin{tabular}{|c|c|c|c|}
\hline
\textbf{Systems} & \textbf{Computational} & \textbf{I/O} & \textbf{Search} \\ 
\textbf{} & \textbf{Efficiency} & \textbf{Efficiency} & \textbf{Performance} \\ 
\hline\hline
DiskANN~\cite{jayaram2019diskann}            & Low  & Median & Low \\  \hline
Starling~\cite{wang2024starling}     & Low  & High   & Median \\ \hline
PipeANN ~\cite{guo2025achieving}            & Low  & High & Median \\ \hline
\sysname{} (this work)     & \textbf{High}  & \textbf{High}   & \textbf{High} \\  \hline
\end{tabular}
\end{table}

\stitle{Summary of the roofline model analysis}
As depicted in Figure~\ref{fig:roofline_model}, all three on-disk graph-based index systems are I/O-bound when the vector dataset has moderate dimensionality, e.g., several tens to a hundred dimensions.
However, the performance bottleneck of on-disk graph-based index systems for efficient ANNS fundamentally shifts from I/O-bound to compute-bound with the rising of vector dimensionality (e.g., several hundreds or thousands).
For example, all existing on-disk graph-based index systems are compute-bound on 960-dimensional GIST1M.

Table~\ref{tab:analysis_existing_studies} summarizes the existing on-disk graph-based index systems from three aspects: (i) computational efficiency, (ii) I/O efficiency, and (iii) search performance.
The computational efficiency of all existing solutions is not good enough as all of them are optimizing the I/O bottlenecks and overlooking the computation bottleneck.
The I/O efficiency of Starling and PipeANN is better than DiskANN as they improve it by devising specific optimization techniques. 
Combing the Computational efficiency and I/O efficiency of existing systems,
it is safe to conclude that this is a substantial performance improvement space in on-disk graph-based index systems.
Hence, the research objective of this work is proposing an efficient on-disk graph-based index system, which achieves high efficiency on all three aspects for large-scale high-dimensional vector retrieval, see the last row in Table~\ref{tab:analysis_existing_studies}.
}



\section{Index Layout of \sysname{}}\label{sec:layout}
As confirmed by the adapted roofline model in Section~\ref{sec:preliminary}, the performance bottleneck of existing on-disk graph-based index systems is the expensive computational costs.
In this section, we devise an effective index layout in \sysname{} to overcome it.
However, it is not trivial to achieve it due to the technical challenges in Section~\ref{sec:challenges}.
We next design the SIMD-friendly layout of \sysname{} in Section~\ref{sec:laserlayout}, and present the node caching scheme of \sysname{} in Section~\ref{sec:lasercaching}.

\rvsn{
\subsection{Technical Challenges}\label{sec:challenges}
We analyze the technical challenges to overcome the compute-bound of existing on-disk graph-based index systems as follows.

\stitle{C1: Limitation of sequential computation pattern}
As discussed in Section~\ref{sec:analysis}, the computation cost of existing on-disk graph-based index systems (i.e., DiskANN, Starling, and PipeANN) is dominated by the approximate distance calculation from the quantization codewords.
In particular, it incurs numerous LUT lookups. 
Taking Figure~\ref{fig:diskann} as an example, it consumes 12 LUT lookups to derive the approximate distances from the query $q$ to the neighbors of data vector $1$.
Although LUT is typically small enough to fit within the L1 cache of CPU, which allows low-latency lookups, the lookup process must proceed step by step, with each stage dependent on the completion of the prior one.
For example, as shown in Figure~\ref{fig:diskann}, 0.9, 2.8, 4.8 and 0.3 are accessed by looking up the LUT step by step then used to estimate the approximate distance between query vector $q$ and data vector $2$.
Thus, the above computation procedure cannot be effectively parallelized and executed by exploiting SIMD instructions in modern CPUs due to the irregular memory access and inefficient vectorization.
Thus, it poses the first technical challenge to overcome the compute-bound of existing on-disk graph-based index systems.

\stitle{C2: The precision-storage dilemma}
Existing in-memory index systems~\cite{ngt_github,gou2025symphonyqg} alleviate the computation performance by arranging the quantization codewords in a contiguous, interleaved manner. 
Thus, it can invoke a single SIMD instruction~\cite{guide2011intel} to perform multiple LUT lookups in parallel. 
Inspired by these in-memory systems,  a straight-forward idea is revising the index layout of DiskANN by appending the quantization codewords vectors of the neighbors after the neighbor ids.
However, the core technical challenge in this idea is \textsf{the precision-storage dilemma} as the size of quantization codewords affects the on-disk index size.
In particular, if a high-precision compression scheme is applied on the vector dataset and the corresponding quantization codewords are stored to maintain the search accuracy, the size of on-disk graph-based index grows quickly.
Moreover, the disk page also will be extremely large as each data vector has tens of neighbors and the quantization codewords of all its neighbors should be contiguously stored in the same page.
As a result, the overall performance will be significant degenerated as the bandwidth of SSD is limited.
Conversely, aggressively compressing the data vectors into short codewords preserves the small page size, but it fails to maintain the precision of quantization codewords.
Thus, a higher number of nodes will be visited via I/O accesses during the search procedure to achieve the high recall,
which increases the end-to-end latency.
}


\begin{figure}
    \small
    \centering
    \includegraphics[width=0.99\linewidth]{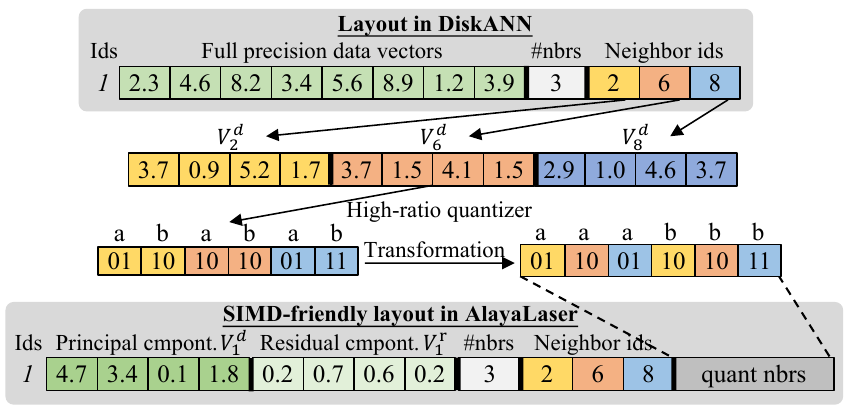}
    \trim 
    \caption{SIMD-friendly layout in \sysname{}}
    \label{fig:simdlayout}
    \trim 
\end{figure}

\subsection{SIMD-friendly Layout of \sysname{}}\label{sec:laserlayout}
With the discussed technical challenges in Section~\ref{sec:challenges}, the research problem is how to design a new index layout with limited page size for on-disk graph-based index systems, which enables parallel approximate distance calculation via SIMD instructions on CPU.
We address it by devising an SIMD-friendly layout in \sysname{}.

\rvsn{
\stitle{Design of SIMD-friendly layout in \sysname{}}
The overview of the SIMD-friendly layout in \sysname{} is illustrated at the bottom of Figure~\ref{fig:simdlayout},
we introduce the designs of it as follows.
The key insight behind the design is that we observed that the quantization scheme (e.g., PQ) treats all dimensions in data vector uniformly and allocates equivalent bits to different dimensions, which results in suboptimal disk page utilization.
To overcome this, we first utilize \textsf{Principal Component Analysis (PCA)} to project the original high-dimensional data vectors into a low-dimensional subspace that captures the majority of the data variance.
In particular, the Principal Component Analysis (PCA) transformer will generate two parts for each data vector $i$: a principal component $V_i^d$, and a residual component $V_i^r$, the exact distance between the query $q$ and data vector $i$ is preserved when the residual component is also included~\cite{bishop2006pattern, jegou2010aggregating, babenko2014inverted}.
Second, we only apply a \textsf{high-ratio quantizer} (e.g., Product Quantization (PQ)~\cite{jegou2010product} or RaBitQ~\cite{gao2024rabitq}) on the principal component of the neighboring data vectors.
For example, the neighboring principal components of data vector $1$ are $V_2^d$, $V_6^d$ and $V_8^d$.
They are compressed via a high-ratio quantizer~\cite{gao2024rabitq}.
We obtain the highly compact quantized codewords, as shown in Figure~\ref{fig:simdlayout}. 
To minimize the quantization error, we apply an orthogonal rotation to balance the variance distribution across the dimensions of principal components, which allows the quantizer to achieve higher precision for a given code size. 
To fully exploit the SIMD instructions on CPU, we re-arrange these quantized codewords in an interleaved manner.
In particular, we decompose the quantized codeword of each data vector into individual sub-codes, then sequentially store the same position sub-codes of different neighbors.
As illustrated in Figure~\ref{fig:simdlayout}, the sub-codes of ``a'' and ``b'' from the neighboring data vectors 2, 6, 8 are stored adjacently after the rearrangement transformation.
Last, the compact and rearranged quantized codewords of the neighbors are appended.

To sum up, in SIMD-friendly layout, the principal component $V_i^d$ and a residual component $V_i^r$ of a data vector are utilized to rerank and refine the final result set of query vector.
The neighbor IDs are used to traverse the graph, and the quantized codewords of neighbors are used to estimate the distance between the query vector and each neighboring data vector.
Comparing to the index layout of DiskANN, \sysname{} slightly introduces extra storage overhead on disk to store the quantized codewords of neighbors and removes the PQ compressed vectors from memory.}

\sstitle{Determining the graph out-degree $R$}
The graph out-degree $R$ directly determines the disk page size in \sysname{}. 
The above SIMD-friendly layout of \sysname{} enables the using of the commonly-used out-degrees (e.g., $R=64$ or $96$) in existing on-disk graph-based index systems (e.g., DiskANN) while keeping small page size.
The core reason is that \sysname{} only stores the quantization codewords of the principal component $V_i^d$, instead of the original data vector.
For example, on 960-dimensional GIST1M, the page size of \sysname{} is 8KB for out-degree $R=64$ and it also is only 12KB if the out-degree is $R=128$. 
It confirms the design of our SIMD-friendly layout ensures the high I/O performance (i.e., IOPS). 
In addition, the performance gains diminish rapidly with the increase of graph out-degree $R$~\cite{fu2017fast}.
Thus, the out-degree $R$ does not exceed 128 in \sysname{}.

\rvsn{
\sstitle{Discussions}
The SIMD-friendly layout used Principal Component Analysis (PCA) to analyze the principal components on data vectors.
It is suitable for the query vectors with the similar distribution of data vectors.
However, the performance of the proposed SIMD-friendly data layout will be degenerate when processing out-of-distribution queries.
We left how to efficiently process out-of-distribution query with on-disk graph-based index system as future work as it is the well-known cross-modal ANNS problem~\cite{chen2024roargraph}, which differs from the ANNS problem in this work as there is a `modality gap' between the data vectors and query vectors.
}

\begin{figure}
    \small
    \centering
    \includegraphics[width=0.99\linewidth]{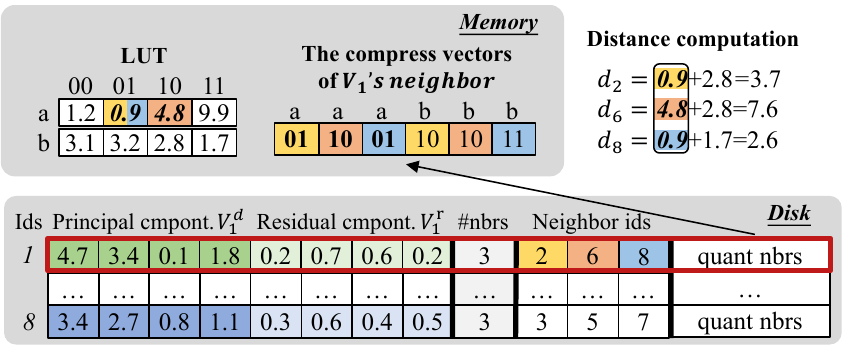}
    \caption{Distance computation in \sysname{}}
    \label{fig:simd_computation}
\end{figure}

\sstitle{Running example} The example in Figure~\ref{fig:simd_computation} illustrates the distance computation procedure with SIMD-friendly layout.
In particular, it loads the quantized codewords of the same position from different neighbors into a SIMD register.
For example, it first loads the sub-codes ``a'' from neighboring data vectors 2, 6, 8 of data vector 1.
Then, the corresponding LUT for the sub-codes ``a'' is loaded into another SIMD register.
After that, the distances between the each sub-code of the query and vectors 2, 6, 8 are obtained via applying SIMD shuffle instruction~\cite{guide2011intel} on these two SIMD registers, which are 0.9, 4.8 and 0.9, respectively, as depicted by colors in LUT ``a''.
Then applying the same procedure on the sub-codes ``b'', and compute the approximate distance between query vector and data vectors $2$, $6$ and $8$.
Compared to DiskANN, \sysname{} transforms the strict serial LUT accesses into a parallel execution procedure via the SIMD-friendly layout, which significantly reduces the computational cost. 
In particular, as shown in Figure~\ref{fig:diskann}, DiskANN computes the approximate distances among the neighboring data vectors $2$, $6$ and $8$ of data vector $1$ with query vector $q$ by invoking 24 scalar loads and 12 adds~\cite{blalock2017bolt}.
However, our \sysname{} computes these approximate distances by 4 SIMD loads, 2 SIMD shuffles and 2 SIMD adds.
It is worth pointing out that \sysname{} works on the quantized codewords of principal component of data vector, which is smaller than the PQ compressed vectors in DiskANN.
Thus, the computation cost of \sysname{} will be further reduced.

\subsection{Node Caching Scheme in \sysname{}}\label{sec:lasercaching}

\begin{table}
\centering
\trim 
\caption{Performance analysis of the SIMD-friendly layout} \label{tab:performance_impact_of_simd_layout}
\begin{tabular}{l|c|c}
\hline
\textbf{Metric} & \textbf{DiskANN layout} & \textbf{\sysname{} layout} \\ \hline \hline
Search Latency (1 thread) & 3223.66 $\mu$s & 2036.44 $\mu$s \\
Computation Cost & 1846.55 $\mu$s & 220.833 $\mu$s \\
Mean I/Os per Query & 77.21 & 131.109 \\
I/O Cost & 1254.09 $\mu$s & 1774.51 $\mu$s \\
Throughput (48 threads) & 7652.19 & 8763.89 \\ \bottomrule
\end{tabular} 
\end{table}

\stitle{Effect of \sysname{} index layout}
To investigate the effect of the proposed SIMD-friendly layout in \sysname{}, 
we measured the performance of DiskANN and a modified version of DiskANN, i.e., it replaces the index layout of DiskANN by our proposed SIMD-friendly layout in Section~\ref{sec:laserlayout} on 960-dimensional GIST1M at Recall@10=0.90.
The search algorithm and all other parameters (i.e., the out-degree $R=64$ and beam search width $BW=8$) are the same in this experiment.
As shown in Table~\ref{tab:performance_impact_of_simd_layout}, the SIMD-friendly layout yielded a significant 88\% reduction of the \textsf{computation cost}, which confirms its effectiveness to resolve the compute-bound of DiskANN.
However, the computational efficiency improvement from the SIMD-friendly layout also introduces extra overhead.
The 960-dimensional data vector in GIST1M is quantized into a 48-byte codeword in the SIMD-friendly layout. 
The high compression ratio losses the precision of data vectors, which in turn requires visiting more nodes to guarantee the same recall, i.e., Recall@10=0.90. 
Thus, the mean I/Os per query of the modified DiskANN with SIMD-friendly layout is higher than the original DiskANN.
Taking both compute cost and mean I/Os per query (resp. I/O cost in Table~\ref{tab:performance_impact_of_simd_layout}) into consideration, the corresponding search latency and throughput of the DiskANN with the proposed SIMD-friendly layout only improve  37\% and 14.5\%, respectively.
Thus, our new research question is: how to reduce the mean I/Os per query then fully unlock the potential of the SIMD-friendly layout in \sysname{}?

\begin{figure}
  \small
  \centering
  \begin{tabular}{cc}
     \includegraphics[width=0.45\columnwidth]{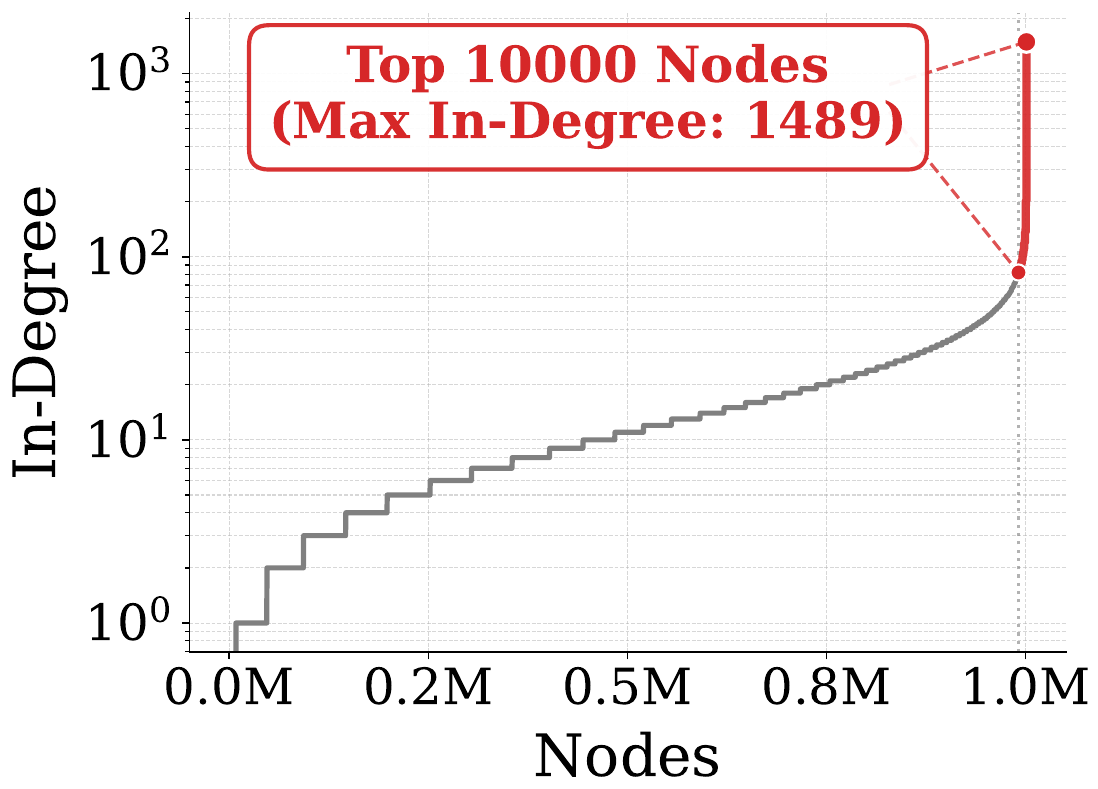}   &  
     \includegraphics[width=0.45\columnwidth]{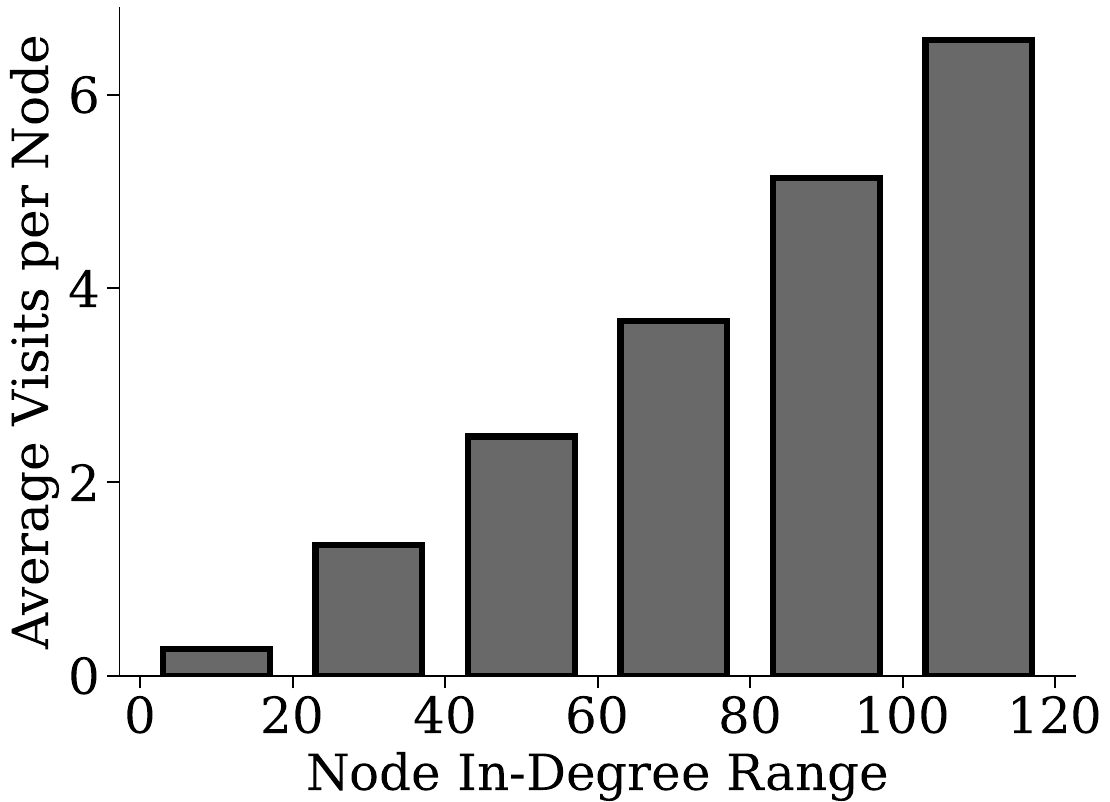} \\
      (a) The distribution of nodes & (b) The number of node visits
  \end{tabular}
  \trim 
  \caption{In-degree of node analysis}   \label{fig:degree_base_cache}
  \trim 
\end{figure}

\rvsn{
\stitle{In-degree based node cache scheme}
Unlike DiskANN, the \sysname{} index layout does not store compressed vectors in memory as the quantized codewords of neighboring data vectors are stored on the SIMD-friendly index layout,
which brings a unique opportunity of \sysname{} to utilize the available memory budget for search performance improvement. 
We designed an in-degree based node cache scheme on \sysname{} by the following key observation.
In particular, we observed that the high in-degree nodes in the graph-based index are visited more frequent than those with low in-degrees.
Figure~\ref{fig:degree_base_cache}(a) illustrates the highly skewed in-degree distribution of the constructed graph-based index of DiskANN on 960-dimensional GIST1M.
Figure~\ref{fig:degree_base_cache}(b) confirms the strong positive correlation between the in-degree of a node and its average number of visits during the search process.
Thus, the in-degree based node cache scheme in \sysname{} first computes in-degree of node in the graph-based index and then caches the top-ranked nodes up to the available memory budget.
The performance gain becomes larger with the rising of the available memory budget.
we use 80\% of the memory budget for the node caching in \sysname{} and the rest 20\% are reserved to handle unexpected out-of-memory failures.}
The cached nodes can be periodically updated when the underlying graph-based index are updated by insertions and deletions.
One side-benefit of our cache scheme is that it overlaps CPU costs and I/O costs in \sysname{}.
For example, consider a search hop that visits 5 nodes: {$V_1, V_2, V_3, V_4, V_5$}, where nodes \{$V_1, V_5$\} are in the node cache and nodes \{ $V_2, V_3, V_4$ \} are not in the cache.
\sysname{} overlaps the compute cost and I/O cost by executing the following tasks concurrently:
(i) issuing asynchronous I/O requests to fetch nodes \{ $V_2, V_3, V_4$ \};
and (ii) computing the distance between the cached data vectors \{ $V_1, V_5$ \} and query.

Until now, it is safe to conclude that the computational improvement brought by the SIMD-friendly layout in \sysname{} changes the game again!
Theoretically, the peak performance $\pi$ in roofline model increases with the SIMD-friendly layout as the cycles to process a FLOP during approximate distance computation will be less than 3 cycles.
The maximum operational intensity $\mathcal{I}_{max} = \pi / \beta$, which also will become larger with the improved of peak performance $\pi$.
When $\mathcal{I}_{max} > \mathcal{I}_{960D}$ (see Figure~\ref{fig:roofline_model}), these on-disk graph-based index systems turn to I/O-bound again.

\section{Search Strategy in \sysname{}}\label{sec:search}
To alleviate the re-emergent I/O bound of \sysname{} with SIMD-friendly layout, we design a suite of optimizations 
to improve the search strategy in it, including (i) cluster-based entry point selection in Section~\ref{sec:searchep},
(ii) adaptive beam expansion strategy in Section~\ref{sec:searchbw}, and (iii) early dispatch mechanism in Section~\ref{sec:longtail}.

\begin{figure}
  \small
  \centering
  \begin{tabular}{cc}
     \includegraphics[width=0.45\columnwidth]{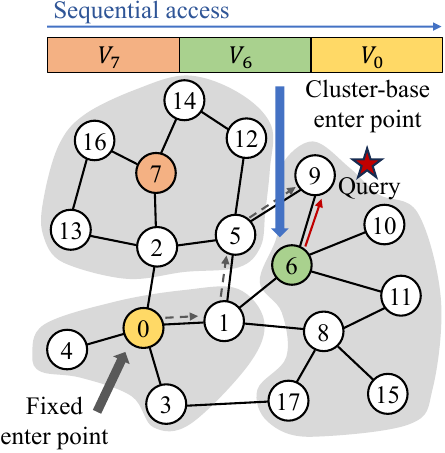}   &  
     \includegraphics[width=0.45\columnwidth]{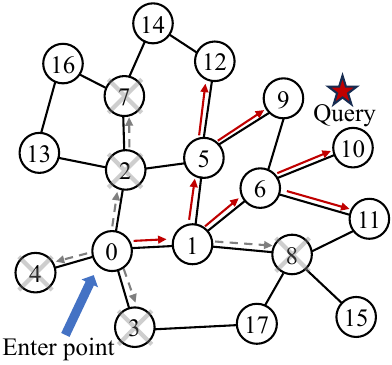} \\
      (a) Cluster-based EP selection & (b) Adaptive beam expansion
  \end{tabular} 
  \trim 
  \caption{Search strategy optimization techniques in \sysname{}} \label{fig:optimation}
  \trim 
\end{figure}

\rvsn{
\subsection{Cluster-based Entry Point Selection}\label{sec:searchep}
Conventional graph-based search strategy initiates graph traversal from a fixed or randomly selected entry point to locate the nearest data vectors for the given query. 
However, the initial entry point often be distant from the actual location of query in the vector space, which results to a long search path.
To address it, we propose a cluster-based entry point selection strategy in \sysname{}. 
The core idea is to identify a region which is close to the query, then commence the search from an entry point in that  region. 
As illustrated in Figure~\ref{fig:optimation}(a), during the index building phase, we employ k-means clustering to partition the set of base vectors into several regions, see these three shaded areas in Figure~\ref{fig:optimation}(a).
The centroid vector of each region is selected as a candidate entry point, and these candidates are stored sequentially in an array, e.g., vectors 0, 7, and 6. 
During the search phase, for a given query vector, \sysname{} sequentially scans these candidate entry points and computes their corresponding distances to the query.
The candidate data vector with the smallest distance to the query is then used as the actual entry point for the graph traversal.
The above idea ensures the search procedure begins in a region which is close to the query, then it significantly shortens the search path.

Our cluster-based entry point selection method in \sysname{} differs from existing  subgraph-based method~\cite{wang2024starling} by our clustering-based method is ``distribution-aware" as it learns the topology of data to maintain a small-yet-effective entry point candidates.
It achieves comparable quality to existing methods which sample a large number of starting points randomly.
Moreover, the online part of \sysname{} is a lightweight sequential scan of all candidate entry points. 
It is faster than the costly in-memory graph traversal of subgraph-based method for every query.}

\subsection{Adaptive Beam Expansion Strategy}\label{sec:searchbw}
\sysname{} follows the beam search strategy of DiskANN to reduce the I/O times during graph traversal. 
The core of beam search is exploring a batch of nodes at each hop, and the batch size is determined by the beam width $BW$.
Generally, a larger beam width $BW$ increases the likelihood to find a shorter search path, but incurs higher computational and I/O overhead.
Existing on-disk graph-based index systems (i.e., DiskANN and Starling) typically configure a fixed beam width $BW$ throughout the entire search process, which is inefficient as it cannot adapt to the dynamic context of the search.
Specifically, in the early stage of search processing, the query vector is far away from the visited nodes in the graph-based index, the fixed wide beam width may waste cost as it visits many unqualified nodes, i.e., which are not be the final result set, by ``over expansion'', see the gray dashed arrows in Figure~\ref{fig:optimation}(b).
However, in the final stage of search processing,  the fixed beam width $BW$ may lead to insufficient exploration as the search procedure enters a dense region of candidates, which leads to long search path, consequently, results to higher search latency.
\rvsn{Inspired by~\cite{peng2023iqan}, we devise an adaptive beam expansion strategy in \sysname{} to overcome this issue. 
In particular, the beam width $BW$ at the i-th hop is determined by $min(2^i, BW_D)$, where $BW_D$ is the default beam width and it is 2-4 times larger than that of DiskANN. 
The core idea of this method is enabling coarse-to-fine search: a narrow initial beam reduces wasteful I/Os in distant regions, while the exponential growth ($2^i$) enables a rapid transition to intensive exploration when the search converges to the targets, see the red arrows in Figure~\ref{fig:optimation}(b).}

\subsection{Early Dispatch Mechanism}\label{sec:longtail}
Until now, the total number of I/O operations per query is reduced in \sysname{}. 
However, we observe that the average search latency for a given target recall increases with the rising of concurrency level.

\begin{figure}
  \small
  \centering
  \begin{tabular}{cc}
     \includegraphics[width=0.5\columnwidth]{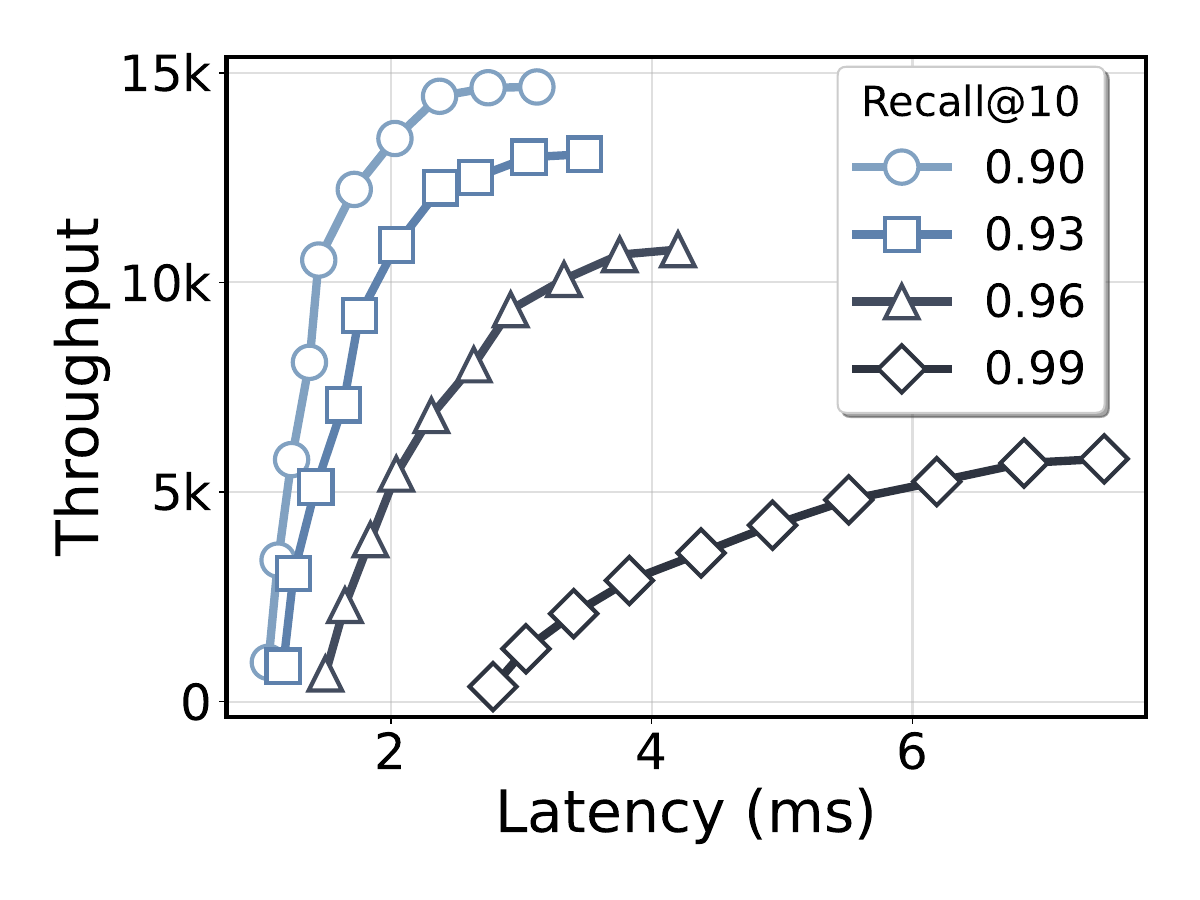}   &  
     \includegraphics[width=0.5\columnwidth]{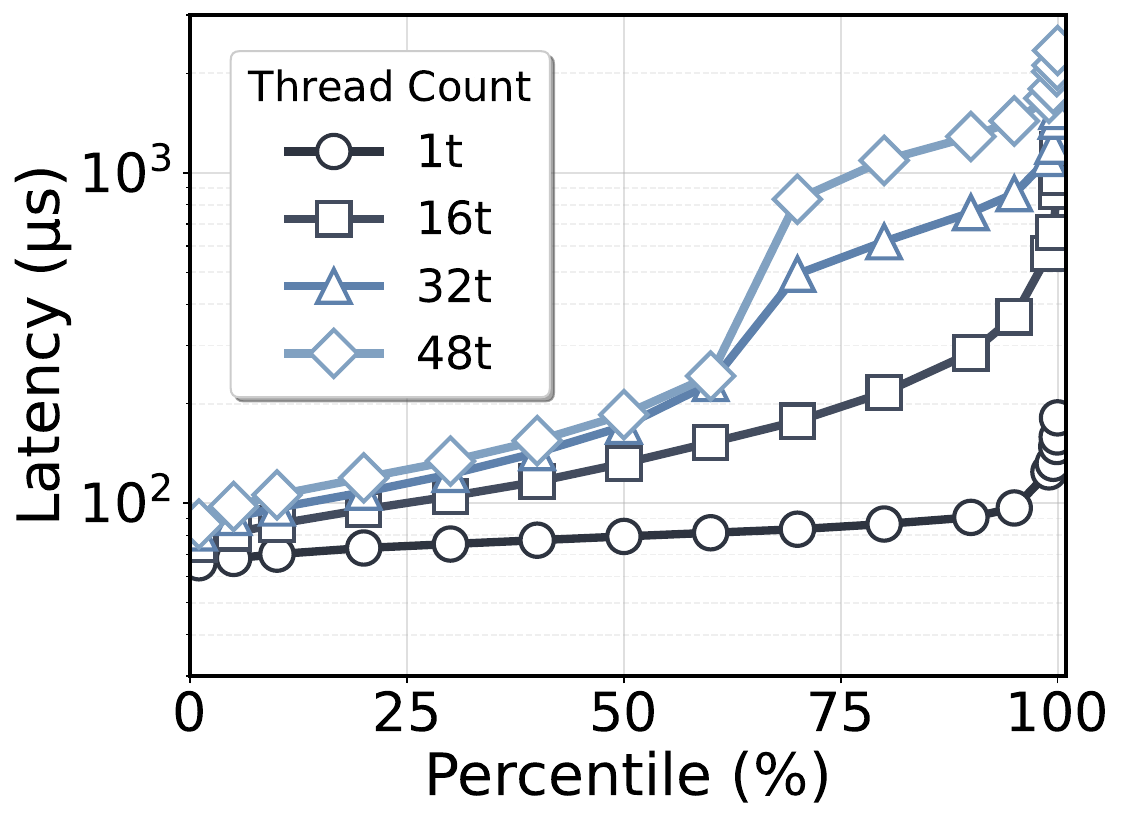} \\
      (a) Throughput-latency curve & (b) CDF of SSD random read latency
  \end{tabular}
  \trim 
  \caption{Long-tail I/O on used SSD}  \label{fig:cdf_of_ssd_random_read_latency}
  \trim 
\end{figure}

Figure~\ref{fig:cdf_of_ssd_random_read_latency}(a) shows the throughput-latency curves of \sysname{} on 960-dimensional GIST1M.
It is obvious the search latency of  Recall\@10=0.99 is significantly higher than that of Recall\@10=0.90 with the same throughput.
Interestingly, this problem is overlooked by almost all (if not all) existing work.
The core reason is that there is a fundamental mismatch between the synchronous nature of the beam search algorithm and the inherent latency variance of modern SSD.
At each search hop, the search procedure waits for all I/O requests in the current beam to complete. 
Consequently, the latency of each step is not determined by the average I/O time, but by the slowest I/O request in the batch, i.e., the long-tail I/O latency. 
Therefore, the slowest I/O access stalls the entire query, which enlarges the end-to-end search latency.
For example, as shown in Figure~\ref{fig:cdf_of_ssd_random_read_latency}(b), 
there is a significant long-tail distribution of the search latency. 
To make the matter worse, the higher concurrency level, the longer search latency.
Specifically, with 48 threads, the 99.9th percentile latency is 2.0ms, which is 10.8x higher than the median latency (i.e., 185.3µs). 
We next design a search process that is robust to the inherent latency variance of modern SSD.

\rvsn{
The key idea is leveraging the structural property of Vamana graphs: \textsf{Path Redundancy}. 
By maintaining a large out-degree (e.g., out-degree $R$ = 64 or 96), the graph index has a high degree of path redundancy. 
Within any local region of the graph, the neighbor lists of adjacent nodes overlap significantly. 
This dense connectivity provides multiple alternative paths to the target nearest neighbors,
thus, the search process is more robust. 
Crucially, a high-quality search direction on the graph can be obtained by using a large-enough subset of the neighbors.
Thus, it does not need to wait for the long tail I/O access in the current beam search.
Hence, \sysname{} decouples the search process from the slowest I/O accesses.
In particular, instead of synchronous waiting, it employs an asynchronous processing and early dispatch mechanism.
At each hop, \sysname{} continuously polls for completed I/O requests. 
As soon as a request finishes, it immediately begins the distance computations for the neighbors. 
The key idea is that \sysname{} does not wait for all beam width nodes to be processed. 
Instead, after a predetermined ratio (e.g., 0.5, targeting the median latency) of the fast-responding nodes have been processed, it decides the next hop and dispatches the next batch of I/O requests immediately. 
Fortunately, this mechanism also overlaps CPU costs and I/O costs effectively.
}

\stitle{Example} considering a search hop which is accessing 4 nodes $V_1, V_2, V_3, V_4$ on disk. 
When any node (e.g., $V_1$) arrives, its computation begins immediately and—overlaps the ongoing I/O for other nodes (e.g., $V_2, V_3, V_4$). 
Once a predetermined ratio of nodes are processed (i.e., 0.5, it processed $V_1$ and $V_2$), it then submits I/O requests for the next hop candidates $V_5, V_6, V_7, V_8$. 
Consequently, the computation for the late-arriving $V_3$ and $V_4$ runs asynchronously, which overlaps their CPU costs with the I/O cost of fetching $V_5, V_6, V_7, V_8$ effectively. 
Finally, the results from these deferred nodes $V_3, V_4$ are used to update the candidate set asynchronously, and maintain the same accuracy of the search procedure.

\section{Experimental Evaluation}\label{sec:exp}
In this section, we first introduce the experimental setup in Section~\ref{sec:exp-setup},
then evaluate the overall performance in Section~\ref{sec:exp-overall}, and conduct the effectiveness study in Section~\ref{sec:exp-effect}.

\subsection{Experimental Setup}\label{sec:exp-setup}

\stitle{Platform} All experiments in this work are conducted on a server with the following configurations:
\squishlist
    \item \textsf{CPU:} Intel(R) Xeon(R) Gold 5318Y @ 2.10GHz, 24 cores/48 threads.
    \item \textsf{Memory:} 128 GB DDR4 @ 2933 MT/s (2 x 64 GB)
    \item \textsf{Storage:} 3 x 1.92 TB SAMSUNG PM9A3 NVMe SSDs
    \item \textsf{Operating System:} Ubuntu 22.04 LTS with Linux kernel 5.15.0
\squishend

\stitle{Datasets and query sets} We used 5 high-dimensional vector datasets in the experiments.
Table~\ref{tab:datasets} provides the detailed statistics of them.
All of these 5 datasets are publicly available at HuggingFace~\cite{hfdatasets2025} and BIGANN Benchmarks~\cite{simhadri2024resultsbigannneurips23}. 
\rvsn{For the query set of each dataset, we directly used the provided query set for GIST1M and MSMARCO.
For the rest three datasets, we randomly sampled 10,000 queries from the corresponding dataset to form the query sets as they do not include the query sets.}
We obtain the ground truth of the queries in each dataset via linear scan method.

\begin{table}
\small
\centering
\trim 
\caption{Vector dataset statistics}\label{tab:datasets}
\trim 
\begin{tabular}{|l|c|c|c|c|} \hline
{Dataset $\mathcal{D}$} & \textbf{ $|\mathcal{D}|$} & d & {Data size} & {Query size} \\  \hline\hline
GIST1M~\cite{TexMexCorpus}            & 1,000,000  & 960 & 3.57GB & 1,000 \\  \hline
Cohere~\cite{cohere}     & 10,123,929  & 768   & 28.96GB & 10,000 \\ \hline
BigCode ~\cite{bigcode}            & 10,403,628  & 768 & 28.98GB & 10,000 \\ \hline
DPR~\cite{aguerrebere2023similarity}     & 101,000,000  & 768   & 288.96GB & 10,000 \\  \hline
MSMARCO ~\cite{msmarco}       & 113,520,750 & 1,024   & 433.04GB & 1,677 \\  \hline
\end{tabular}
\end{table}

\begin{figure*}
  \small
  \centering
  
  \begin{tabular}{cc}
     \includegraphics[width=0.85\columnwidth]{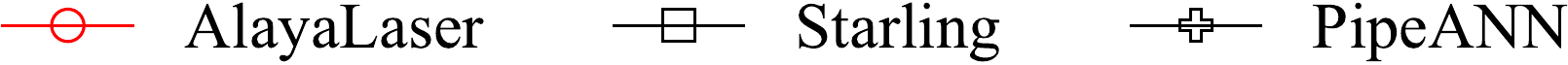}
     \includegraphics[width=0.48\columnwidth]{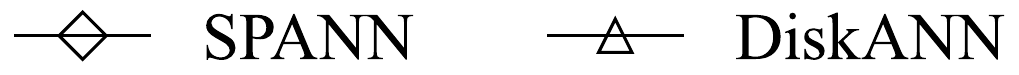}
  \end{tabular}
  
  \begin{tabular}{ccccc}
     \includegraphics[width=0.384\columnwidth]{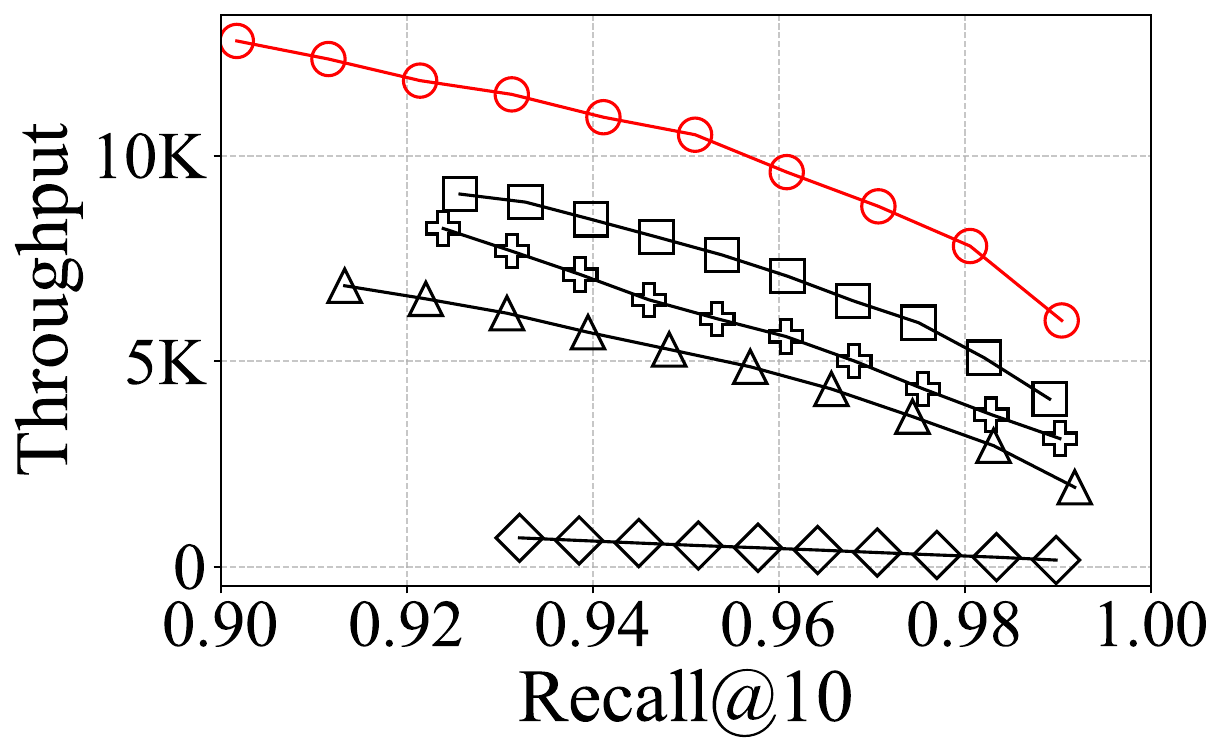}   &  
     \includegraphics[width=0.384\columnwidth]{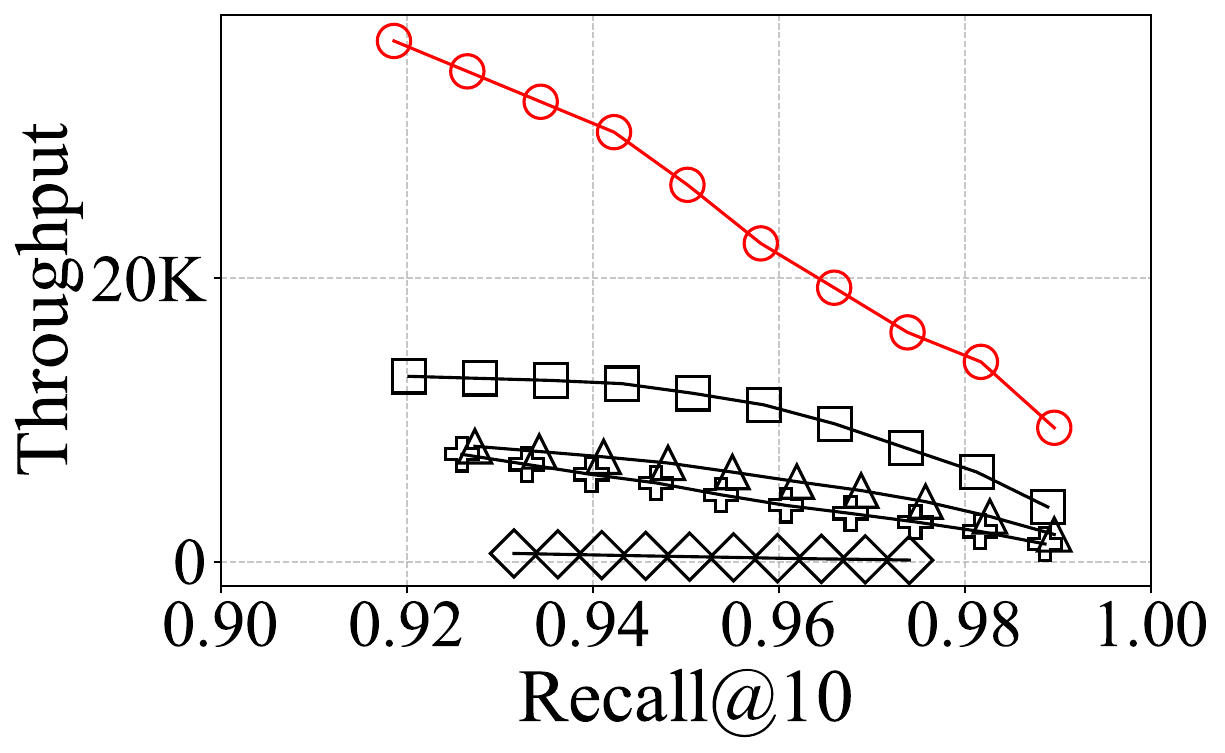} &
     \includegraphics[width=0.384\columnwidth]{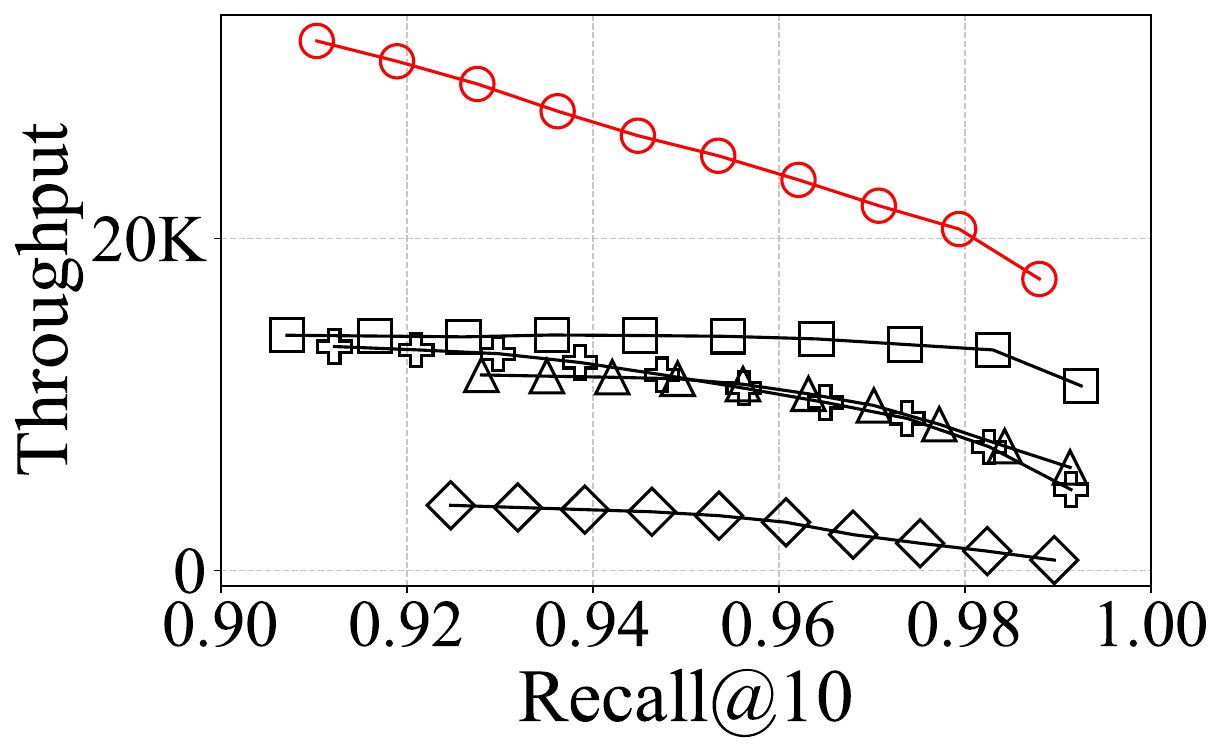} &
     \includegraphics[width=0.384\columnwidth]{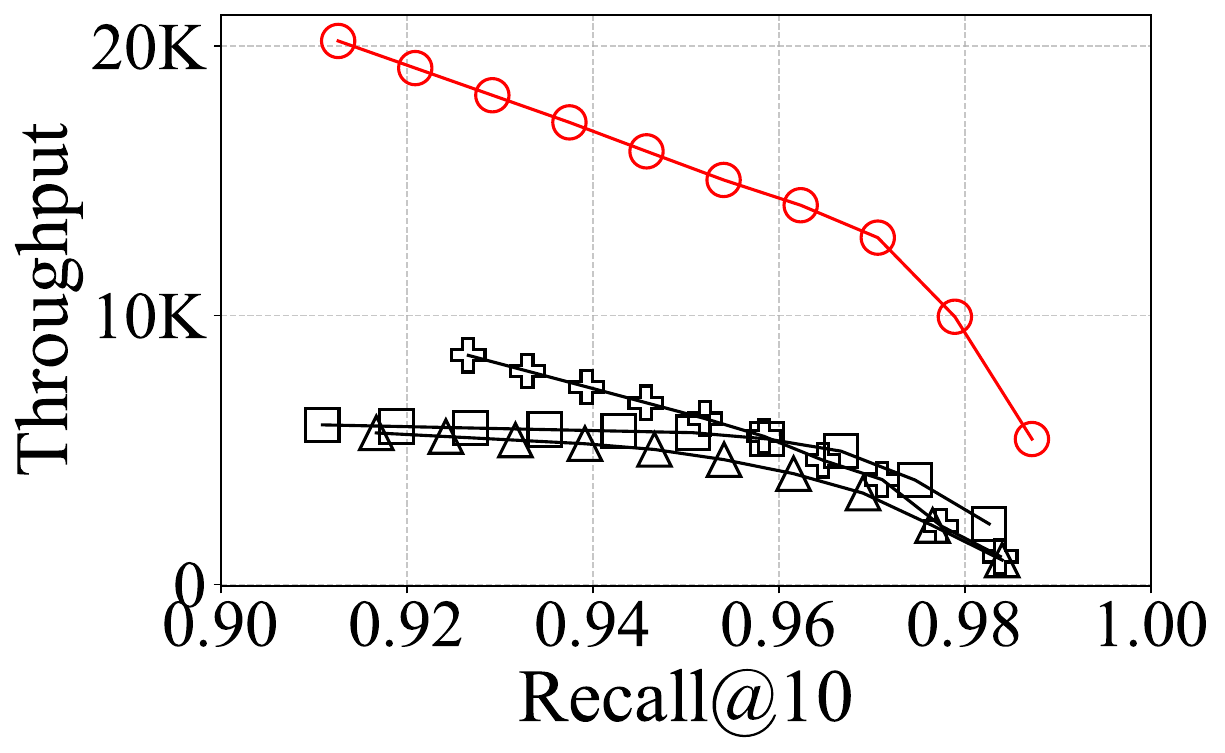} &
     \includegraphics[width=0.384\columnwidth]{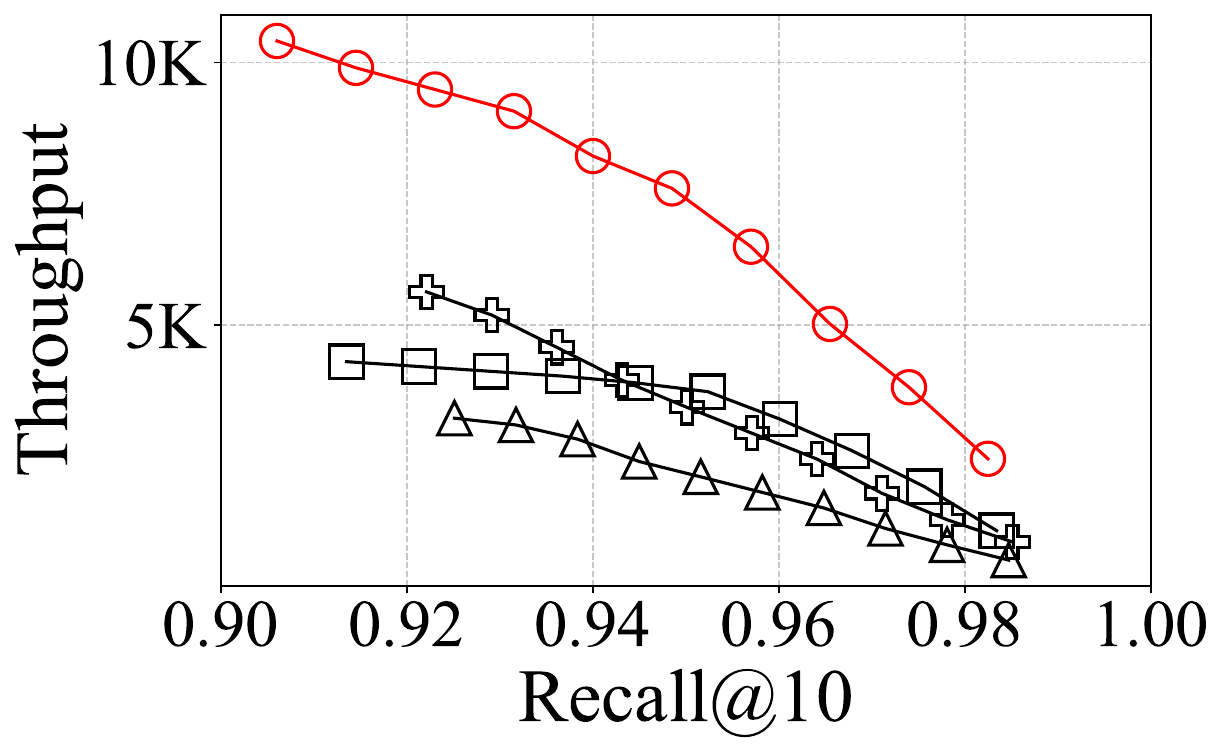} 
     \\
      (a) GIST1M & (b) Cohere & (c) BigCode & (d) DPR & (e) MSMARCO
  \end{tabular}
  \trim 
  \caption{Throughput-Recall@10 evaluation on all on-disk index systems}
  \label{fig:allthroughput}
  \vspace{3mm}
   \begin{tabular}{ccccc}
     \includegraphics[width=0.384\columnwidth]{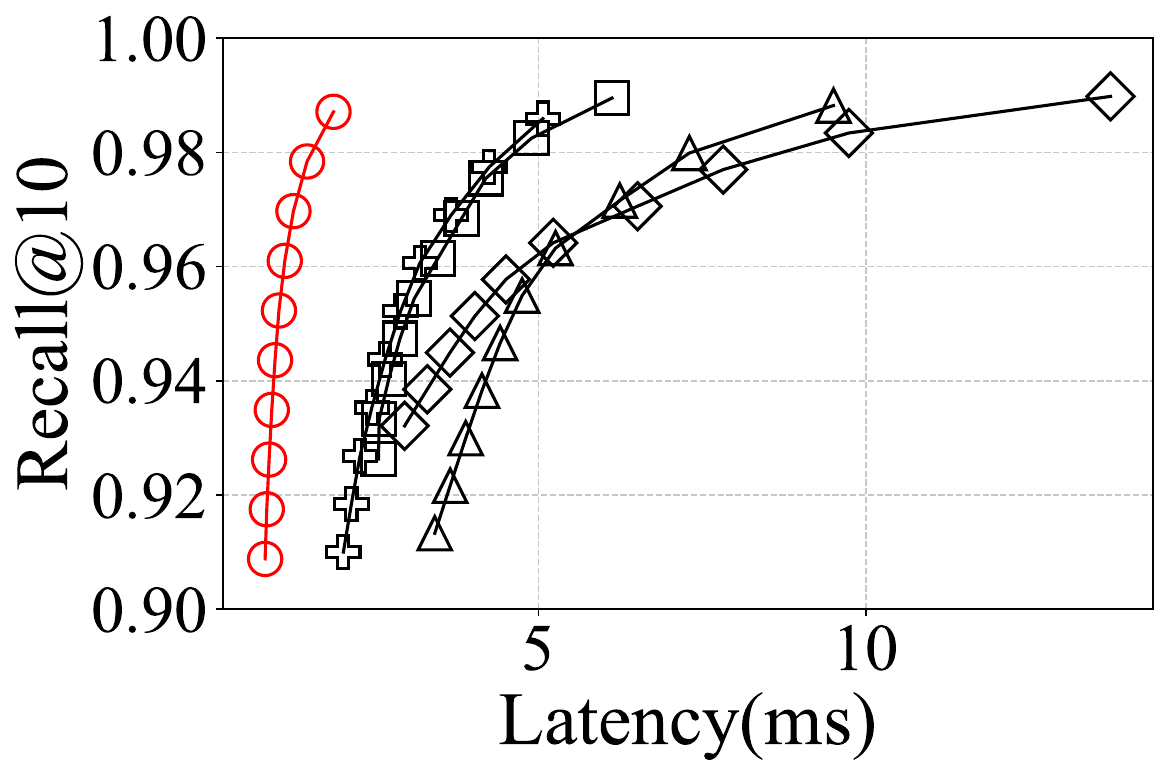}   &  
     \includegraphics[width=0.384\columnwidth]{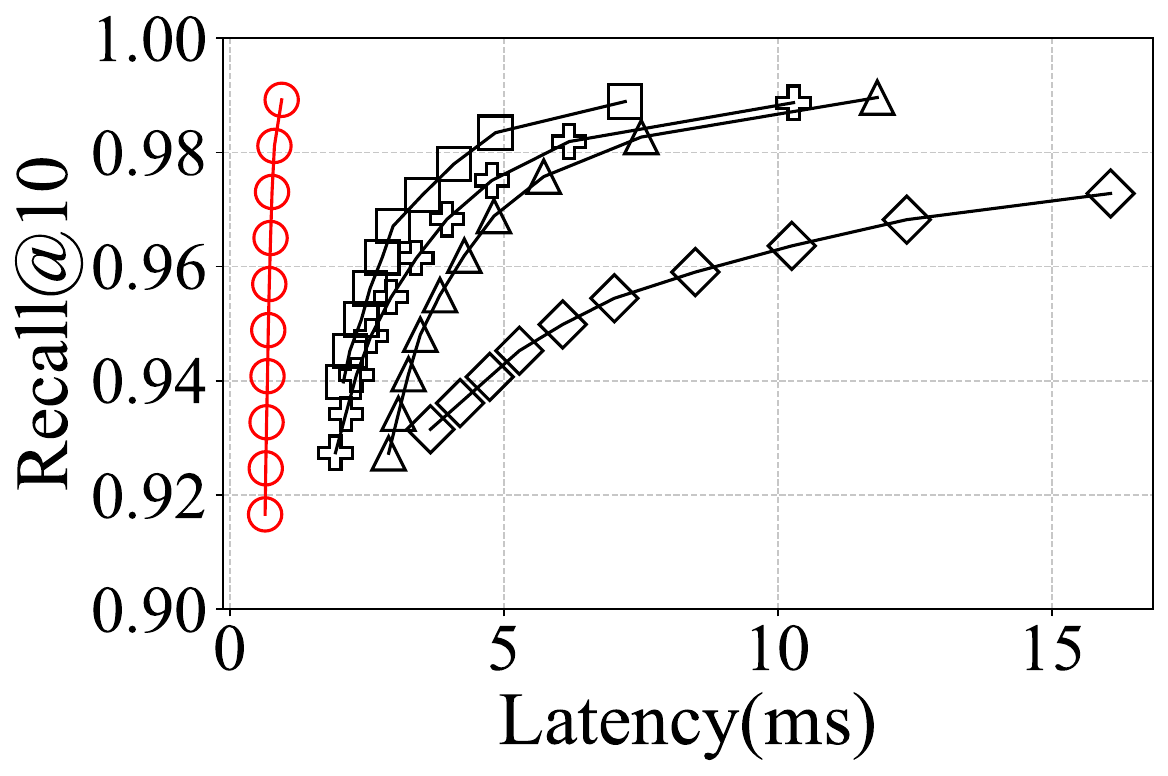} &
     \includegraphics[width=0.384\columnwidth]{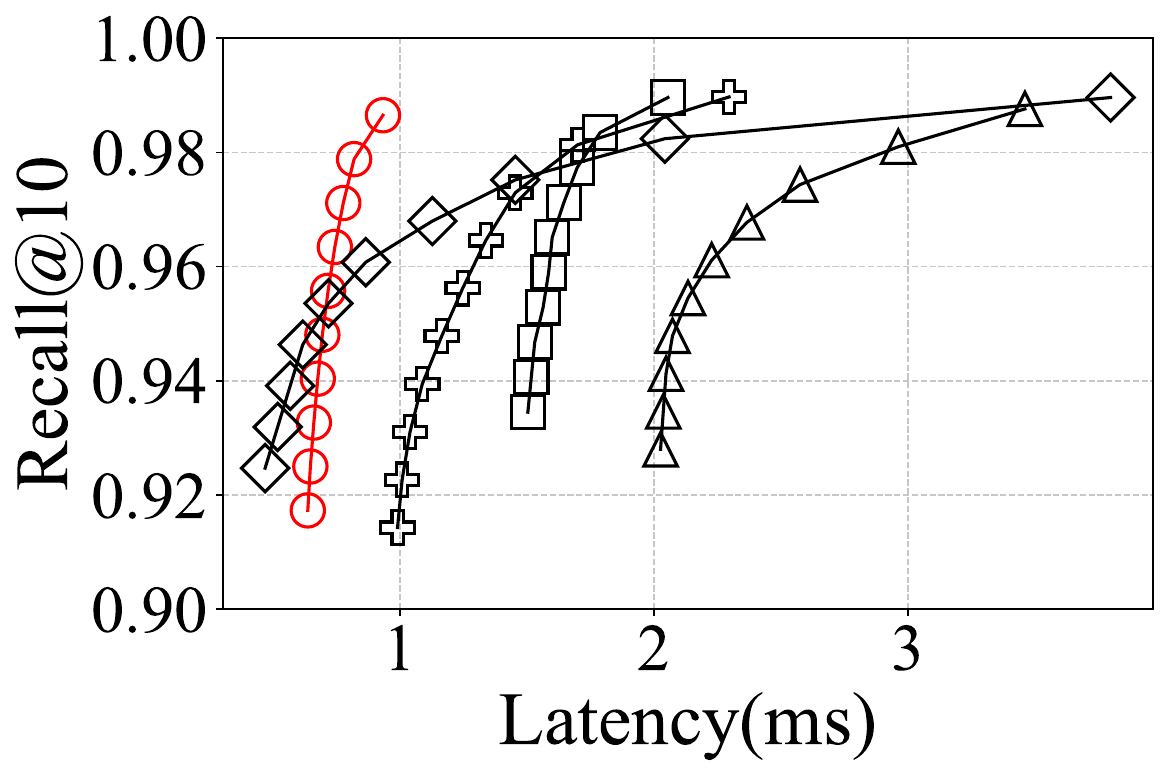} &
     \includegraphics[width=0.384\columnwidth]{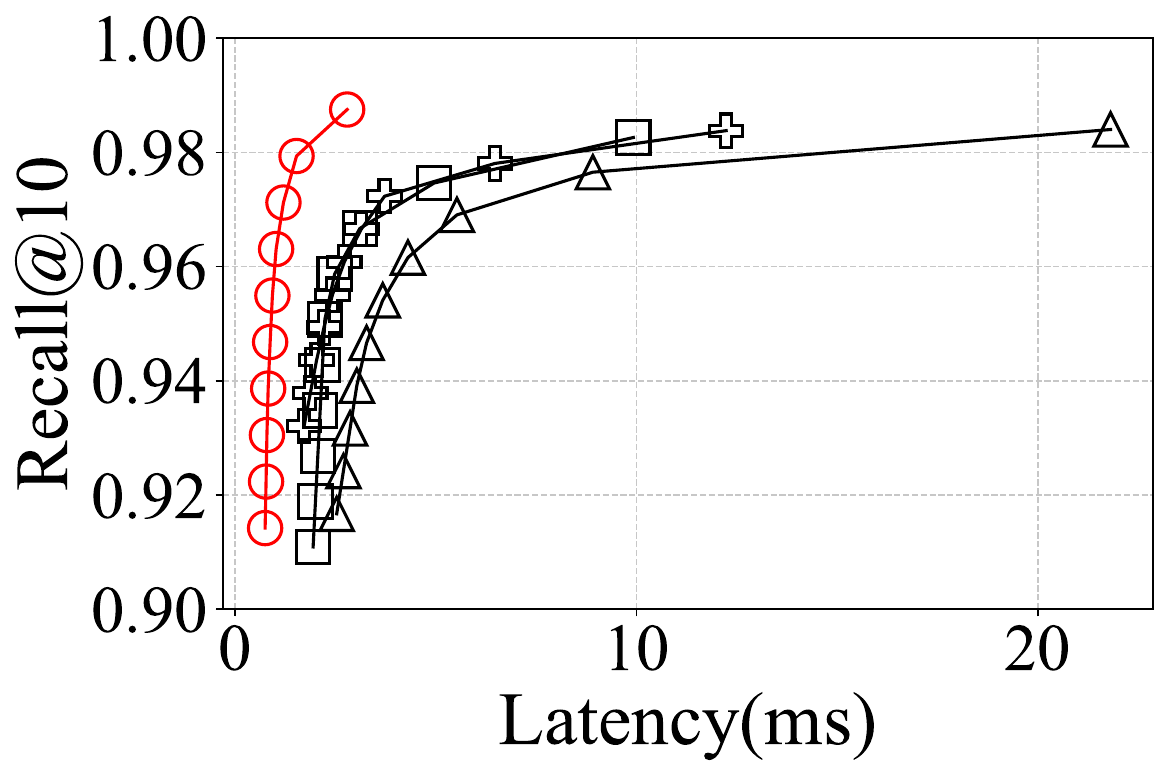} &
     \includegraphics[width=0.384\columnwidth]{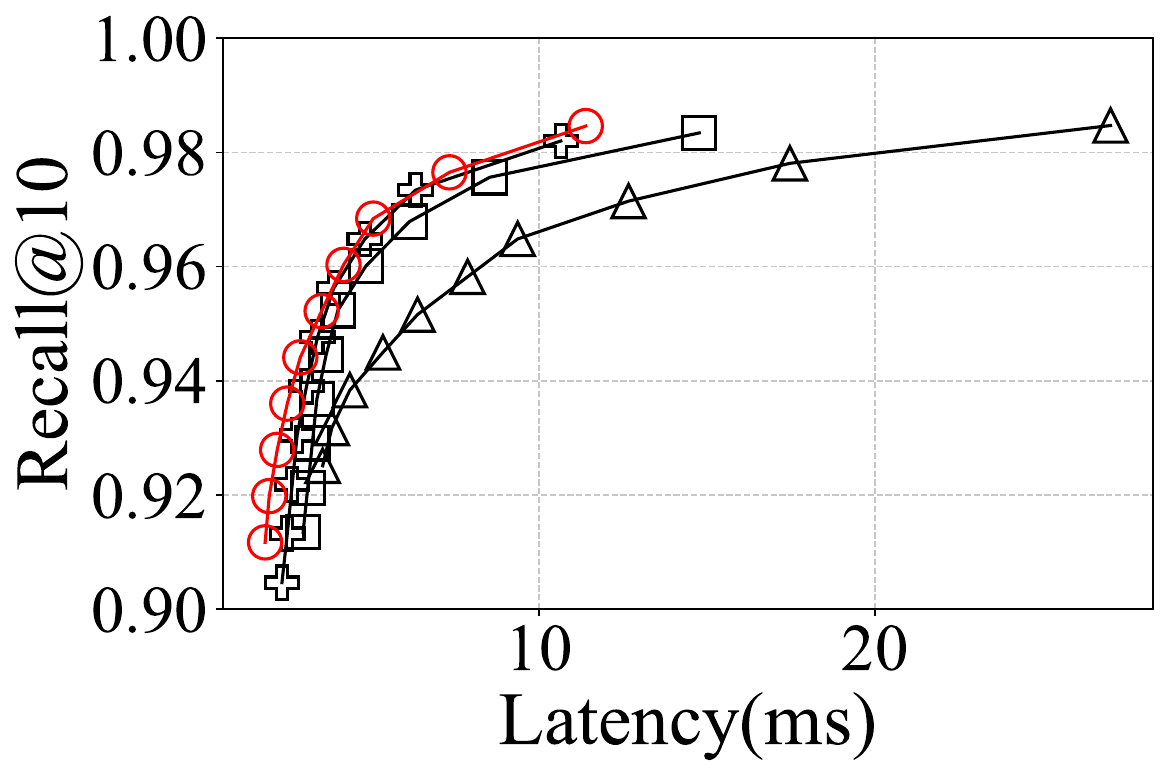} 
     \\
      (a) GIST1M & (b) Cohere & (c) BigCode & (d) DPR & (e) MSMARCO
  \end{tabular}
  \trim 
  \caption{Recall@K-Latency evaluation on all on-disk index systems}
  \label{fig:alllatency}
  \trim 
\end{figure*}

\stitle{Compared methods}
We compare our proposed \sysname{} with two categories of competitors: (i) on-disk systems, and (ii) in-memory solutions, which are introduced as follows:

\stitle{On-disk systems} We evaluate 4 representative on-disk systems: 3 graph-based index systems and 1 cluster-based index system.

\squishlist
    \item \textbf{DiskANN}~\cite{jayaram2019diskann} is the first on-disk graph-based index system that stores the full graph structure and vectors on SSD, maintains PQ compressed vectors in memory to guide the search.
    \item \textbf{Starling}~\cite{wang2024starling} improves the I/O efficiency of DiskANN by optimizing its on-disk data layout via ``block shuffling'' and using an in-memory navigation graph to shorten the search path.
    \item \textbf{PipeANN}~\cite{guo2025achieving} is an algorithmic optimization that pipelines I/O and computation of DiskANN during the execution, which hides the disk latency and reduces the query latency.
    \item \textbf{SPANN}~\cite{chen2021spann} is the cluster-based index, which stores cluster centroids in memory and posting lists with vectors on disk.
\squishend

\rvsn{
\stitle{\textit{In-memory solutions}} We compare the performance of our \sysname{} against 3 in-memory index solutions.
\squishlist
    \item \textbf{HNSWlib}~\cite{malkov2018efficient} is a widely adopted in-memory graph-based index, which is a \emph{de facto} in-memory solution for ANNS.
    \item \textbf{Vamana}~\cite{jayaram2019diskann} is the underlying graph structure of DiskANN, we evaluate its in-memory performance by storing the entire Vamana graph of DiskANN in memory.
    \item \textbf{SymphonyQG}~\cite{gou2025symphonyqg} is the state-of-the-art in-memory graph-based index for ANNS. It leverages a compact layout to minimize random memory accesses and exploits SIMD instructions to accelerate distance computations.
\squishend
}

\rvsn{
\stitle{Parameter configurations}
The source code of the compared methods are obtained from the original authors.
All the parameters of the evaluated methods are configured by following the settings in existing work~\cite{jayaram2019diskann, chen2021spann, wang2024starling, guo2025achieving}.
The memory budget $\mathcal{B}$ of all these on-disk systems are set as follows: 1GB for GIST1M, 10GB for Cohere and BigCode, and 128GB for DPR and MSMARCO.
To bypass the effect of operating system cache, we use the \textsf{libaio} library with the \textsf{O\_RDONLY} and \textsf{O\_DIRECT} flags on all on-disk methods as the setting in~\cite{jayaram2019diskann, chen2021spann, wang2024starling, guo2025achieving}.
For all graph-based index construction, the maximum out-degree is $R=64$ and the construction list size is 200.
For entry point selection, Starling and PipeANN constructed an in-memory navigation graph from {1\%} data samples, which is built with maximum out-degree $R=48$ and the construction list size 128.
In \sysname{}, we store 300 candidate entry points for GIST1M, Cohere and BigCode, and 500 for DPR and MSMARCO by default. 
The dimension of the principal components in \sysname{} for all five tested datasets is {256}. 
During the search phase, the search beam width is 8, 8, 32 and 16 for DiskANN, Starling, PipeANN, and \sysname{}, respectively.
We vary the size of the candidate list (i.e., \textsf{search\_L}) for all compared graph-based methods to achieve the targeted recall.
}

\stitle{Measured metrics} We measure both search performance and result accuracy on all compared methods.
In particular, the search performance metrics include: \textsf{throughput} (i.e.,  queries per second, QPS) and \textsf{latency} (time cost to process a query).
In all our experiments, the \textsf{throughput} of a method is measured by using all available threads (i.e., 48 threads) and the latency is measured on a single thread.
We also report the \textsf{mean I/Os per query}, which measures the average number of disk I/O operations to process a query. 
The result accuracy is measured  by \textsf{Recall@K}, which shows the fraction of the true K-nearest neighbors found within the top-K results returned by the algorithm. It is a standard metric for ANNS.

\subsection{Overall Performance Evaluation}\label{sec:exp-overall}
We plot (i) throughput-recall@K curve (where top-right is better), and (ii) recall@K-latency curve (where top-left is better)
to evaluate the overall search performance of all compared systems. 

\stitle{Compared with on-disk systems}
Figure~\ref{fig:allthroughput} depicts the throughput-recall@10 curves of all compared on-disk index systems on 5 tested datasets.
Visually, our proposal \sysname{} outperforms all existing on-disk systems at all tested cases.
The throughput of \sysname{} is 2.0x-3.6x, 1.4x-2.7x, 1.6x-4.2x, 7.9x-60.6x faster than DiskANN, Starling, PipeANN and SPANN among all these tested datasets, which confirmed the superiority of the designs in \sysname{}.
In particular, SPANN has the lowest throughput as it is cluster-based index, which needs to access a large amount of data vectors to achieve a high recall. We ignored SPANN at DPR and MSMARCO, see Figures~\ref{fig:allthroughput}(d) and (e), as the index construction of SPANN is out-of-memory at large vector datasets.
Starling outperforms DiskANN as it employs an in-memory navigation graph to shorten the search path and reduce overall computation.
PipeANN performs better than DiskANN as it effectively hides the latency in DiskANN via asynchronous I/O.
However, it is slightly worse than Starling in some cases.
The major reason is that the PIPESearch strategy in PipeANN may visit more nodes when comparing with the visited nodes in Starling.

Figure~\ref{fig:alllatency} shows the recall@10-latency curves of all evaluated on-disk index systems.
It is no doubt that the latency of \sysname{} is the overall winner among all these tested systems on all 5 datasets.
In particular, the search latency of \sysname{} is up to 91\% lower than that of DiskANN on Cohere where recall@10 is 0.99. 
In all these tested cases, the search latency of \sysname{} is approximately 24-70\% and 22-74\% lower than that of Starling and PipeANN, respectively.
The performance gain of \sysname{} are from three-fold: (i) the SIMD-friendly layout reduced the approximate distance computation time in memory, (ii) the in-degree based node cache scheme avoids I/O accesses as the frequently accessed nodes are cached in memory,
and (iii) the optimization techniques in search strategy of \sysname{}, e.g., the asynchronous early dispatch strategy mitigates the effect of long-tail I/Os.
Besides, the search latency of  PipeANN is comparable to Starling as it overlaps computation cost and I/O cost. 

\begin{figure}
\small
\centering
  \includegraphics[width=0.85\columnwidth]{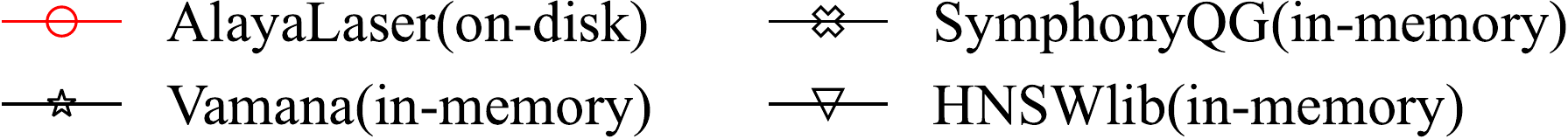}%
  \\
\begin{tabular}{cc}
    \includegraphics[width=0.45\columnwidth]{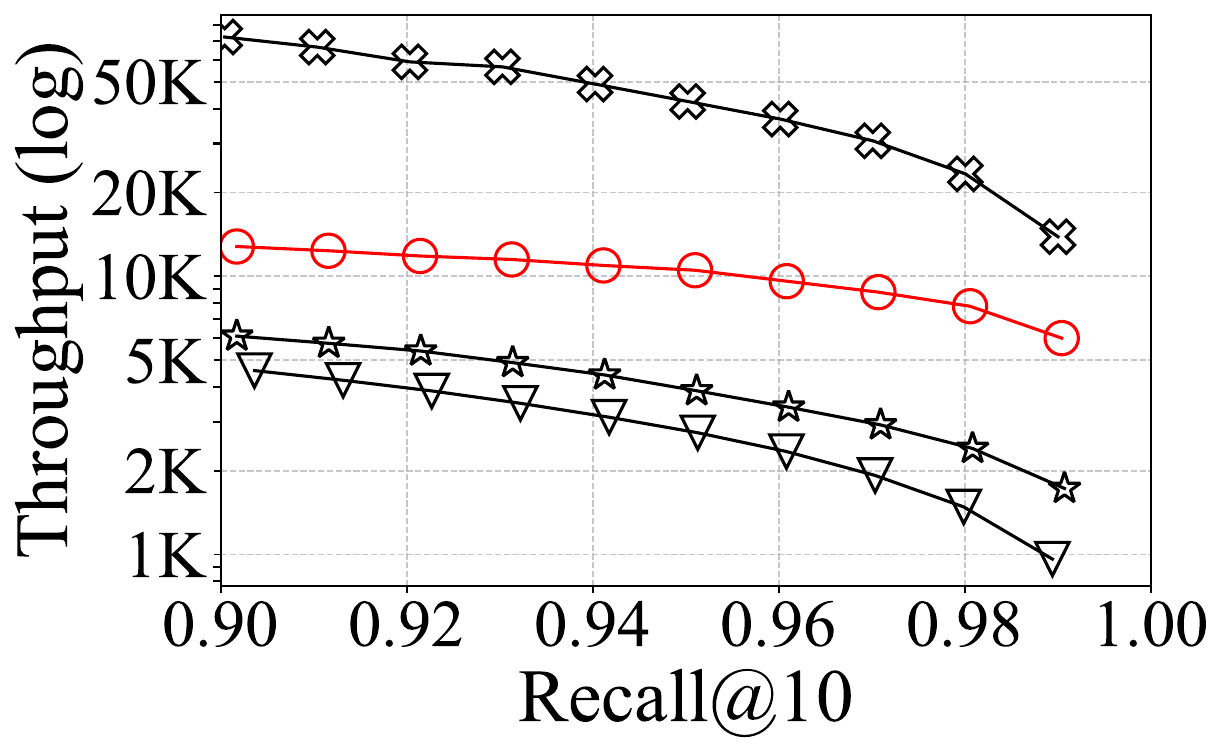}
    &  
    \includegraphics[width=0.45\columnwidth]{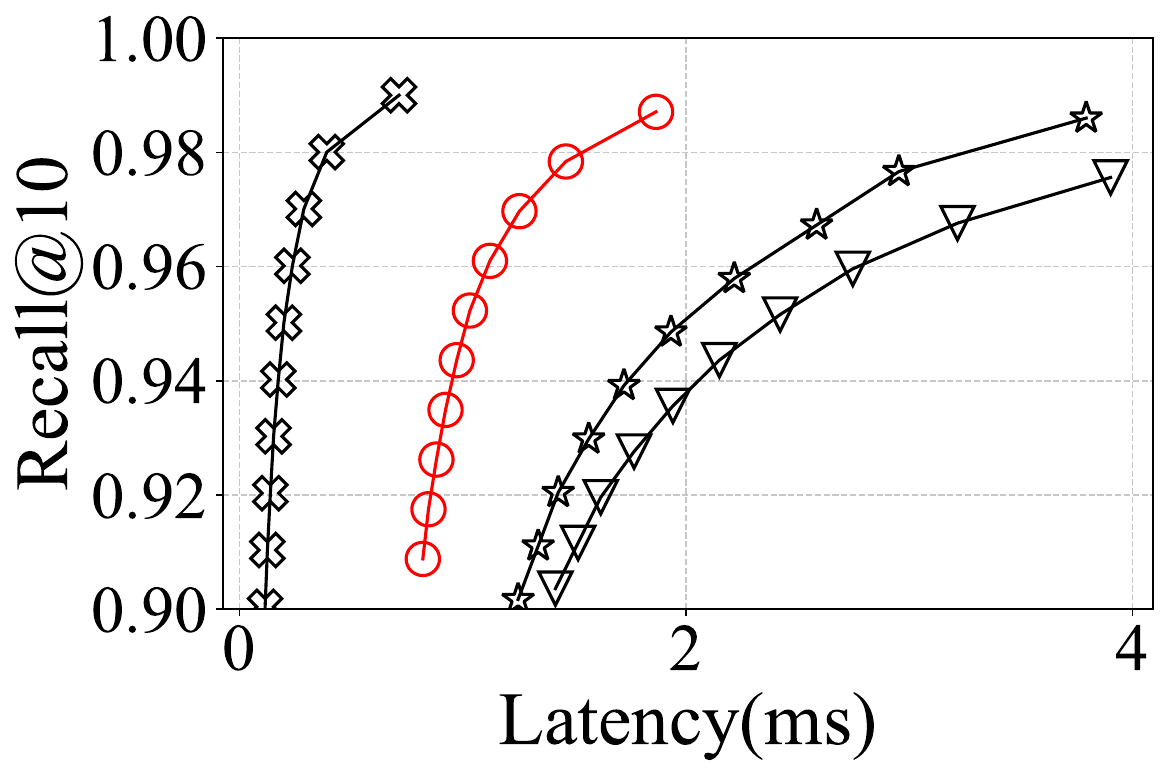}
     \\
    (a) Throughput on GIST1M & (b) Latency on GIST1M
    \\
    \includegraphics[width=0.45\columnwidth]{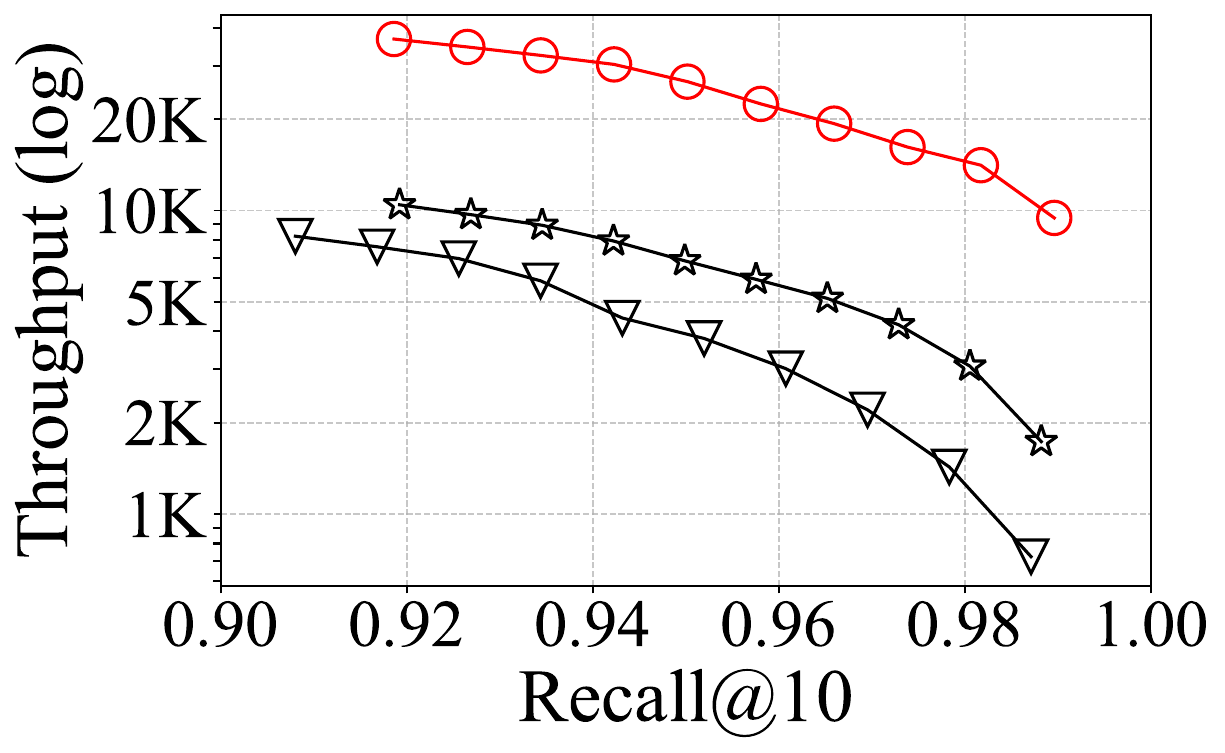}
    &  
    \includegraphics[width=0.45\columnwidth]{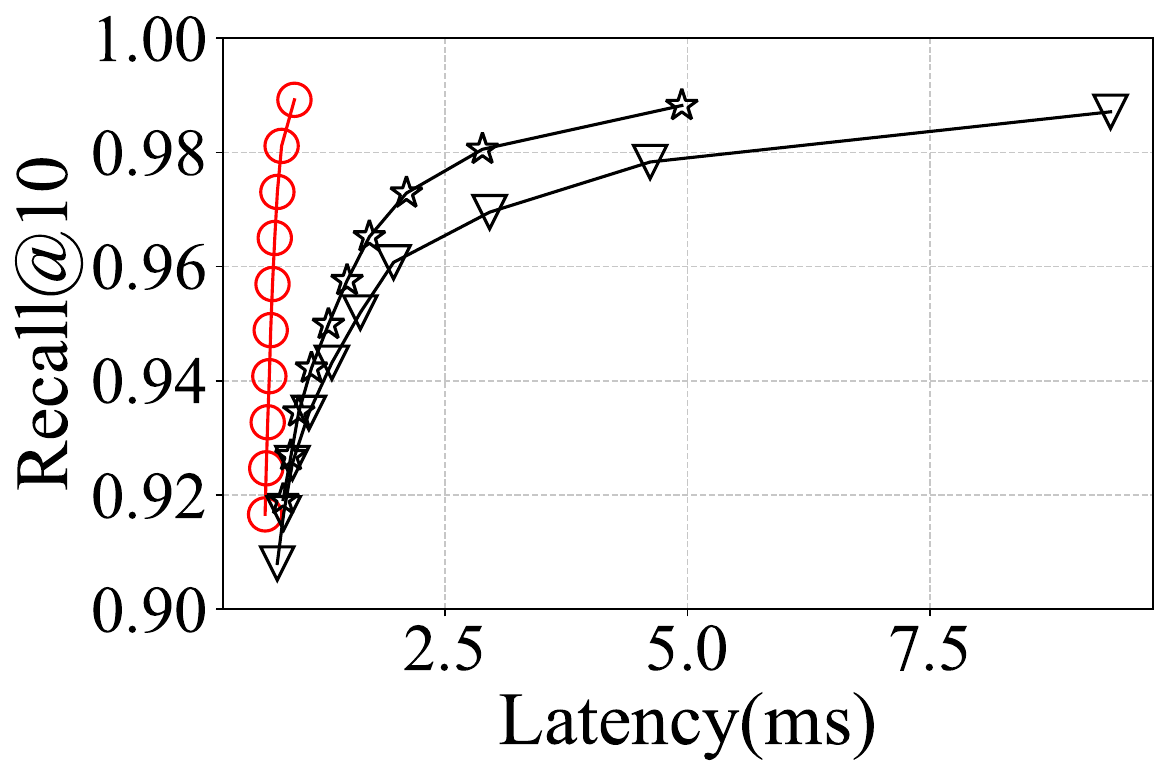}
     \\
     (c) Throughput on Cohere & (d) Latency on Cohere
     \\
    \includegraphics[width=0.45\columnwidth]{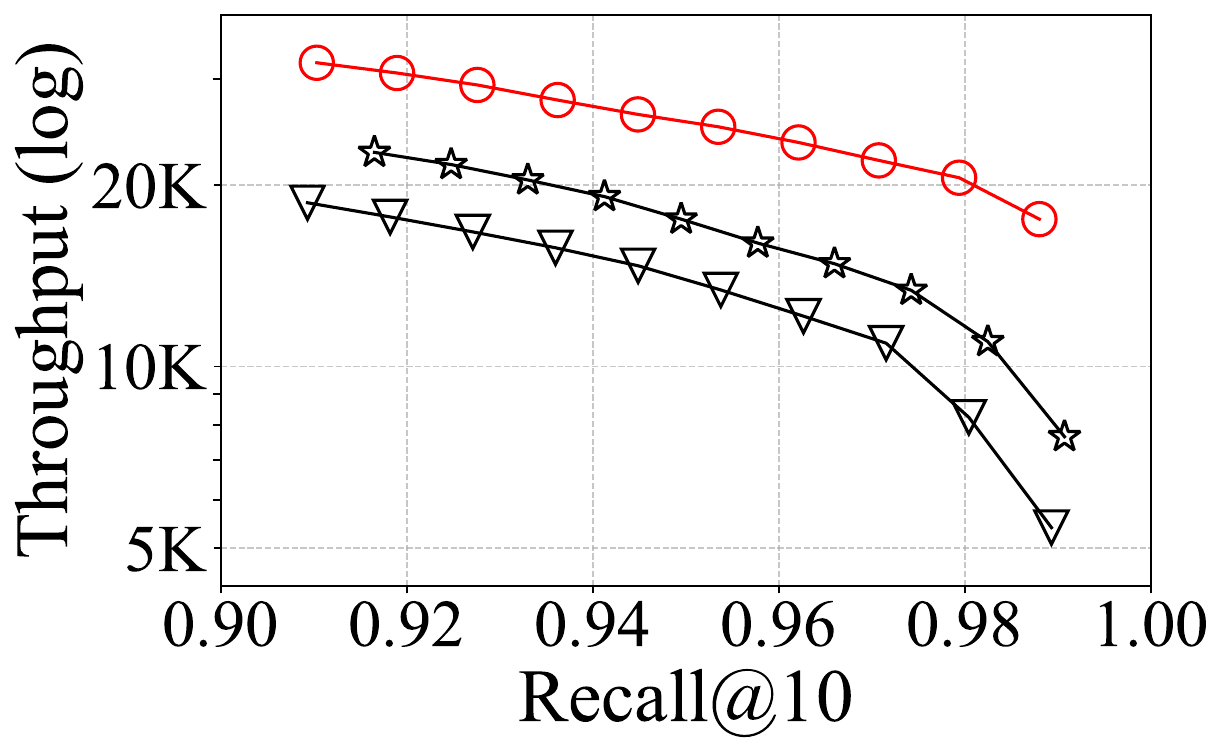}
    &  
    \includegraphics[width=0.45\columnwidth]{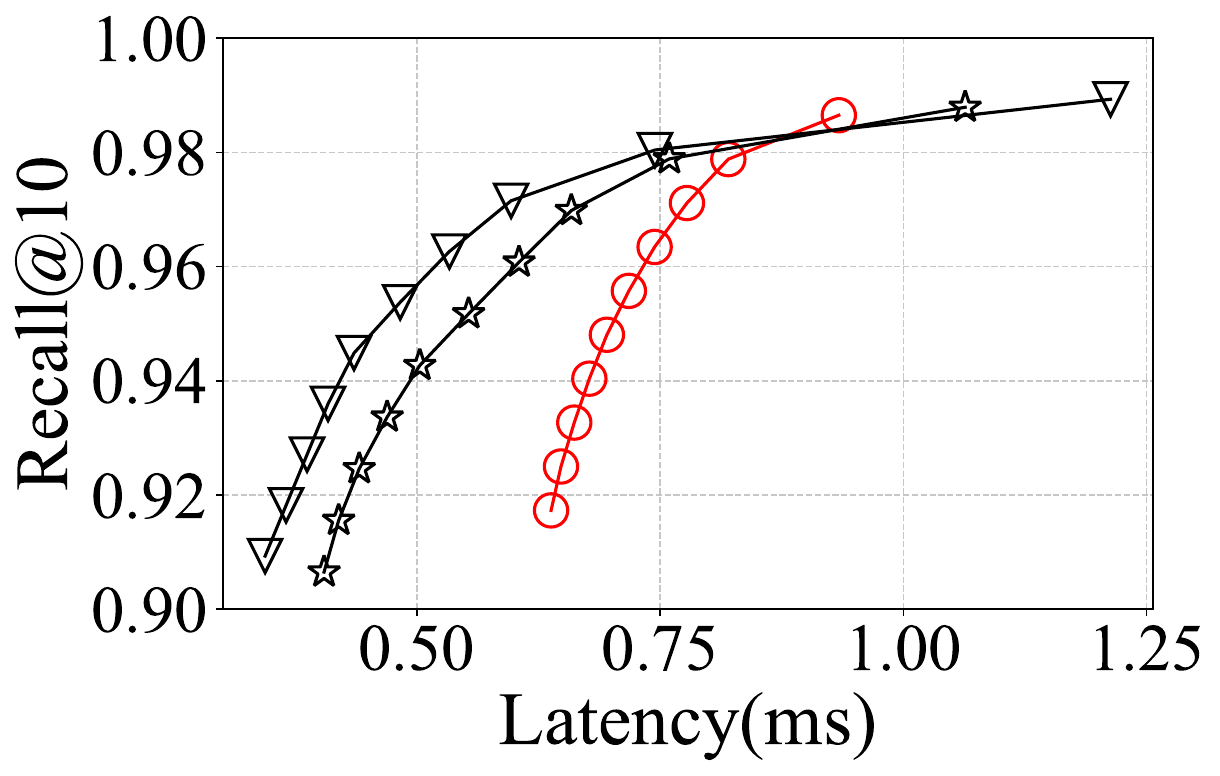}
     \\
     (e) Throughput on BigCode & (f) Latency on BigCode
\end{tabular}
\trim 
\caption{In-memory index solutions evaluation} \label{fig:inmemory}
\trim \trim 
\end{figure}

\rvsn{
\stitle{Compared with in-memory solutions}
Even though \sysname{} is an on-disk graph-based index system, we compare the performance of \sysname{} with existing in-memory graph-based index methods in Figure~\ref{fig:inmemory}. We ignore two large vector datasets (i.e., DPR and MSMARCO) in this experiment as their data sizes exceed the memory size of our machine (128GB).
First of all, as shown in Figures~\ref{fig:inmemory}(a), (c) and (e),
the throughput-recall@10 of Vamana (in-memory) and HNSWlib (in-memory) are obviously below the corresponding curve of \sysname{} (on-disk).
It confirms that \sysname{} not only outperforms its own in-memory graph-based index version, i.e., Vamana (in-memory), but also the widely-used in-memory index HNSWlib (in-memory) on these three datasets.
Secondly, the state-of-the-art in-memory index SymphonyQG~\cite{gou2025symphonyqg} is better than \sysname{} on GIST1M as it is careful-designed for in-memory scenario.
However, the index construction of SymphonyQG is failed on Cohere (29GB) and BigCode (29GB) due to out-of-memory (OOM) errors on our 128GB server.
The reason is that SymphonyQG stores the quantized codewords of the neighbors in each data vector to fully exploit SIMD instructions for superior performance. However, it also has the precision-storage dilemma, as we discussed in Section~\ref{sec:challenges}.
In particular, the index size of SymphonyQG is 4 times larger than the original dataset on Cohere (29GB) and BigCode (29GB) with the settings in~\cite{gou2025symphonyqg}, which incur OOM during SymphonyQG index construction on both datasets.
The limitation of the scalability of SymphonyQG further confirms the efficiency of \sysname{} as it is an on-disk system.
Thirdly, considering the search latency, \sysname{} also demonstrates strong competitiveness against Vamana (in-memory) and  HNSWlib (in-memory) in all three datasets, as shown in Figures~\ref{fig:inmemory}(b), (d) and (e).
The performance of \sysname{} is consistently better than its of Vamana and HNSWlib on GIST1M and Cohere.
However, its performance is worse than its of Vamana and HNSWlib when Recall@10 is smaller than 0.98 on BigCode.
The major reason is that Vamana and  HNSWlib achieve the target recall (i.e, $\leq$ 0.98) by visiting much few nodes than \sysname{} on BigCode.
However, both Vamana and HNSWlib need to visit more nodes to achieve a high recall (e.g., $\geq$ 0.98), which enabling \sysname{} to surpass them by unlocking its superior computation ability.
As illustrated in Figure~\ref{fig:inmemory}(f), the latency of \sysname{} is smaller than Vamana and  HNSWlib when the Recall@10 is higher than 0.98 on BigCode.}

\begin{figure}
\small
\centering
  \includegraphics[width=0.85\columnwidth]{method_figure/new_fig_v2/disk_qps_recall/label1_alayalaser.pdf}\\
  \includegraphics[width=0.50\columnwidth]{method_figure/new_fig_v2/disk_qps_recall/label2.pdf}\\
\begin{tabular}{cc}
  \includegraphics[width=0.45\columnwidth]{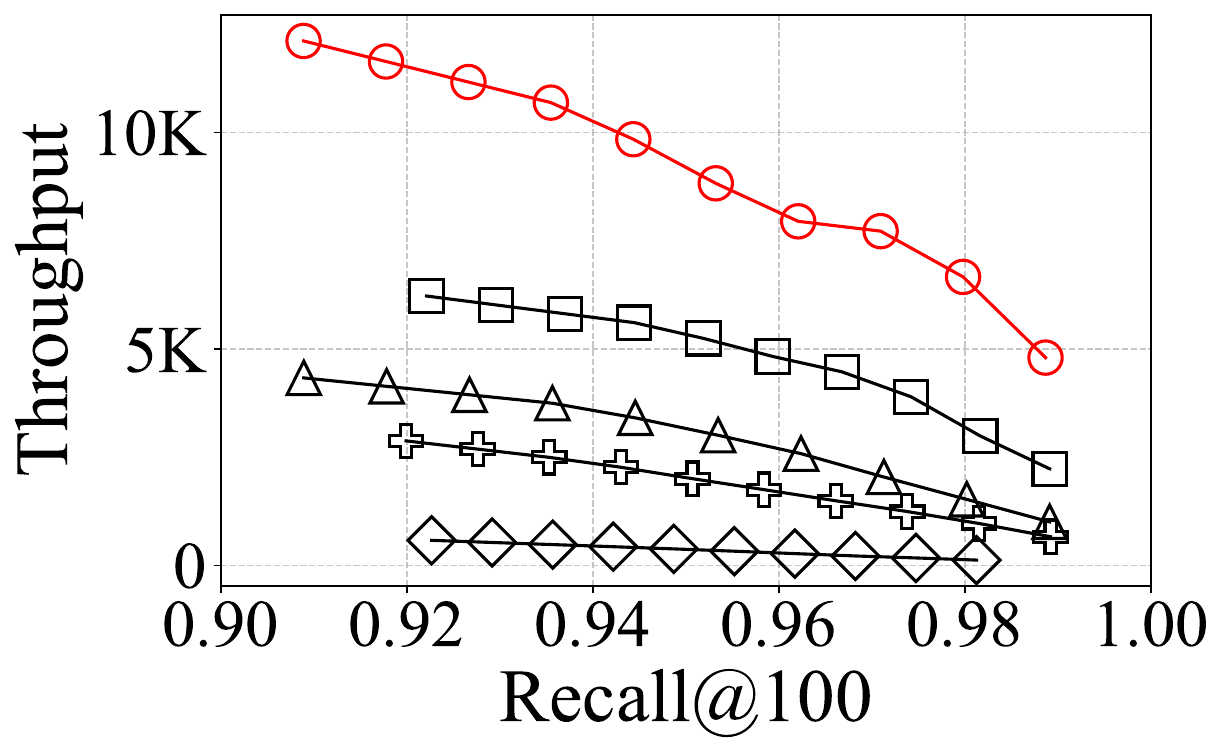}
  &  
  \includegraphics[width=0.45\columnwidth]{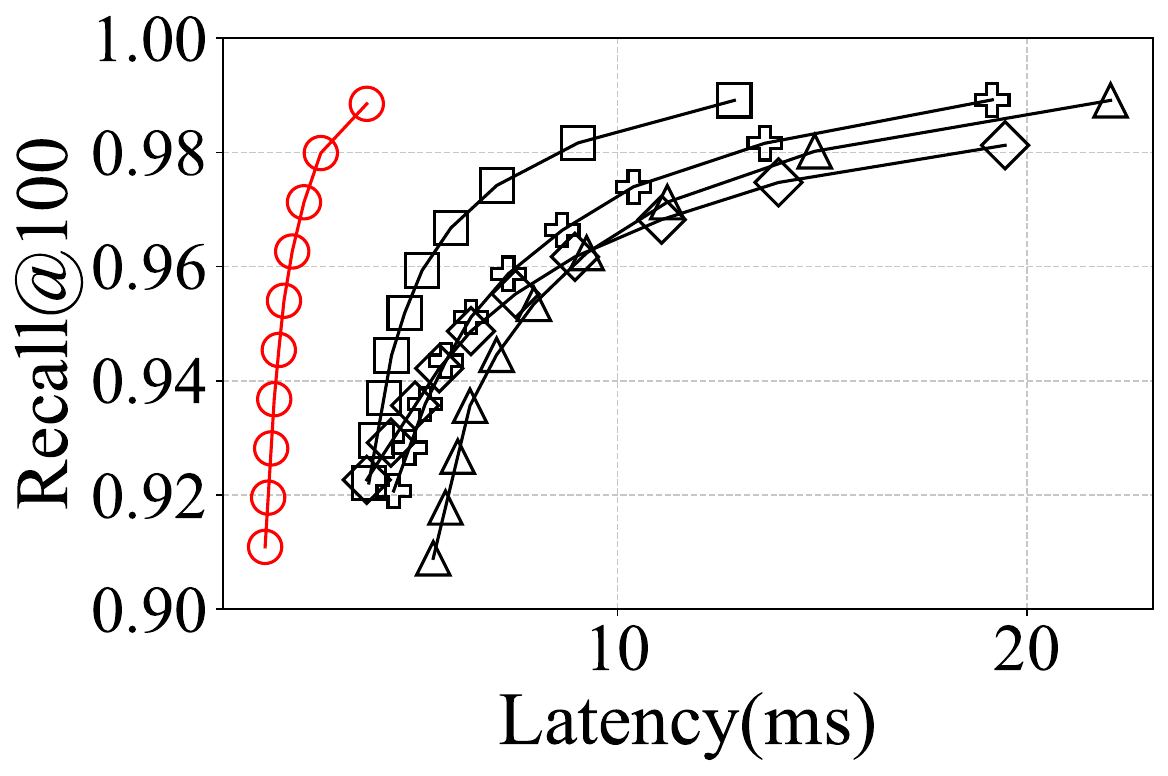}
   \\
   (a) Throughput with K=100 & (b) Latency with K=100
  \\
  \includegraphics[width=0.45\columnwidth]{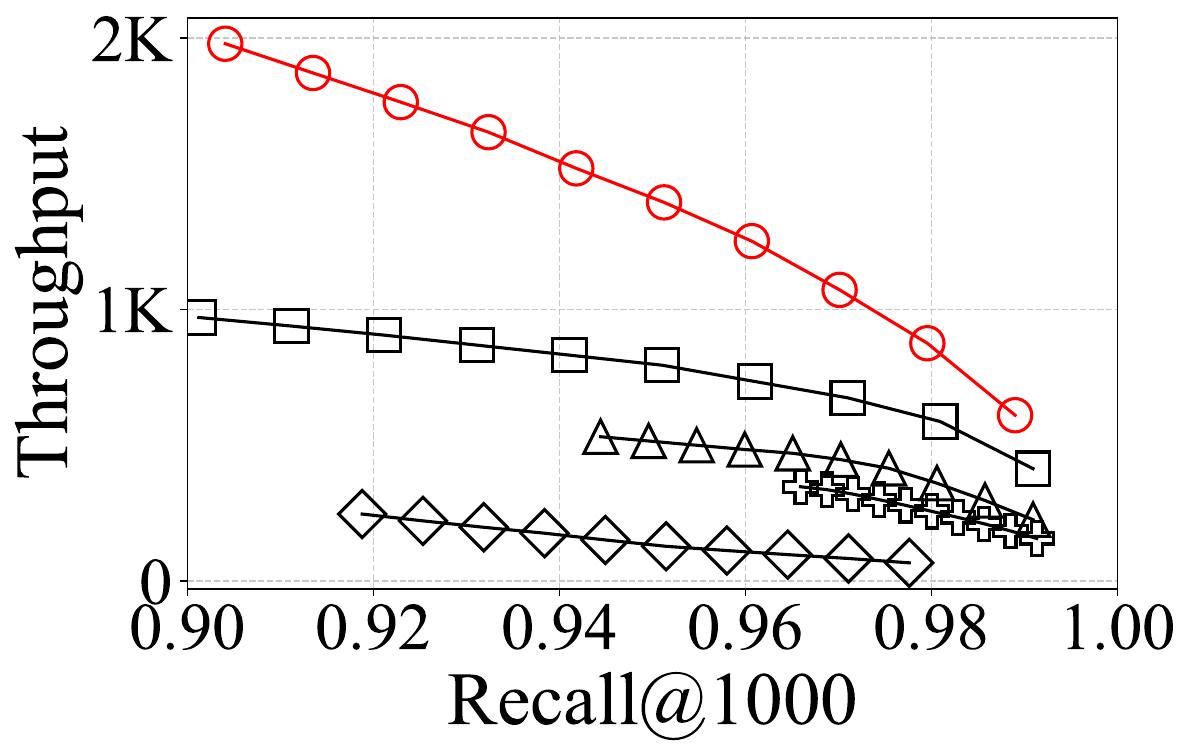}
  &  
  \includegraphics[width=0.45\columnwidth]{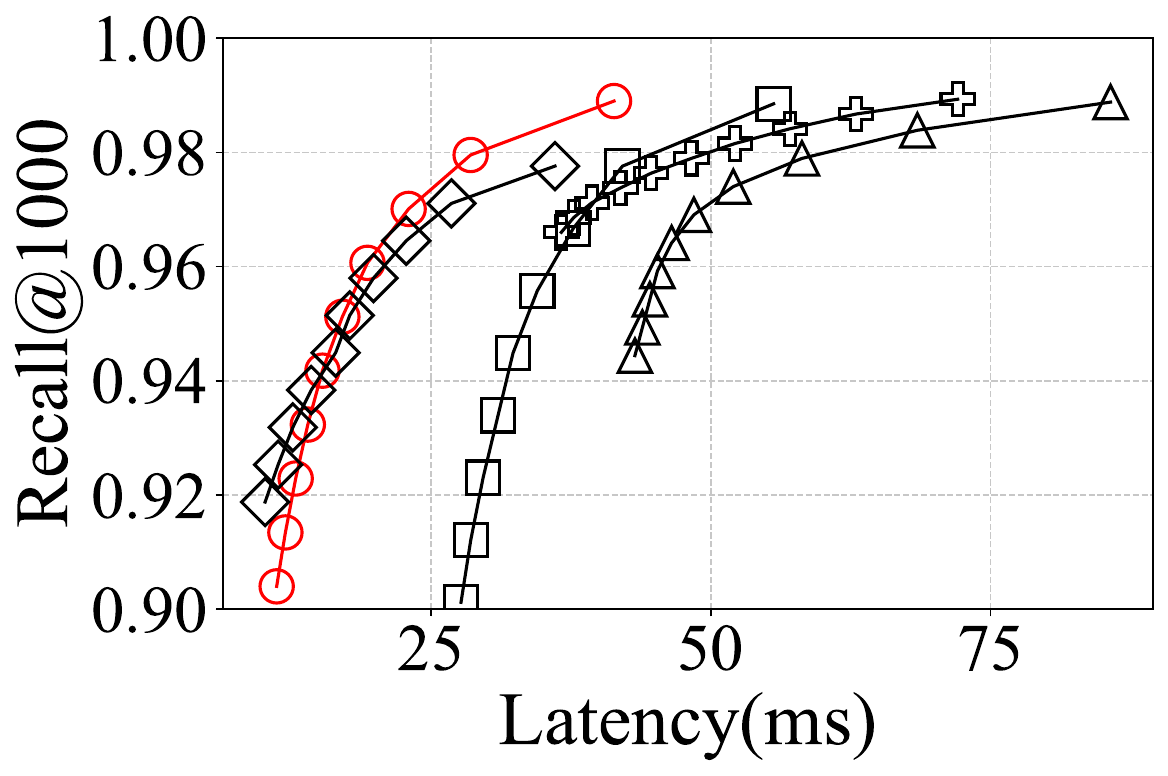}
   \\
   (c) Throughput with K=1000 & (d) Latency with K=1000
\end{tabular}
\trim 
\caption{Performance evaluation by varying K, Cohere} \label{fig:varyingk}
\trim 
\end{figure}

\stitle{Varying K at Recall@K}
We evaluate the performance of all on-disk systems by varying K in this experiment.
Figure~\ref{fig:varyingk} shows the measured result of Recall@100 and Recall@1000 on Cohere, respectively.
We omit the experimental results on other datasets as they share similar trend.
It is no doubt our \sysname{} is the overall winner among all these compared methods with different Recall@Ks.
For example, the throughput of \sysname{} is nearly 2x of Starling, and 4x of DiskANN when Recall@100 is 0.95.
Interestingly, the latency of cluster-based index SPANN is better than existing graph-based indexes (e.g., DiskANN, Starling, PipeANN) when K is 1000, see Figure~\ref{fig:varyingk}(d).
The core reason is that SPANN only needs 332 I/Os to visit clusters and compute a high quality top-1000 result set, 
but the on disk graph-based methods incur thousands I/Os (e.g., DiskANN uses 1115 I/Os) to achieve the similar recall at top-1000 result set.
However, our \sysname{} is the only on disk graph-based index method, which performs better than SPANN w.r.t latency when the Recall@1000 is larger than 0.95.
This is because the computation cost of \sysname{} is significantly smaller than SPANN to compute top-1000 results with high recall (i.e., Recall@1000 $\geq$ 0.95) as SPANN computes all the candidates data vectors in these accessed clusters and \sysname{} obtains a tight yet high quality candidate set via the fast-computed approximate distances.

\rvsn{
\stitle{Index construction cost}
Table~\ref{tab:Index_construction_times} presents the Vamana index construction cost of the compared on-disk graph-based index systems on the 5 datasets. 
Comparing with DiskANN, Starling, and PipeANN, \sysname{} introduces extra overhead for PCA transformation and data clustering during index construction.
However, it significantly reduces cost during quantization phase as it only quantizes the  principal components of data vector, but the other index methods need to quantize the original vectors.
For example, the dimensionality of MSMARCO is 1,024, which is significantly larger than the used dimension of principal component 256.
Interestingly, the gained benefit is slightly outweighs the introduced overhead.
Thus, the index construction cost of \sysname{} is slightly faster than its of existing on-disk graph-based index systems on all datasets.
}

\begin{table}
\small
\centering
\caption{Index construction cost (minutes)} \label{tab:Index_construction_times}
\trim 
\resizebox{\columnwidth}{!}{
\begin{tabular}{|l|c|c|c|c|c|}
    \hline
     & {GIST1M} & {Cohere} & {BigCode} & {DPR} & {MSMARCO} \\ \hline \hline
    DiskANN & 22.8 & 68.8 & 51.7 & 1,464.5 & 1,532.4 \\ \hline
    Starling & 18.4 & 70.6 & 71.0 & 1,420.5 & 1453.7 \\ \hline
    PipeANN & 19.4 & 78.2 & 68.3 & 1,472.8 & 1,448.3 \\ \hline
    \sysname{} & \textbf{14.2} & \textbf{52.3} & \textbf{49.8} & \textbf{1,532.4} & \textbf{1,436.4} \\ \hline
\end{tabular}
}
\end{table}

\begin{table}
\small
\centering
\caption{Index size (GB)}
\trim 
\resizebox{\columnwidth}{!}{
\begin{tabular}{|c|c|c|c|c|c|}
    \hline
     & {GIST1M} & {Cohere} & {BigCode} & {DPR} & {MSMARCO} \\ \hline \hline
    DiskANN & 7.7 & 39 & 40 & 386 & 867 \\ \hline
    Starling & 7.7 & 39 & 40 & 386 & 867 \\ \hline
    PipeANN & 7.7 & 39 & 40 & 386 & 867 \\ \hline
    \sysname{} & \textbf{7.7} & \textbf{78} & \textbf{80} & \textbf{771} & \textbf{867} \\ \hline
\end{tabular}
}
\label{tab:Disk_cost}
\end{table}

\stitle{Index size}
Table~\ref{tab:Disk_cost} reports the on-disk index size of all compared on-disk graph-based index methods on five datasets.
As expected, \sysname{} is typically larger than other on-disk graph-based methods as it stores quantized codewords in the SIMD-friendly layout.
In particular, the node size of \sysname{} is $S_{node} = 4d + 4R + R(\frac{d_{PCA}}{8}+12)$ bytes, where $d$, $R$, and $d_{PCA}$ are the dimensionality of data vectors, the out-degree of graph, and the dimensionality of principal components.
However, the node size of DiskANN (resp. Starling and PipeANN) is  $S_{node} = 4d + 4R + 4$ bytes.
Consequently, \sysname{} exhibits the largest index size on Cohere, BigCode, and DPR.
However, on GIST1M and MSMARCO, \sysname{} has the same size as the other methods as \sysname{} effectively utilizes the space wasted in existing methods due to I/O alignment padding.

\begin{figure}
\small
\centering
    \includegraphics[width=0.90\linewidth]{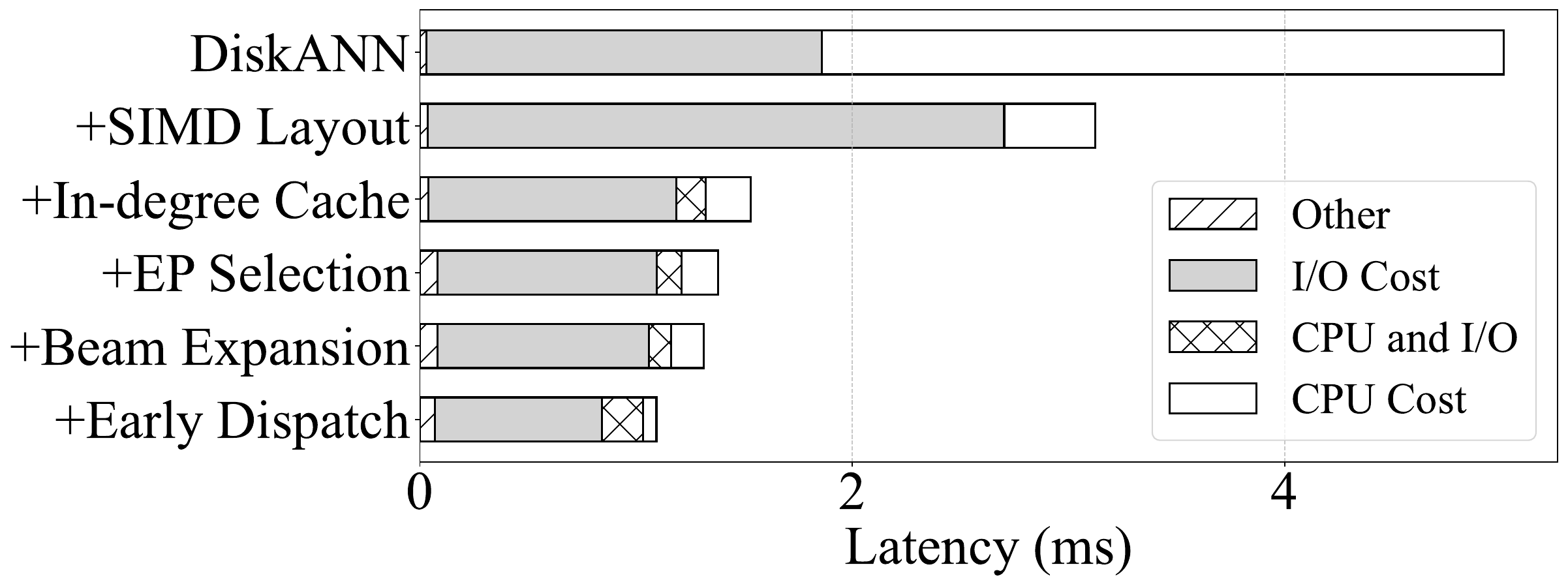}
    \\
    (a) Latency evaluation on GIST1M
    \\
    \vspace{2mm}
    \includegraphics[width=0.94\columnwidth]{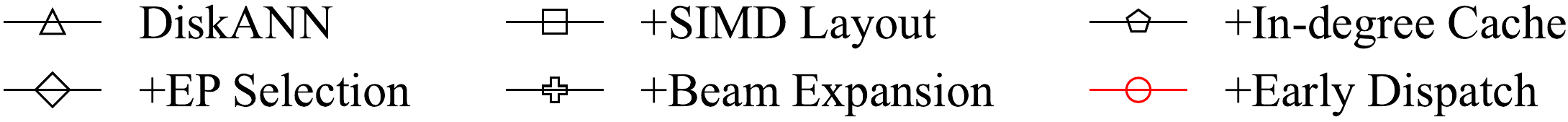}%
    \\
    \begin{tabular}{cc}
    \includegraphics[width=0.45\columnwidth]{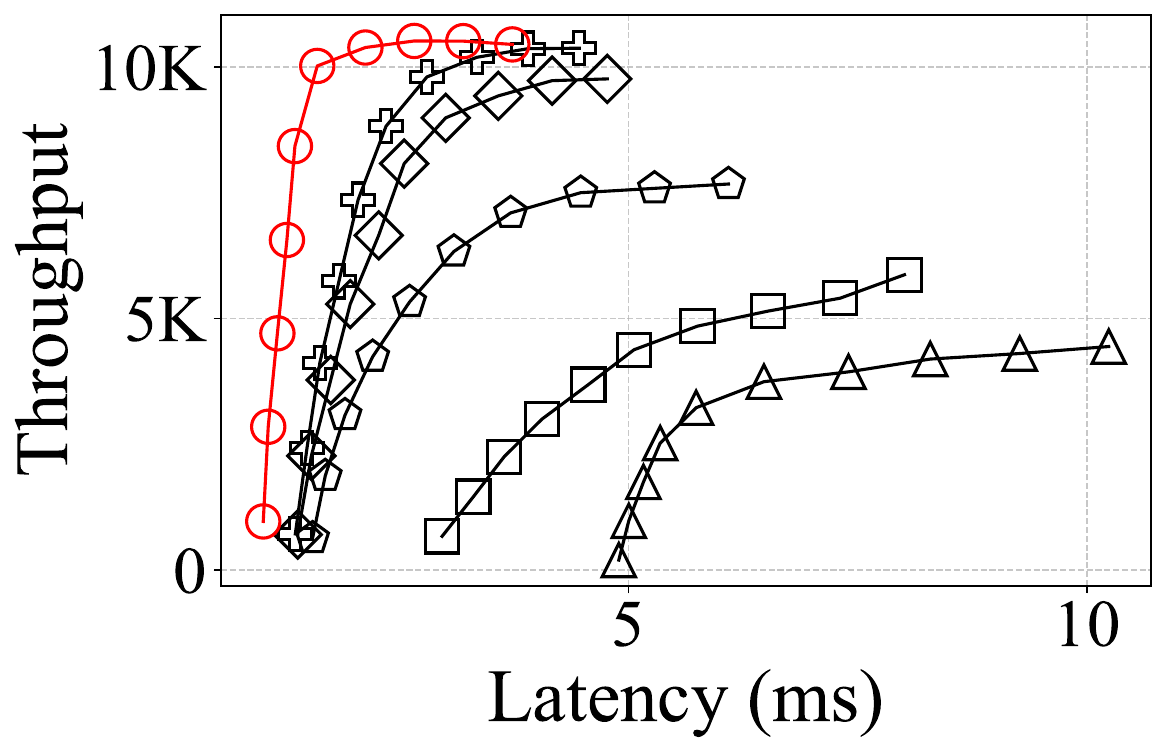}
    &  
    \includegraphics[width=0.45\columnwidth]{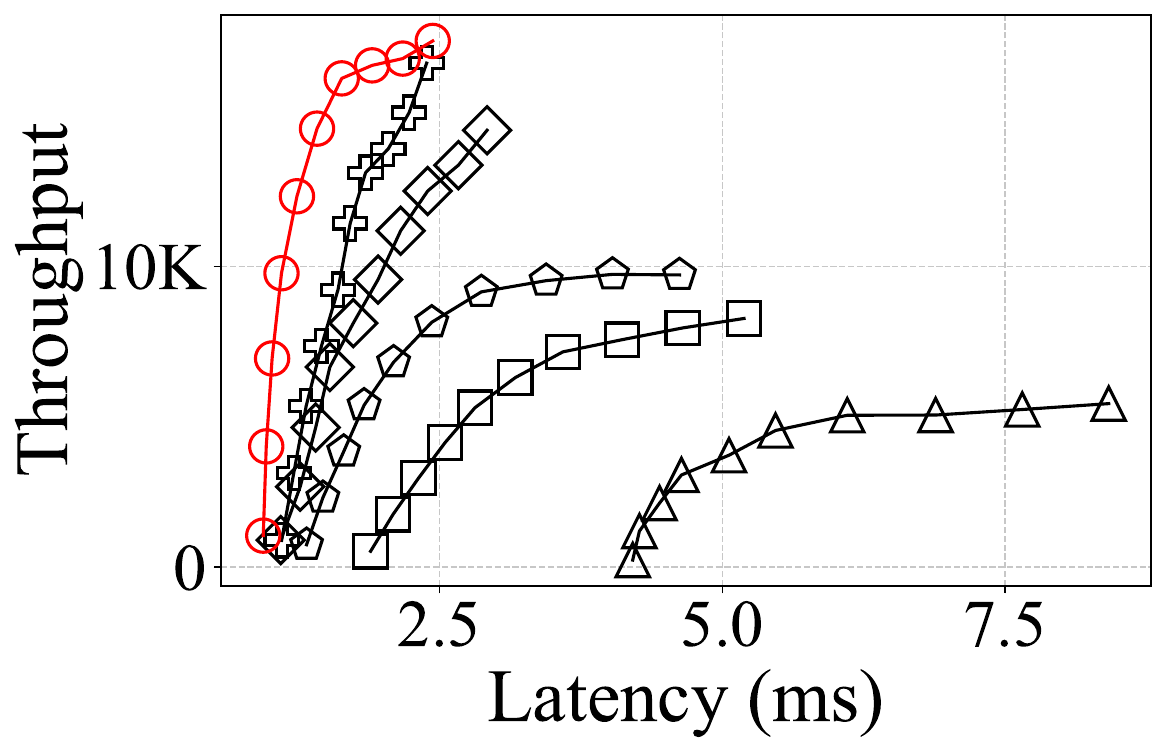}
     \\
    (b) GIST1M & (c) Cohere
    \end{tabular}
    \trim 
    \caption{\sysname{} ablation study, Recall@10=0.95} \label{fig:exp-ablation-experiment}
    \trim \trim 
\end{figure}

\stitle{Memory footprint}
Table~\ref{tab:Memory_cost} presents the peak memory footprint of each on-disk graph-based index system during the search phase.
The peak memory consumption is measured by using the \textbf{psutil} library~\cite{psutil}.
The results confirm that \sysname{} effectively utilizes the available memory budget.
In particular, the SIMD-friendly layout of \sysname{} eliminates the need to store a full set of compressed vectors in memory, and \sysname{} utilized the available memory budget by proposing in-degree based node cache scheme, which significantly reduced the I/O cost, see the third bar in Figure~\ref{fig:exp-ablation-experiment}(a).

\begin{table}
\small
\centering
\caption{Peak memory footprint during search phase} \label{tab:Memory_cost}
\trim 
\resizebox{\columnwidth}{!}{
\begin{tabular}{|c|c|c|c|c|c|}
    \hline
     & {GIST1M} & {Cohere} & {BigCode} & {DPR} & {MSMARCO} \\     \hline \hline
    DiskANN & 518MB & 5.0GB & 4.9GB & 77GB & 108GB \\ \hline
    Starling & 797MB & 8.1GB & 7.7GB & 78GB & 112GB \\ \hline
    PipeANN & 734MB & 9.7GB & 8.0GB & 78GB & 110GB \\ \hline
    \sysname{} & \textbf{769MB} & \textbf{7.7GB} & \textbf{7.7GB} & \textbf{89GB} & \textbf{98GB} \\ \hline
\end{tabular}
}
\end{table}

\rvsn{
\subsection{Effectiveness Study}\label{sec:exp-effect}
We conduct effectiveness study on \sysname{} in this section.

\stitle{Ablation study}
To verify the effectiveness of \sysname{}, we measure the latency of \sysname{} by including each proposed technique step by step.
Figure~\ref{fig:exp-ablation-experiment}(a) illustrates the measured latencies on GIST1M with Recall@10=0.95.
As analyzed in Section~\ref{sec:lasercaching}, the SIMD-friendly layout of \sysname{} successfully addresses the compute-bound of DiskANN by slightly introducing I/O overheads, see the second bar in  Figure~\ref{fig:exp-ablation-experiment}(a).
With SIMD-friendly layout, the in-degree node cache scheme reduces almost half of the I/O cost and overlaps computation and I/Os, as the third bar shown in Figure~\ref{fig:exp-ablation-experiment}(a).
The overall latency of \sysname{} is further optimized by the proposed cluster-based entry point selection method, adaptive beam expansion strategy and early dispatch idea, as the last three bars shown in Figure~\ref{fig:exp-ablation-experiment}(a).

We next evaluate the effectiveness of each proposed technique in \sysname{} by measuring the corresponding throughput-latency curves.
In particular, we tune the candidate list size parameter ($search\_L$) to maintain the Recall@10 at 0.95.
For each curve, we vary the number of used threads from 1 to 48.
The experimental results on GIST1M and Cohere are plotted in Figures~\ref{fig:exp-ablation-experiment}(b) and (c), respectively.
It is clear that each proposed technique moves the throughput-latency curve of \sysname{} closer to the top-left corner, which means \sysname{} achieves the higher throughput and the lower latency by integrating the proposed techniques one by one.
One interesting observation is that the early dispatch method on \sysname{} mitigates the impact of long-tail I/O in multi-thread setting, as the red curves shown in Figures~\ref{fig:exp-ablation-experiment}(b) and (c).

\stitle{Effect of entry point selection}
We investigate the effectiveness of the cluster-based entry point selection method of \sysname{} by comparing it with the graph-based entry point selection method in~\cite{wang2024starling, guo2025achieving}.
In particular, we replace the cluster-based method by the existing graph-based method in \sysname{}.
Figure~\ref{fig:exp-ep-selection}(a) demonstrates that the cluster-based method of \sysname{} achieves an approximate 10\% latency reduction over the graph-based method.
The performance gain is from that the cluster-based method only computes 300 candidates but the graph-based method computes 10,000 nodes in the in-memory navigation graph to identify the top-1 entry point for ANNS query processing on GIST1M.
We next measure the mean I/Os during query processing to quantify the quality of the identified entry points by both methods.
As shown in Figure~\ref{fig:exp-ep-selection}(b), the mean I/Os curve of the cluster-based method is slightly larger than that of the graph-based method at the same recall.
It confirms the effectiveness of our cluster-based method as it visits few nodes (i.e., 300 nodes) and achieves comparable quality of graph-based method, which visits many more nodes (i.e., 10,000 nodes).

\begin{figure}
\small
\centering
    \begin{tabular}{cc}
        \includegraphics[width=0.45\columnwidth]{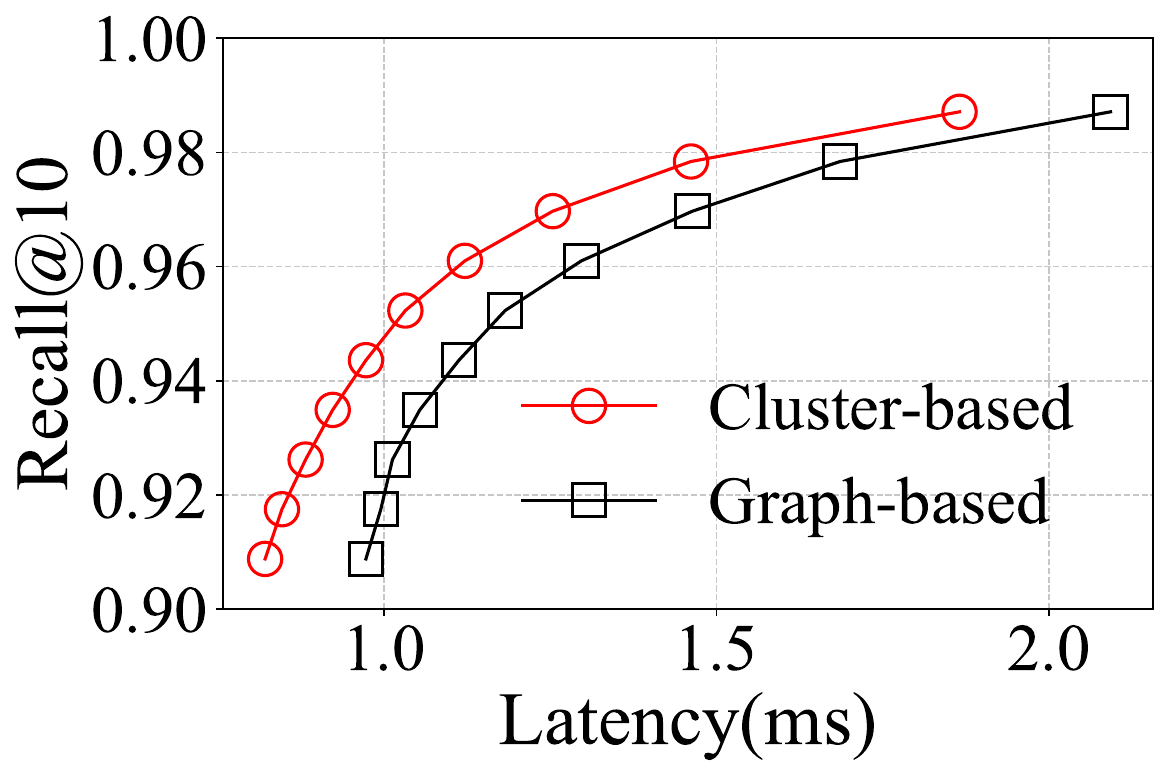}
        &
        \includegraphics[width=0.45\columnwidth]{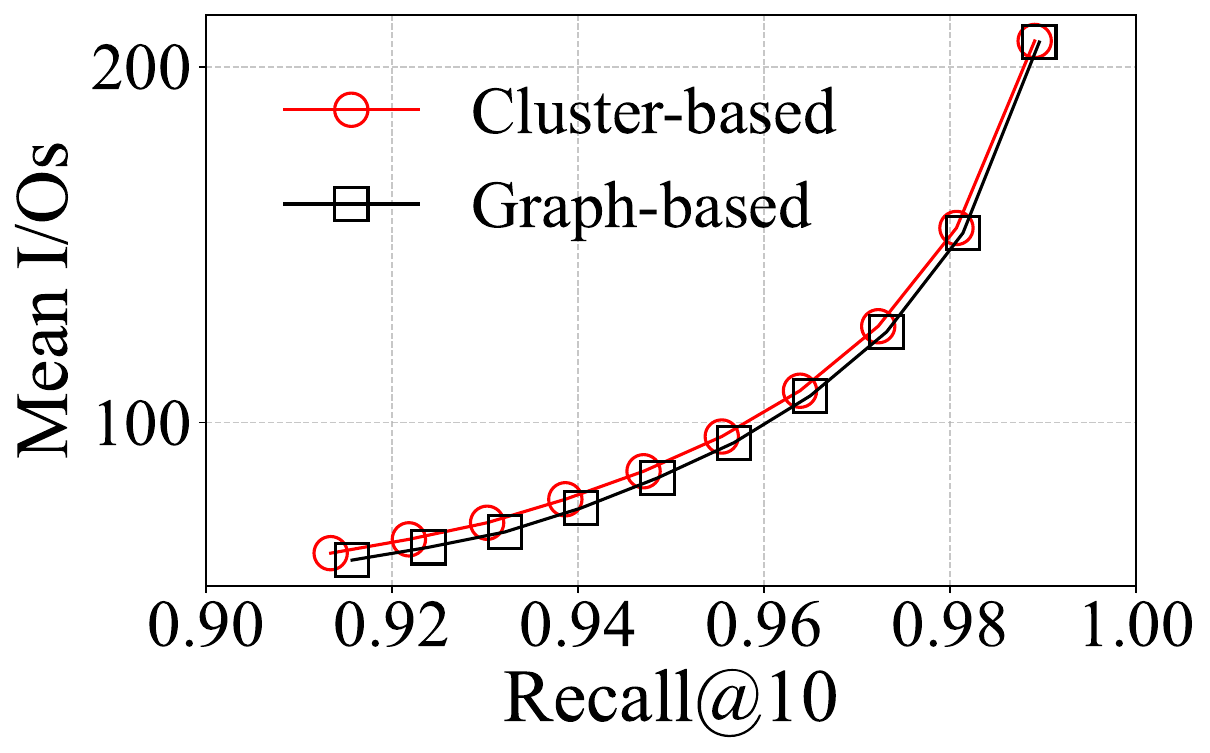}
        \\
        (a)  Latency & (b) Mean I/Os
\end{tabular}
\trim 
\caption{Effect of entry point selection, GIST1M} \label{fig:exp-ep-selection}
\trim 
\end{figure}

\begin{figure}
\small
\centering
\begin{tabular}{cc}
    \includegraphics[width=0.47\columnwidth]{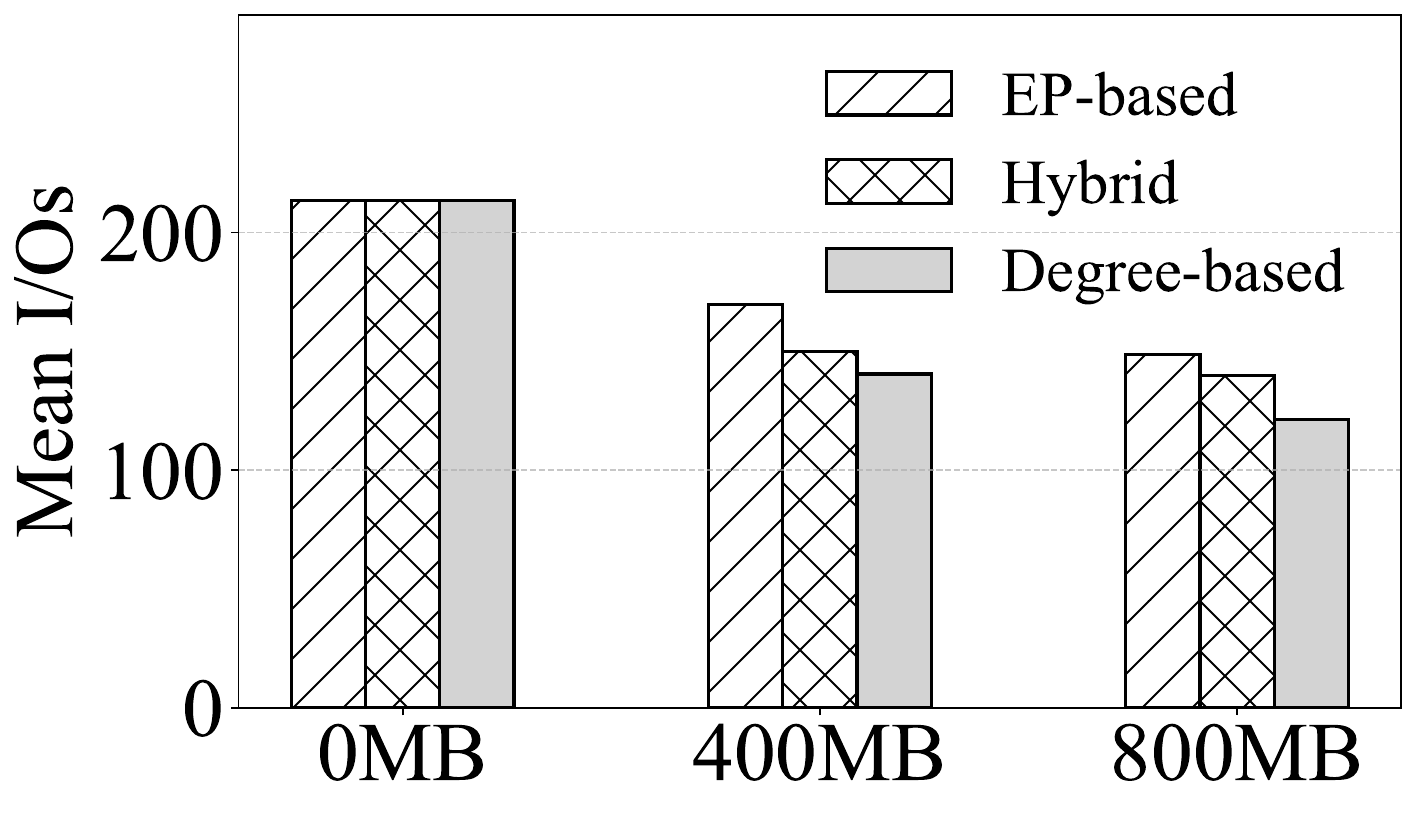}
    &  
    \includegraphics[width=0.47\columnwidth]{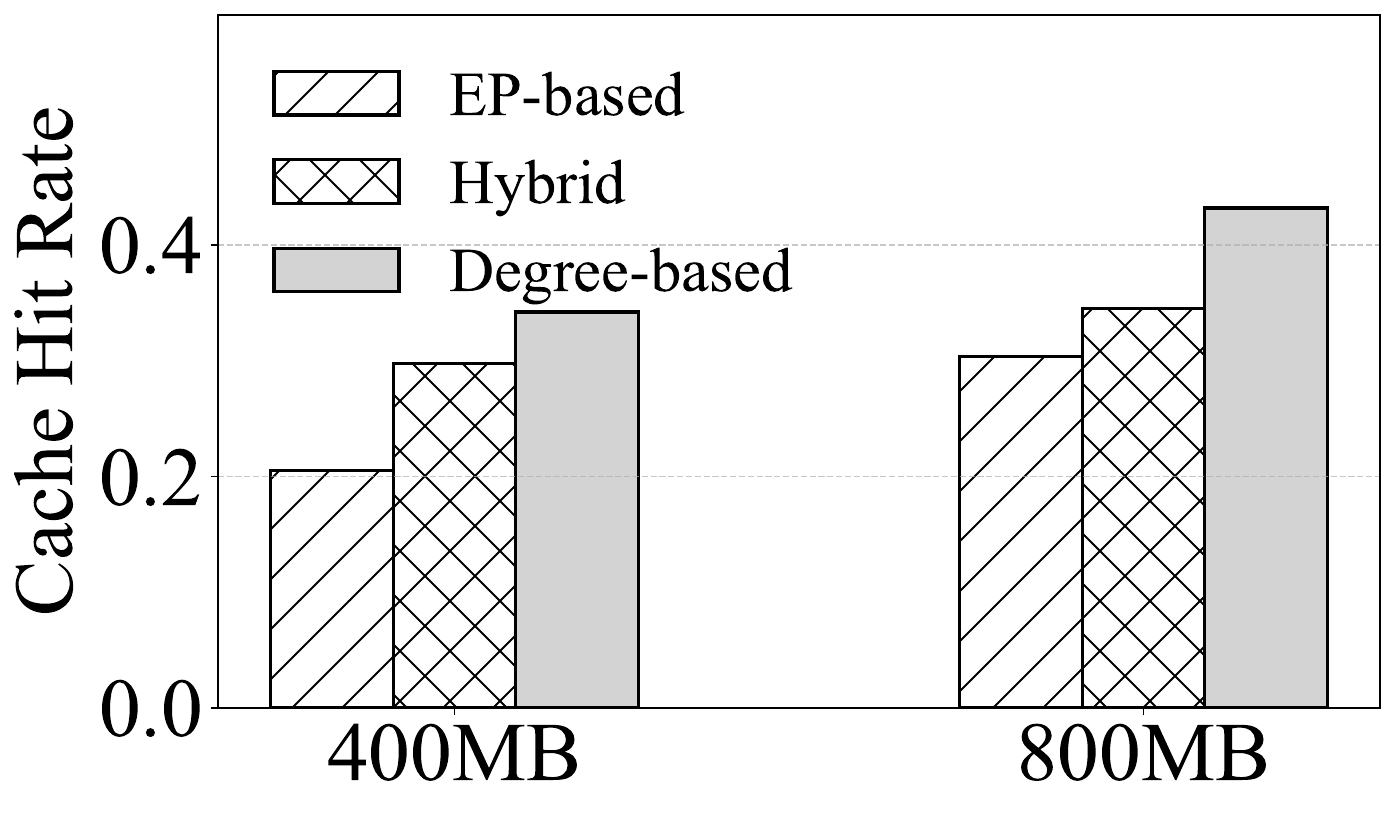}
    \\
    (a) Mean I/Os & (b) Cache hit ratio
    \\
\end{tabular}
\trim 
\caption{Effect of cache scheme, Recall@10=0.95 on GIST1M}
\label{fig:exp-caching-strategy}
\trim
\end{figure}

\stitle{Effect of cache scheme}
To fully utilize the available memory budget, different cache schemes have been proposed in existing on-disk graph-based solutions.
In particular, DiskANN~\cite{jayaram2019diskann} caches the frequently accessed entry points and their neighbors. 
The hybrid cache scheme is devised in~\cite{zhou2025govector} which combines static cache (with 20\% budget) and dynamic cache (with 80\%  budget). The static cache stores the same content as DiskANN and the dynamic cache employs Least Frequently Used (LFU) policy to dynamically update the cache budget by the visited nodes during query processing.
We compare these existing cache schemes against our proposed in-degree based cache scheme for \sysname{} in this experiment.
Specifically, we replace the degree-based cache scheme of \sysname{} by (i) EP-based cache scheme in~\cite{jayaram2019diskann} and (ii) hybrid cache scheme in~\cite{zhou2025govector}, respectively.
The tested cache budgets are  0\% (0MB), 5\% (400MB) and 10\% (800MB) of the size of GIST1M.
Figures~\ref{fig:exp-caching-strategy}(a) and (b) reported the mean I/Os and cache hit ratio of tested cache schemes.
We can observe: (i) the larger cache budget, the lower mean I/Os and the higher cache hit ratio of all cache schemes;
and (ii) in-degree-based node cache strategy in \sysname{} consistently outperforms the alternative cache schemes (e.g., EP-based and Hybrid).

\stitle{Effect of quantization method}
In this experiment, we investigate the effect of different quantization methods on the performance of \sysname{} on GIST1M.
In particular, we test 4 different quantization methods on \sysname{}: (i) PQ with 240 bytes, (ii) PQ with 48 bytes,
(iii)  PCA+PQ with 48 bytes, and (iv) PCA+RabitQ with 48 bytes, which is the default setting of \sysname{}.
Figure~\ref{fig:exp-pca-quant} reports the mean I/Os and throughput of different quantization methods on \sysname{} at different Recall@10.
Firstly, the mean I/Os of PQ (48) is the largest as it severely loses precision due to high compression ratio.
Secondly, PQ (240) has the smallest mean I/Os among all these 4 tested methods but its throughput is obvious below our PCA based methods.
Thirdly, PCA-based methods are better than PQ only method and PCA+RabitQ(48) slightly outperforms PCA+PQ(48), which demonstrates (i) the generality of our PCA-based method and (ii) the efficiency of RabitQ.

\begin{figure}
\small
\centering
\begin{tabular}{cc}
    \includegraphics[width=0.45\columnwidth]{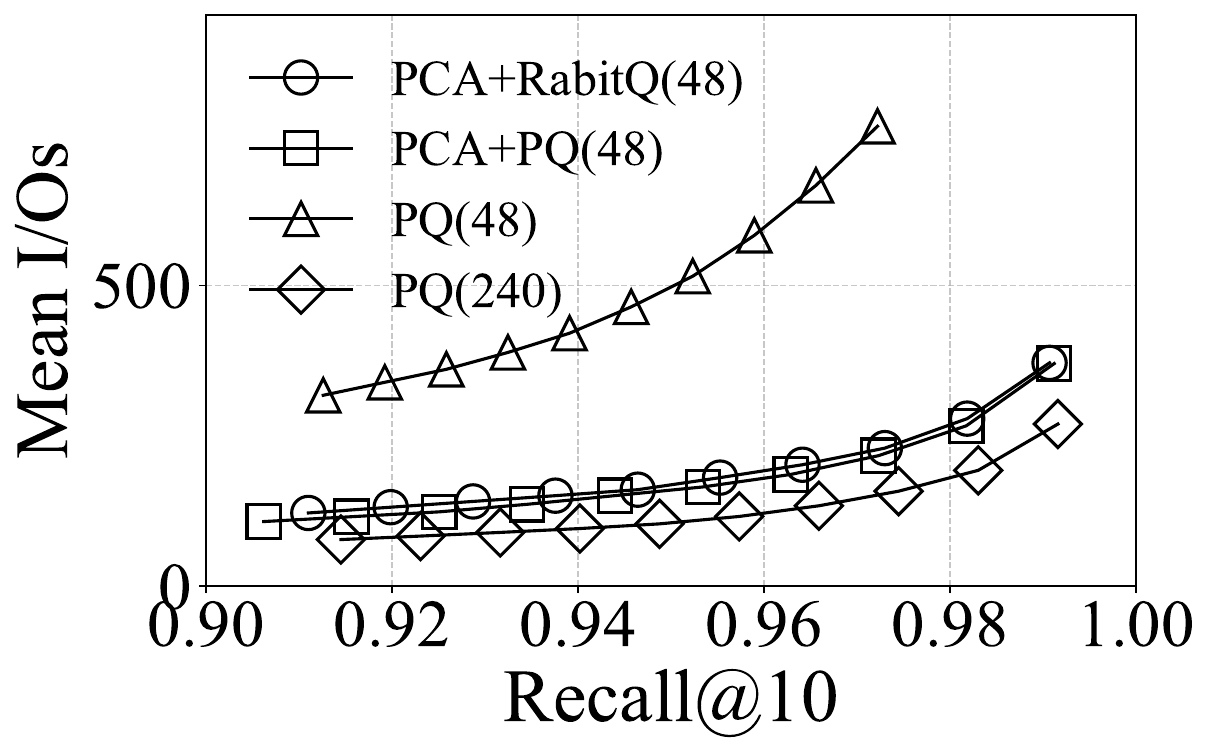}
    &
    \includegraphics[width=0.45\columnwidth]{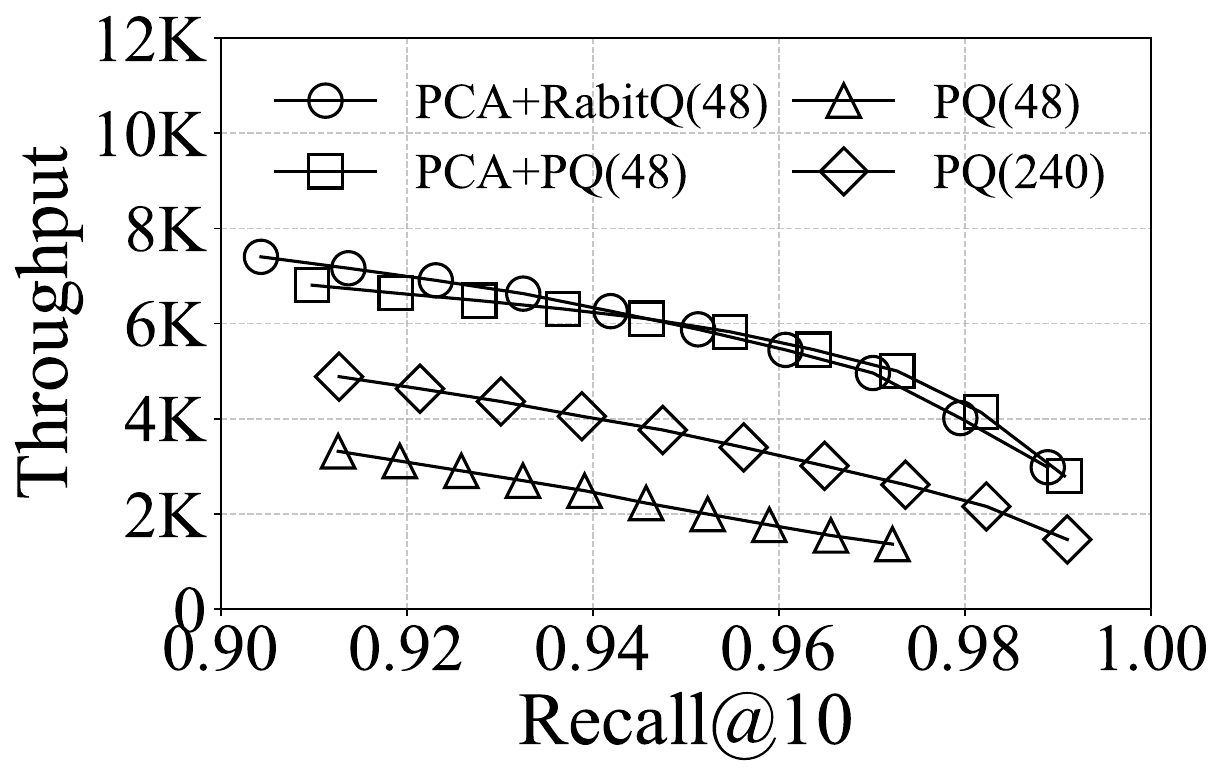}
    \\
    (a) Mean I/Os & (b) Throughput
\end{tabular}
\trim 
\caption{Effect of quantization method, GIST1M} \label{fig:exp-pca-quant}
\trim 
\end{figure}

\begin{figure}
\small
\centering
\begin{tabular}{cc}
    \includegraphics[width=0.45\columnwidth]{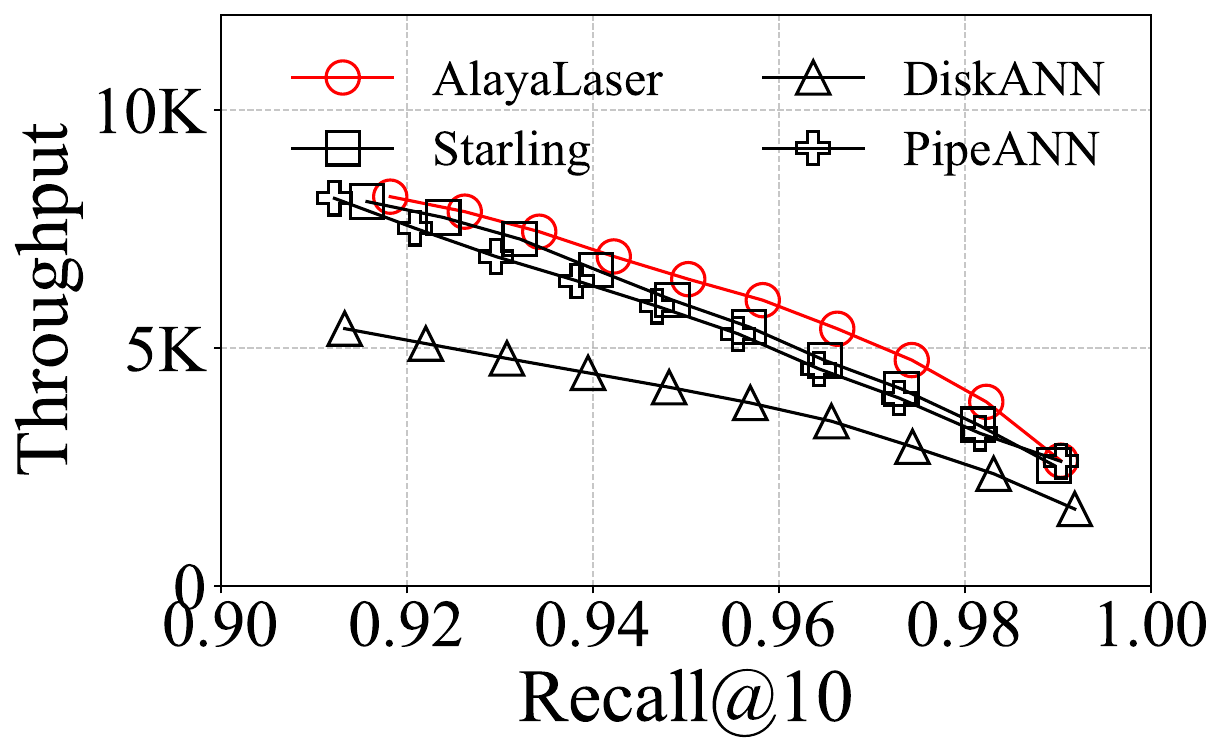}
    &  
    \includegraphics[width=0.45\columnwidth]{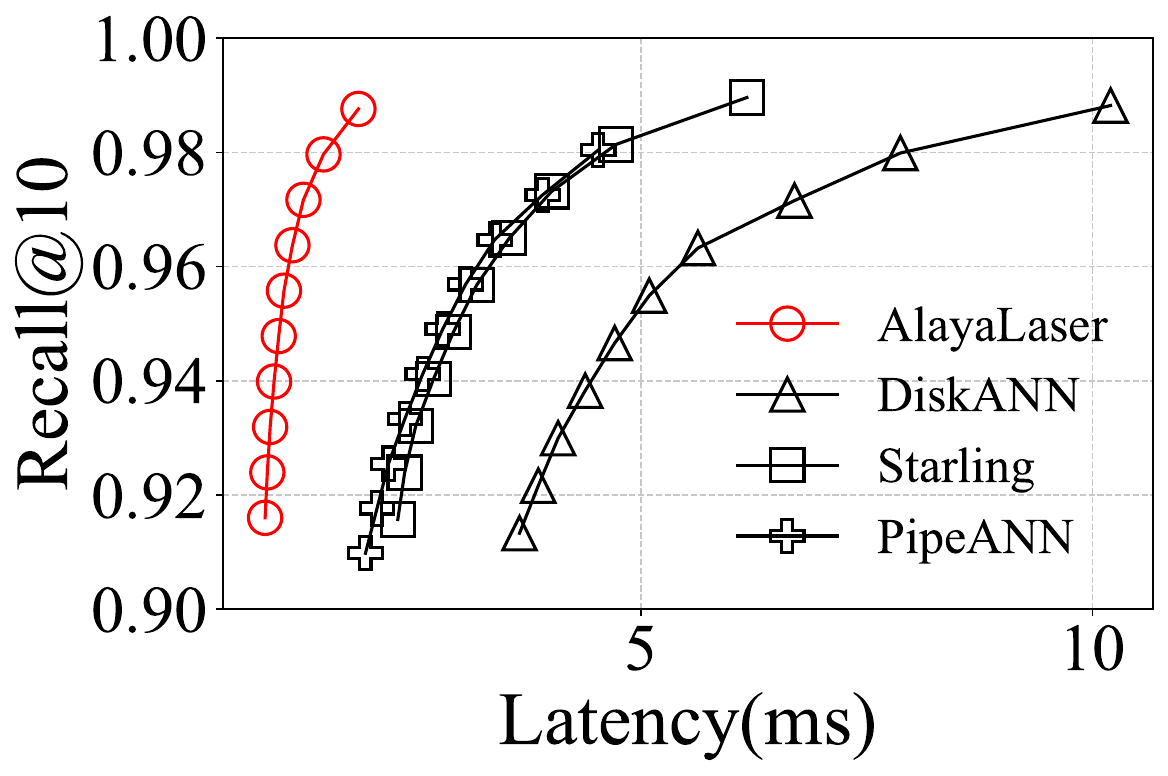}
 \\
      (a) Throughput & (b) Latency
\end{tabular}
\trim 
\caption{Effect of hardware configuration, GIST1M} \label{fig:exp-sensitivity-hardware}
\trim 
\end{figure}

\stitle{Effect of hardware configurations}
We agree that the roofline model analysis is sensitive to the CPU and SSD configurations.
To evaluate the effectiveness of our designs in \sysname{}, we compare all on-disk graph-based index systems on a machine with small I/O bandwidth in this experiment.
In particular, the I/O bandwidth is 3.7GB/s (i.e., we only use a single 1.92 TB SAMSUNG PM9A3 NVMe SSD), which is 11.1GB/s in our default setting.
As depicted in Figures~\ref{fig:exp-sensitivity-hardware}(a) and (b), \sysname{} outperforms all existing on-disk graph-based index systems in terms of throughput and latency, respectively.
Thus, it is safe to conclude \sysname{} enjoys high I/O efficiency, computational efficiency and search performance simultaneously, 
which does not rely on any specific hardware configurations.
We also agree that the performance improvement times of \sysname{} over existing on-disk graph-based index systems may be different with different hardware configurations.


{
\stitle{Effect of data updates}
To evaluate the robustness of \sysname{}, we divide GIST1M into two subsets: base set and insertion set.
\sysname{} constructs the on-disk graph-based index with the base set.
Then \sysname{} updates the index with the vectors in insertion set by following the same PCA matrices. 
The Vamana graph structure of \sysname{} is updated with the HNSW insertion strategy in~\cite{malkov2018efficient}.
Figure~\ref{fig:exp-data-updates} depicts the throughput and latency of \sysname{} in three cases: (i) rebuild index, 
(ii) 10\%$|\mathcal{D}|$ inserts, and (iii) 20\%$|\mathcal{D}|$ inserts.
Firstly, we recommend periodic rebuilding the index of \sysname{} to offer optimal performance, as rebuild index enjoys the best throughput and latency.
Secondly, \sysname{} is robust to data updates as the performance of 10\%$|\mathcal{D}|$ inserts and  20\%$|\mathcal{D}|$ inserts are slightly worse the rebuild index.
}

\begin{figure}
\small
\centering
\begin{tabular}{cc}
    \includegraphics[width=0.45\columnwidth]{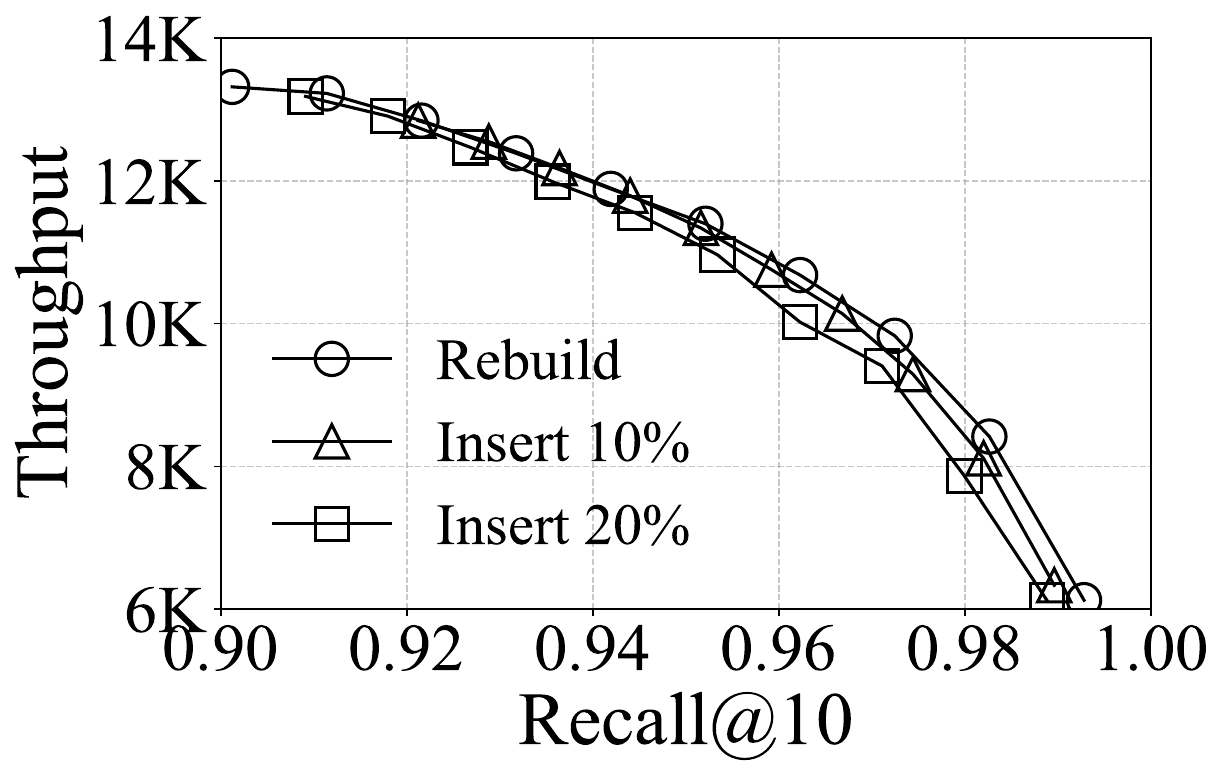}
    &  
    \includegraphics[width=0.45\columnwidth]{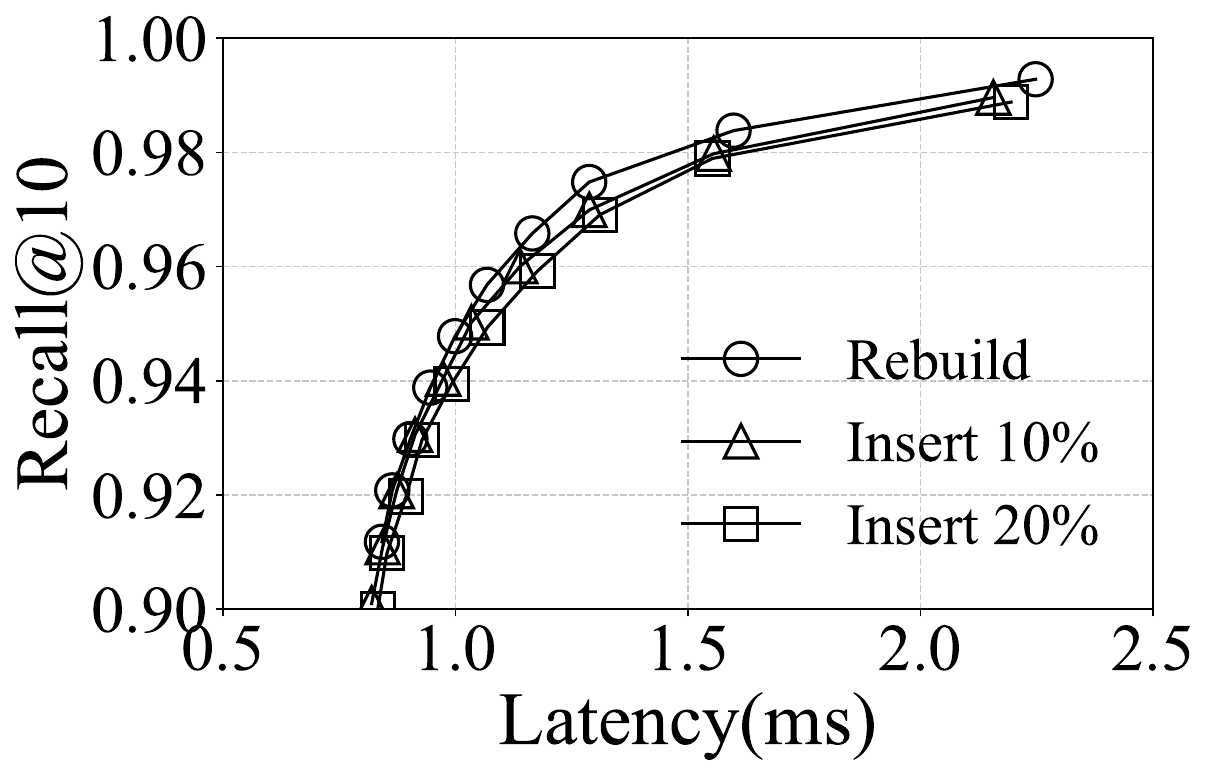}
    \\
    (a) Throughput & (b) Latency
\end{tabular}
\trim 
\caption{Effect of data updates, GIST1M} \label{fig:exp-data-updates}
\trim \trim 
\end{figure}

\stitle{Effect of OS cache}
We evaluate the effect of OS cache in Figure~\ref{fig:exp-impact-iosetup}.
Specifically, the effect of OS cache is disabled in our default setting.
In this experiment we enable it on \sysname{} and DiskANN by removing \textsf{O\_DIRECT} flag.
As illustrated in Figure~\ref{fig:exp-impact-iosetup}, both \sysname{} and DiskANN without OS cache perform better than with OS cache.
The major reason is that the OS cache is ineffective (i.e., low cache hit rate) due to the highly random access pattern of graph-based search on both \sysname{} and DiskANN.

\begin{figure}
\small
\centering
\begin{tabular}{cc}
    \includegraphics[width=0.45\columnwidth]{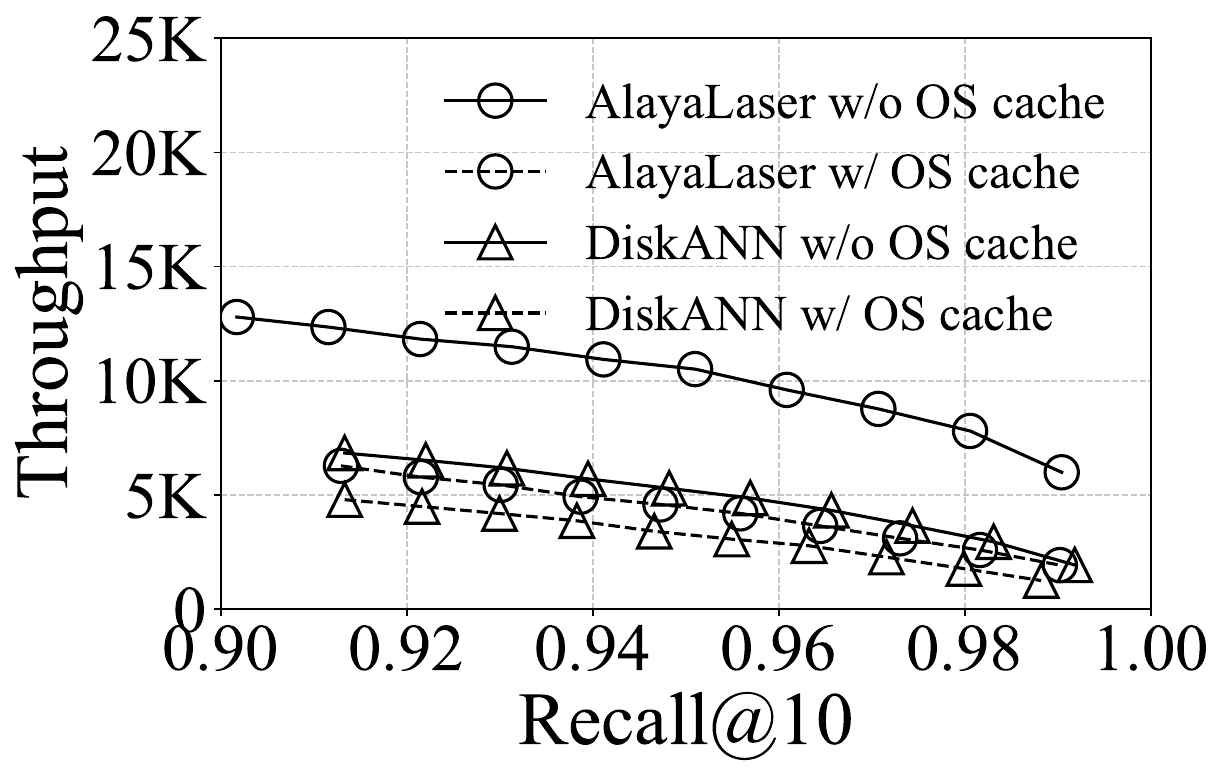}
    &
    \includegraphics[width=0.45\columnwidth]{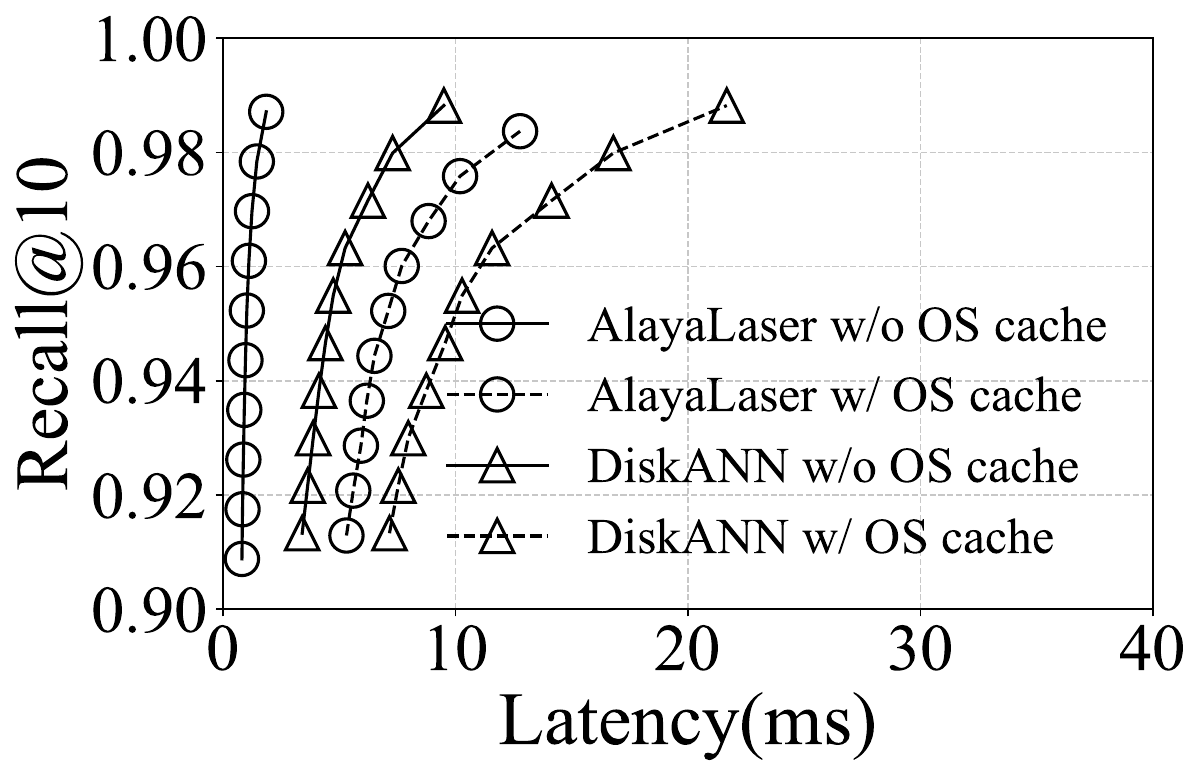}
     \\
     (a) Throughput & (b) Latency
\end{tabular}
\trim 
\caption{Effect of OS cache, GIST1M}
\label{fig:exp-impact-iosetup}
\trim \trim 
\end{figure}

}

\section{Related Work}\label{sec:related_work}  

\stitle{In-memory graph-based ANNS solutions}
In-memory graph-based index solutions, such as HNSW~\cite{malkov2018efficient}, NSG~\cite{fu2017fast}, and Vamana~\cite{jayaram2019diskann}, are the state-of-the-art solutions for high-performance approximate nearest neighbor search. 
These methods leverage various optimization strategies: HNSW enhances search speed through hierarchical small-world graphs, NSG ensures strong connectivity via monotonic search paths, and Vamana improves robustness with two-phase graph pruning. 
Other notable approaches, such as KGraph~\cite{kgraph} and DPG~\cite{li2019approximate}, focus on optimizing graph construction through neighborhood propagation and angle-based diversification, respectively, while SPTAG~\cite{ChenW18} adopts a divide-and-conquer strategy for better scalability. 
However, a fundamental limitation of these approaches is their requirement to store the entire index—comprising the graph structure and the original vectors—in main memory, which becomes a critical scalability bottleneck for large-scale high-dimensional vector data.

\stitle{On-disk ANNS solutions} 
Besides the graph-based indexing adopted in DiskANN~\cite{jayaram2019diskann}, Starling~\cite{wang2024starling} and PipeANN~\cite{guo2025achieving}, there are also solutions following a two-level architecture.
In particular, GRIP~\cite{zhang2019grip} and FusionANNS~\cite{tian2025towards} employ a two-stage ``preview and validate'' strategy: they first use in-memory quantized indexes to identify candidate vectors, and then use the original vectors stored on disk to rerank. While this approach ensures high final accuracy, the reranking step incurs significant random I/O costs, as fetching numerous small, non-contiguous vectors from an SSD leads to severe read amplification.
SPANN~\cite{chen2021spann} mitigates this overhead by clustering vectors and enabling sequential access to relevant clusters. 
However, maintaining high recall in SPANN still incurs high latency as it incurs substantial I/O accesses.


\stitle{Acceleration techniques} 
To accelerate vector similarity search, two approaches are widely used.
First, many studies compress vectors to reduce memory access costs.
The representative techniques include Product Quantization (PQ)~\cite{jegou2010product}, Scalar Quantization~\cite{aguerrebere2023similarity}, and RaBitQ~\cite{gao2024rabitq}.
Second, many methods have been studied, such as Quick ADC~\cite{andre2017accelerated}, Bolt~\cite{blalock2017bolt}, Quicker ADC~\cite{andre2019quicker}, and SymphonyQG~\cite{gou2025symphonyqg}, which utilized SIMD instructions for the distance computation acceleration
However, these techniques require excessive data storage and are specifically designed for in-memory solutions.





\section{Conclusion}\label{sec:conclusion}  

In this work, we revisited the design paradigm of on-disk graph-based index system for efficient large-scale, high-dimensional vector similarity search. 
We revealed a key paradigm shift, contrary to the prevailing wisdom, proving both theoretically and empirically that as vector dimensionality increases, the primary performance bottleneck transitions from disk I/O to CPU computation. 
To address this new compute-bound paradigm, we propose \sysname{},
which consists of a novel SIMD-friendly data layout and an optimized search strategy.
In particular, the novel SIMD-friendly index layout of \sysname{} first resolves the computational bottleneck through parallelized computation with SIMD instructions, then enables in-degree based node cache scheme in \sysname{}. 
The optimized search strategy addresses the re-emergent I/O bottleneck via a suite of optimization techniques, including cluster-based entry point selection, adaptive beam expansion strategy, and early dispatch mechanism. 
Extensive experiments demonstrate that the performance of \sysname{} not only significantly surpasses existing on-disk index systems but also, is comparable to or slightly better than that of in-memory index solutions, e.g., HNSWlib.
The promising directions for further work is (i) processing out-of-distribution queries on on-disk graph-based index system efficiently, 
and (ii) accelerating the search performance of on-disk index system on heterogeneous computing hardware.

\bibliographystyle{ACM-Reference-Format}
\bibliography{ref}

\end{document}